\documentclass[11pt,a4paper]{article}
\usepackage{jheppub}
\usepackage[T1]{fontenc}
\usepackage[utf8]{inputenc}
\usepackage[british]{babel}
\usepackage{mathtools,amssymb,amsthm,booktabs,array,xurl}
\usepackage[svgnames]{xcolor}
\usepackage{tikz}
\usepackage{placeins}
\usetikzlibrary{arrows.meta}
\makeatletter\renewcommand\@fpheader{}\makeatother
\renewcommand{\beforetochook}{\pagestyle{myplain}\pagenumbering{roman}%
 \begingroup\renewcommand{\baselinestretch}{1}\fontsize{10}{11.3}\selectfont}
\renewcommand{\afterTocRuleSpace}{\par\endgroup\bigskip}
\hypersetup{pdftitle={Doubly-spinning Black Saturns},pdfauthor={Jay Armas},colorlinks=true,
 linkcolor=MidnightBlue,citecolor=MidnightBlue,urlcolor=MidnightBlue}
\newcommand{\ancillarylink}[1]{\href{https://jacomearmas.org/research/BlackSaturns/}{#1}}
\newcommand{\dd}{\mathrm d}
\newcommand{\diag}{\operatorname{diag}}

\newcommand{\R}{\mathrm R}
\newcommand{\Hh}{\mathrm H}
\newcommand{\cG}{\mathcal G}
\newcommand{\cA}{\mathcal A}
\newtheorem{proposition}{Proposition}[section]
\numberwithin{equation}{section}
\allowdisplaybreaks[2]
\title{\bf Doubly-spinning Black Saturns}
\author[1,2,3,4]{Jay Armas}
\affiliation[1]{Institute for Theoretical Physics, University of Amsterdam, \\ 1090 GL Amsterdam, The Netherlands}
\affiliation[2]{Dutch Institute for Emergent Phenomena,\\
1090 GL Amsterdam, The Netherlands}
\affiliation[3]{Institute for Advanced Study, University of Amsterdam,\\
Oude Turfmarkt 147, 1012 GC Amsterdam, The Netherlands}
\affiliation[4]{Niels Bohr International Academy, The Niels Bohr Institute,\\
University of Copenhagen, Blegdamsvej 17, DK-2100 Copenhagen \O, Denmark}
\emailAdd{j.armas@uva.nl}
\abstract{We construct a six-parameter family of stationary,
asymptotically flat solutions of five-dimensional vacuum Einstein gravity
describing a spherical black hole surrounded by a black ring, with rotation in both independent planes.
In a specific regime both horizons can become extremal, thereby providing the first example of a regular asymptotically flat vacuum Einstein solution with two extremal disconnected horizons. We also show that while non-extremal black Saturns can be in thermodynamic equilibrium, the doubly extremal Saturn cannot. The solution admits the doubly spinning Myers-Perry, the Pomeransky-Sen'kov black ring, and the singly spinning black Saturn as regular limits. We also show agreement with earlier blackfold predictions for the phase diagram of these solutions.}

\renewcommand{\afterEmailSpace}{%
  \par\smallskip
  \begin{center}
    \includegraphics[width=.72\textwidth,height=4.6cm,keepaspectratio]{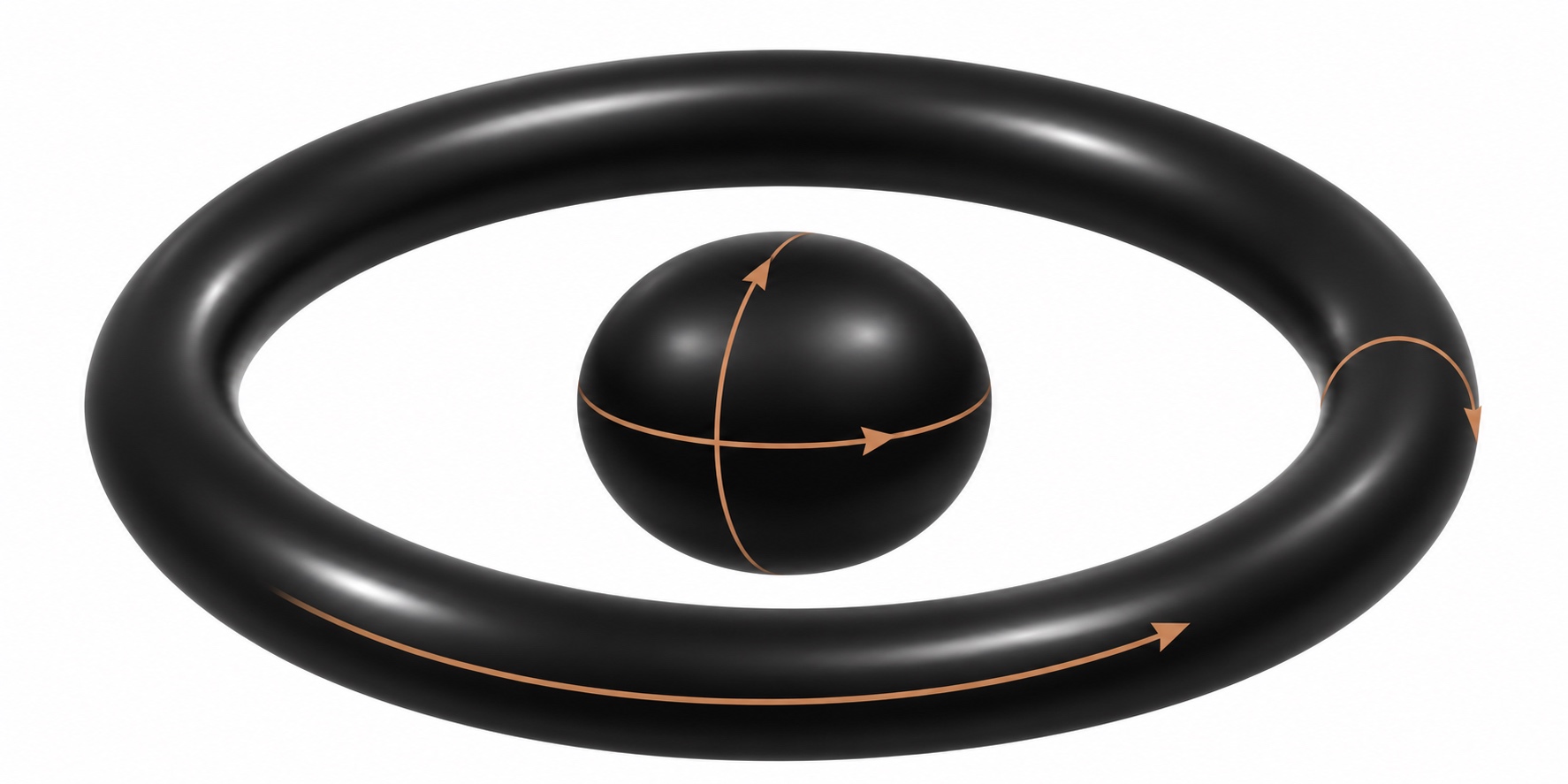}\\[-.2em]
  \end{center}
  \medskip
}
\begin{document}
\maketitle

\section{Introduction}
\label{sec:intro}

The space of black holes in higher-dimensional vacuum gravity is richer
than its four-dimensional counterpart. Independent rotation planes and
non-spherical horizon topology allow geometrically distinct solutions
to carry the same mass and angular momenta. In five dimensions, the
Myers--Perry black hole~\cite{MP}, the singly spinning black
ring~\cite{ER} and the doubly spinning Pomeransky--Sen'kov
ring~\cite{PS} provide explicit examples of this richer structure.

Disconnected horizons enlarge this space further. Black Saturn consists
of a spherical black hole surrounded by a black ring. Its singly spinning
solution exhibits continuous non-uniqueness and frame dragging between
the two components~\cite{EF}. Other exact configurations include
concentric black di-rings, whose rings lie in the same
plane~\cite{IMDiring,EKDiring}, and orthogonal or bicycling black
rings, whose circular directions lie in different
planes~\cite{Izumi,ElvangRodriguez}. Constructions and phase-space
arguments also extend to configurations with several rings, with or
without a central hole~\cite{IMDiring,Izumi,EEF,ElvangRodriguez}.
These solutions show how the freedom to distribute mass and angular
momentum among several horizons produces non-uniqueness at fixed
asymptotic charges.

For stationary five-dimensional vacuum spacetimes with two commuting
axial symmetries, the Einstein equations reduce to an integrable system
on a two-dimensional orbit space~\cite{Harmark,Pomeransky}. Inverse
scattering then generates exact solutions by algebraic transformations
of a simpler seed metric~\cite{BZ,Pomeransky}. This method underlies
the constructions of the doubly spinning ring and singly spinning
Saturn~\cite{PS,EF,CHT}. Extending the Saturn solution to independent
rotation in both planes on each horizon is nevertheless difficult.
The resulting expressions must satisfy the axis identifications,
conical balance and horizon regularity conditions simultaneously.

The second spin is particularly important for extremality. Both the
five-dimensional Myers--Perry black hole and the balanced doubly spinning
ring require nonzero angular momenta in both planes to reach a regular
extremal horizon of finite area~\cite{MP,PS}. These examples motivate
retaining both spins on each component of a Saturn. The blackfold
construction already provides perturbative evidence for a thin extremal
doubly spinning ring surrounding an extremal doubly spinning central
hole~\cite{AHOExt}. An exact solution is needed to determine how this
configuration extends beyond the thin-ring approximation and to study
its full exterior geometry.

In this paper we construct a six-parameter family of doubly spinning
black Saturns in five-dimensional vacuum gravity. We extend the
inverse-scattering construction of the doubly spinning ring
in~\cite{CHT} to a seed with the rod structure appropriate to a ring
surrounding a spherical hole, realising the Saturn generalisation
proposed there. We use AI tools to assist with the symbolic manipulations
and numerical checks. We exhibit balanced configurations in which
both horizons are extremal and have finite area, with a regular,
causal exterior in the parameter domains established below.
To our knowledge, this gives the first regular exact solution with
standard asymptotic flatness and disconnected extremal horizons of
topologies $S^3$ and $S^1\times S^2$ in vacuum gravity.

We organise the paper as follows. Section~\ref{sec:metric} presents the
metric, its parameters and the balance conditions.
Section~\ref{sec:properties} studies the geometry, thermodynamics and
nonextremal regularity. Section~\ref{sec:extremal} constructs the
doubly extremal metric and its horizon extensions.
Section~\ref{sec:phases} discusses phase diagrams, comparisons with
blackfolds and frame dragging. We conclude in
section~\ref{sec:discussion}. Appendix~\ref{app:limits} derives the single-object and singly spinning
limits, together with the two metrics in which only one horizon is
extremal, including the required parameter scalings and coordinate
continuations. Appendix~\ref{app:explicit} gives the explicit metric
coefficients and thermodynamic formulas, followed by the regularity
criteria and the domains covered by our checks.
Appendix~\ref{app:blackfold} derives the blackfold balance equations,
mutual-equilibrium solutions and thermodynamics, and explains their
comparison with the exact solution.
Appendix~\ref{app:extremequilibrium} proves the obstruction to mutual
equilibrium for the doubly extremal family of
section~\ref{sec:extremal}, and distinguishes the exact boundary
exclusions from the evidence supplied by wider numerical searches.

\section{The doubly-spinning Saturn metric}
\label{sec:metric}

The construction of the doubly-spinning black Saturn has three ingredients. We first generate a stationary
vacuum metric with two candidate horizon rods. We then choose the angular
identifications and impose balance, so that the axes describe smooth
rotation axes without conical defects. Finally, we write the result
in scalar functions suitable for reading off its geometry.

\subsection{Geometry and inverse-scattering construction}

We seek a five-dimensional solution of vacuum Einstein gravity, $R_{\mu\nu}=0$, with one stationary
and two commuting rotational Killing fields. The rotations act in the
ring plane and in the transverse plane. Away from the axes and horizons,
generalised Weyl coordinates separate the three Killing directions from
the two-dimensional orbit space~\cite{Weyl,Harmark,HarmarkOlesen}:

\begin{equation}
 ds^2=G_{ab}\dd x^a\dd x^b+f(\dd\rho^2+\dd z^2),\qquad
 x^a=(t,\psi,\phi),\qquad \det G=-\rho^2.
 \label{eq:rawmetric}
\end{equation}
Here $G$ is a symmetric $3\times3$ matrix containing the time and angular
components, while $f$ is the conformal factor of the orbit-space metric.
The condition $\det G=-\rho^2$ fixes the Weyl radial coordinate. The
coordinates $(t,\psi,\phi)$ are the Killing coordinates used in the construction. Their
physical normalisation and angular periods will be fixed in
section~\ref{sec:balance}.

The boundary $\rho=0$ is divided into intervals called rods~\cite{Harmark,HarmarkOlesen}.
Rod structure is the one-dimensional instance of the more general domain
structure~\cite{Domain,DomainLambda}. On each
open rod, a linear combination of Killing fields lies in the kernel
of $G$. A spacelike kernel is a shrinking angular circle while a horizon
kernel is its null generator. Our target is the sequence
axis-ring horizon-axis-spherical horizon-axis shown in
figure~\ref{fig:rods}. Two additional points facilitate the construction and will subsequently be made smooth.

The vacuum equations for $G$ reduce to the integrable matrix equation~\cite{Harmark,Pomeransky}
\begin{equation}
 \partial_\rho(\rho G^{-1}\partial_\rho G)
 +\partial_z(\rho G^{-1}\partial_zG)=0.
 \label{eq:reducedvacuum}
\end{equation}
Once $G$ is known, define $U=\rho(\partial_\rho G)G^{-1}$ and
$V=\rho(\partial_zG)G^{-1}$. The remaining Einstein equations are~\cite{Harmark,Pomeransky}
\begin{equation}\begin{aligned}
 \partial_\rho\log f&=-\frac1\rho+
             \frac{1}{4\rho}\operatorname{Tr}(U^2-V^2),&
 \partial_z\log f&=\frac{1}{2\rho}\operatorname{Tr}(UV).
\end{aligned}\label{eq:conformalequations}\end{equation}
Their compatibility follows from~\eqref{eq:reducedvacuum}. Integrating
these equations determines $f$ up to a multiplicative constant. We use the Belinski--Zakharov
inverse-scattering method~\cite{BZ}, in its higher-dimensional
form~\cite{Pomeransky,CHT}, to generate $G$ from a simpler diagonal
solution called the seed. The method replaces the nonlinear matrix
equation by an auxiliary linear system, then transforms a known solution
of that system algebraically. The resulting metric solves the vacuum
field equations.

To proceed further, we choose six ordered points and their associated pole functions
\begin{equation}
 0=a_1<a_2<a_3<a_4<a_5<a_6,\quad
 u_i=\sqrt{\rho^2+(z-a_i)^2}-(z-a_i),\quad v_i=-\rho^2/u_i.
 \label{eq:poles}
\end{equation}
Fixing $a_1=0$ chooses the origin of $z$ without removing a physical
parameter. For $\rho>0$, $u_i$ and $v_i$ are the positive and negative
roots of $\lambda^2+2(z-a_i)\lambda-\rho^2=0$. They encode the rod
endpoints and make the metric functions algebraic. We define $x=\rho^2$
and $d_{ij}=a_i-a_j$ and choose the diagonal seed to be

\begin{equation}
 G_s=\diag\left(-\frac{u_1u_4}{u_3u_6},\frac{u_2u_5}{u_1},
             \frac{x u_3u_6}{u_2u_4u_5}\right),
 \label{eq:seed}
\end{equation}
whose determinant is $-x$. The powers of the $u_i$ specify which
Killing components vanish or diverge on each axis interval. The seed is
an intermediate solution and regularity is imposed on the final metric.

To introduce rotation while retaining $\det G=-x$, we use the
two-step soliton transformation of~\cite{Pomeransky,CHT}. First, we remove six
poles with polarisations along single Killing directions. Then we insert
the same poles with polarisations that mix those directions. A
polarisation is a constant vector specifying the combination of Killing
directions used in a soliton transformation. The first step gives a
diagonal matrix $H$ while the second generates the time-angle and
angle-angle components in the metric responsible for rotation. Because the pole
locations are the same in both steps, their determinant factors cancel.

The six removed poles and their trivial directions
are
\begin{center}
\begin{tabular}{c|cccccc}
Pole & $u_1$ & $u_2$ & $u_3$ & $v_4$ & $u_5$ & $v_6$\\
Slot & $\psi$ & $\phi$ & $t$ & $t$ & $\phi$ & $\phi$
\end{tabular}
\end{center}
The auxiliary linear system uses a complex variable $\lambda$, called
the spectral parameter, in addition to $\rho,z$. Its matrix solution
$\Psi_0(\lambda,\rho,z)$ encodes $H$ through $\Psi_0(0,\rho,z)=H$.
The pole functions $u_i,v_i$ specify the moving poles of the soliton
transformation in this auxiliary variable~\cite{BZ,Pomeransky}. The quantities needed below are
\begin{align}
 H&=\diag\left(-\frac{u_1u_3}{u_4u_6},-\frac{u_1u_2u_5}{x},
                     -\frac{u_2u_3u_5}{u_4u_6}\right),\\
 P_b(\lambda)&=x-2(z-b)\lambda-\lambda^2,
 \qquad\mathcal T(\lambda)=(u_4-\lambda)(u_6-\lambda),\qquad b=-1.
\end{align}
We place $b$ to the left of $a_1=0$ and choose the unit of the Weyl
coordinates so that $a_1-b=1$, hence $b=-1$. This auxiliary separation
is not a physical rod length and fixes none of the gaps $a_{i+1}-a_i$.
A different negative $b$ gives the same family after redefining the
polarisation constants. The factor $P_b$ is
a convenient way to represent the seed factor $x$ in the auxiliary
system, since $P_b(0)=x$. Its freedom is a choice of representation for
$\Psi_0$, accompanied by a redefinition of the constant polarisations. Indeed, at a pole,
$P_b(u_i)=2(b-a_i)u_i$, so changing $b$ only rescales the corresponding
constant coefficients. Keeping $b$ to the left of all $a_i$ avoids a
zero of this factor. For this choice a diagonal generating matrix is
\begin{align}
 (\Psi_0)_{tt}&=-\frac{(u_1-\lambda)(u_3-\lambda)}{\mathcal T(\lambda)},
 & (\Psi_0)_{\psi\psi}&=-\frac{(u_1-\lambda)(u_2-\lambda)(u_5-\lambda)}{P_b(\lambda)},\nonumber\\
 (\Psi_0)_{\phi\phi}&=-\frac{(u_2-\lambda)(u_3-\lambda)(u_5-\lambda)}{\mathcal T(\lambda)},
 & (\Psi_0)_{ab}&=0\quad(a\ne b).
\end{align}
This matrix satisfies the auxiliary linear equations, usually called
the Lax system~\cite{BZ,Pomeransky,CHT}. Their compatibility is
\eqref{eq:reducedvacuum}: this is why transforming the generating matrix
produces another vacuum solution. Inserting the six poles with new
polarisations then gives the following position-dependent rows $m_i$.
They contain the result of solving the auxiliary system once and for
all. Evaluating the final metric only requires these explicit rows.

The constants $(C_1,C_2,C_3,B_4,B_5,C_6)$ are polarisation parameters. Let $\mathcal T_i=\mathcal T(u_i)$,
$\widehat{\mathcal T}_i=\mathcal T(v_i)$ and $P_i=P_b(u_i)$, then

\begin{align}
 m_1&=\left(\frac{C_1\mathcal T_1}{u_3-u_1},
       \frac{P_1}{(u_2-u_1)(u_5-u_1)},0\right),\nonumber\\
 m_2&=\left(0,\frac{C_2P_2}{(u_1-u_2)(u_5-u_2)},
       \frac{\mathcal T_2}{(u_3-u_2)(u_5-u_2)}\right),\nonumber\\
 m_3&=\left(\frac{\mathcal T_3}{u_1-u_3},0,
       \frac{C_3\mathcal T_3}{(u_2-u_3)(u_5-u_3)}\right),\nonumber\\
 m_4&=\left(-\frac{\widehat{\mathcal T}_4}{(u_1-v_4)(u_3-v_4)},0,
       -\frac{B_4\widehat{\mathcal T}_4}{(u_2-v_4)(u_3-v_4)(u_5-v_4)}\right),\nonumber\\
 m_5&=\left(0,\frac{B_5P_5}{(u_1-u_5)(u_2-u_5)},
       \frac{\mathcal T_5}{(u_2-u_5)(u_3-u_5)}\right),\nonumber\\
 m_6&=\left(-\frac{C_6\widehat{\mathcal T}_6}{(u_1-v_6)(u_3-v_6)},0,
       -\frac{\widehat{\mathcal T}_6}{(u_2-v_6)(u_3-v_6)(u_5-v_6)}\right).
 \label{eq:rows}
\end{align}
The soliton ansatz has only six pole residues. The equations fixing
those residues form a finite linear system, whose coefficients are
$\Gamma_{ij}$ below. Solving it gives the correction to $H$.
With
$(p_i)=(u_1,u_2,u_3,v_4,u_5,v_6)$, we can form the symmetric $6\times6$
matrix $\Gamma$ and the $3\times6$ matrix $K$ according to:
\begin{equation}
 \Gamma_{ij}=\frac{m_iHm_j^{\mathsf T}}{x+p_ip_j},\qquad
 K_{ai}=\frac{H_{aa}(m_i)_a}{p_i},\qquad
 G=H-K\Gamma^{-1}K^{\mathsf T}.
 \label{eq:G}
\end{equation}
The term $K\Gamma^{-1}K^{\mathsf T}$ incorporates the interaction of
all six poles. Together with the rows above,~\eqref{eq:G} specifies
every time and angular component.

It remains to determine $f$, the coefficient of $\dd\rho^2+\dd z^2$
in~\eqref{eq:rawmetric}. This coefficient fixes proper distances in the
orbit space and enters the balance and temperature formulas. Let $f_s$
be the corresponding coefficient for the diagonal seed $G_s$.
Integrating~\eqref{eq:conformalequations} with $G=G_s$ gives $f_s$ as a product of
$R_{ij}=x+u_i u_j$. To state the exponents compactly, define $w_i$ so
that its three entries are the powers of $u_i$ in the $tt,\psi\psi,
\phi\phi$ seed components. For example $u_1$ has powers $(1,-1,0)$.
The six exponent vectors are

\begin{equation}
 w_1=(1,-1,0),\quad w_2=w_5=(0,1,-1),\quad
 w_3=w_6=(-1,0,1),\quad w_4=(1,0,-1).
\end{equation}
The seed coefficient and the coefficient of the final metric are
respectively
\begin{equation}
 f_s=\frac{u_2u_5}{u_1}\prod_i R_{ii}^{-1}
                         \prod_{i<j}R_{ij}^{-w_i\cdot w_j},\qquad
 f=f_s\frac{\det\Gamma(C_1,C_2,C_3,B_4,B_5,C_6)}
                         {\det\Gamma(0,0,0,0,0,0)}.
 \label{eq:f}
\end{equation}
We set the multiplicative constant in $f_s$ to one as an initial
normalisation convention. A different positive constant rescales proper
distances and is accounted for when fixing angular periods and the
overall length scale in sections~\ref{sec:balance} and~\ref{sec:parameters}.
In the denominator of~\eqref{eq:f}, all six polarisation constants are
zero. This is the transformation that inserts the poles along the
original single Killing directions and therefore recovers the seed.
The determinant ratio measures the change in $f$ when those original
polarisations are replaced by the rotating ones.

Equations~\eqref{eq:G} and~\eqref{eq:f} now specify the full vacuum
metric wherever their denominators are nonzero. The next step is to
select parameters and angular identifications for which its axes and
horizons are regular.

\subsection{Physical rods, normalisation and balance}
\label{sec:balance}

An exact vacuum metric that results from this procedure is not necessarily a regular black hole spacetime. We
must remove the two auxiliary rod points, fix the physical angular
circles, and ensure that the finite axis between the horizons has a
smooth identification. These requirements have different roles. First, $a_1$ and $a_6$ are auxiliary points, not ends of physical
horizons. Requiring their apparent turning points to disappear fixes
two polarisation parameters. On the choice of signs used here,

\begin{equation}
 C_1=-\sqrt{\frac{d_{31}}{2d_{21}d_{41}d_{51}d_{61}}},\qquad
 C_6=\sqrt{\frac{d_{61}}{2d_{62}d_{65}}}.
 \label{eq:phantoms}
\end{equation}
In fact the complete metric is invariant under each root involution
$u_g\mapsto-x/u_g$, $g=1,6$, after imposing these squared relations.
Odd dependence on the auxiliary square roots therefore cancels.
Regularity at their centres additionally requires the nonzero
denominator and signature conditions discussed below.

Second, the outer axes determine the angular identifications. Let
$v_L,v_R$ be their kernels in the original $(t,\psi,\phi)$ coordinates, normalised by
$v_L^\phi=1$ and $v_R^\psi=1$. If an orbit of $v_i$ is provisionally
given period $2\pi$, its circumference divided by its proper radius
approaches $2\pi\sqrt{q_i}$, with $q_i$ as below. Thus the smooth
$2\pi$-periodic generator is $v_i/\sqrt{q_i}$, and the physical Killing
basis is

\begin{equation}
 q_i=\lim_{\rho\to0}\frac{v_i^{\mathsf T}Gv_i}{\rho^2 f},\qquad
 A=\left(e_t,\frac{v_R}{\sqrt{q_R}},\frac{v_L}{\sqrt{q_L}}\right),
 \qquad \delta=\det A>0.
 \label{eq:basis}
\end{equation}
For the branch considered here $q_L,q_R>0$. The physical metric is
\begin{equation}
 ds_A^2=\cG_{ab}\dd X^a\dd X^b+f(\dd\rho^2+\dd z^2),\qquad
 \cG=A^{\mathsf T}GA,\quad X=(T,\Psi,\Phi),\quad
 \det\cG=-\delta^2\rho^2.
 \label{eq:physical}
\end{equation}
The columns of $A$ mean $x^a=A^a{}_bX^b$ locally. Globally we take
$\Psi$ and $\Phi$ to be independently $2\pi$-periodic at fixed $T$.
The time components of the outer-axis kernels are absorbed into the
definition of the angular generators. The stationary vector remains $\partial_T=\partial_t$
and has unit norm at infinity. Since the basis change multiplies the
Killing determinant by $\delta^2$, the canonical Weyl coordinates for
this \emph{same} metric are
$\widehat\rho=\delta\rho$ and $\widehat z=\delta(z-z_0)$, with base
factor $f/\delta^2$.

Third, the finite axis $a_3<z<a_4$ must shrink the same primitive
circle as the lower axis. Here ``primitive'' means a cycle traversed
once, not an integer multiple of that cycle. With $v_M^\phi=1$,
matching the time and $\psi$ components fixes its direction, and
matching $q_M=q_L$ fixes its circumference-to-radius ratio:

\begin{equation}
 \mathcal B=\left(v_M^t-v_L^t,\ v_M^\psi-v_L^\psi,
                                      q_M/q_L-1\right)=0.
 \label{eq:balance}
\end{equation}
The first two equations align the degenerating cycles while the third is
mechanical balance, excluding a conical strut or membrane that would
otherwise support the ring. These conditions are algebraic. To see how
they are obtained, we expand $G=G_0+O(\rho^2)$ and $f=f_0+O(\rho^2)$ on
an open spacelike rod. If the kernel is normalised by $v^k=1$ and $I$
denotes the other two Killing directions, solving $G_0v=0$ gives

\begin{equation}
 v^I=-G_{0,II}^{-1}G_{0,Ik},\qquad
 q=-\frac1{f_0\det G_{0,II}}.
 \label{eq:axisalgebra}
\end{equation}
For the second identity, the determinant in the basis $(v,e_I)$ is
$(v^{\mathsf T}Gv)\det G_{0,II}+O(x^2)=-x$. Dividing by $xf_0$
gives $q$, the square of the circumference-to-radius ratio divided by
$2\pi$. Thus $q=1$ means that a $2\pi$-periodic angle has no conical
deficit or excess. In subsection~\ref{app:analyticthermo} we give finite
algebraic formulas for the rod coefficients, so these relations also
specify how to evaluate~\eqref{eq:balance}.

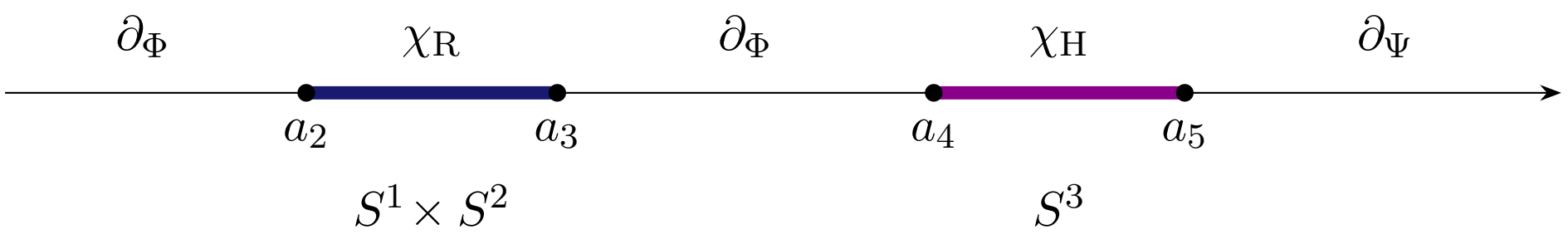
\begin{figure}[htbp]\centering
\begin{tikzpicture}[x=1cm,y=1cm,>=Stealth]
 \draw[->](-.4,0)--(12.0,0);
 \draw[line width=3pt,MidnightBlue](2,0)--(4,0);
 \draw[line width=3pt,DarkMagenta](7,0)--(9,0);
 \foreach \x/\lab in {2/a_2,4/a_3,7/a_4,9/a_5}
  {\fill(\x,0)circle(2pt);\node[below]at(\x,-.08){$\lab$};}
 \node[above]at(.7,.15){$\partial_\Phi$};
 \node[above]at(3,.15){$\chi_\R$};
 \node[above]at(5.5,.15){$\partial_\Phi$};
 \node[above]at(8,.15){$\chi_\Hh$};
 \node[above]at(10.6,.15){$\partial_\Psi$};
 \node[below]at(3,-.6){$S^1\!\times S^2$};
 \node[below]at(8,-.6){$S^3$};
\end{tikzpicture}
\caption{Physical rods after auxiliary-point removal. The horizons
have generators $\chi_h=\partial_T+\Omega_h^A\partial_{\varphi^A}$.
The points $a_1,a_6$ are smooth auxiliary seams, not physical turning
points. Rod lengths in this drawing are schematic.}
\label{fig:rods}
\end{figure}

After removing the two auxiliary points the physical rod diagram is
shown in figure~\ref{fig:rods}. Equal adjacent cycles on the ring give
$S^1\times S^2$ topology while the different primitive cycles bordering the
central rod give $S^3$. The outer cycles generate the standard angular
lattice, so the asymptotic end has $S^3$, with no lens-space quotient.

\subsection{Explicit scalar form}
\label{sec:scalarform}

The matrix construction is compact and explains the origin of the
solution. To read the line element directly, it is useful to package the
same components into the scalar-function form familiar from the singly
spinning Saturn \cite{EF}.
The second spin requires two rotational potentials and a mixed
angular coefficient. We fix the positive auxiliary function
\begin{equation}
 P=R_{34}^{\,2}R_{12}R_{23},\qquad R_{ij}=\rho^2+u_i u_j,
 \label{eq:scalarP}
\end{equation}
and introduce seven scalar functions
$H_x,H_y,\omega_\psi,\omega_\phi,V_{\psi\psi},V_{\psi\phi},V_{\phi\phi}$.
Their complete definitions as finite sums of scalar products are given in
\eqref{eq:scalarsums}-\eqref{eq:scalarseven} in
subsection~\ref{app:scalar}. The \ancillarylink{ancillary Mathematica file} supplies
all seven expanded numerator polynomials and an evaluator, with the
symbol dictionary given in the same appendix. In the original Killing
coordinates $(t,\psi,\phi)$ of~\eqref{eq:rawmetric}, before imposing the
physical angular periods, the full line element is

\begin{equation}\begin{aligned}
 ds^2={}&\frac1{H_x}\bigl[-H_y\dd t^2
       -2\omega_\psi\dd t\dd\psi-2\omega_\phi\dd t\dd\phi
       +V_{\psi\psi}\dd\psi^2\\
       &\hspace{30mm}+2V_{\psi\phi}\dd\psi\dd\phi
                          +V_{\phi\phi}\dd\phi^2\bigr]
       +H_xP(\dd\rho^2+\dd z^2).
\end{aligned}\label{eq:scalarraw}\end{equation}
Thus~\eqref{eq:scalarraw} explicitly displays all six Killing-metric
components and the base factor. The functions are rational in the
pole functions and polynomial in the six polarisation constants,
with a common, positive, polarisation-independent denominator
specified in~\eqref{eq:scalarcommonF}.

Another closed form of the metric is possible and isolates the stationary one-form by completing
the square in $\dd t$. The angular coefficients left over from this
operation are expressed through
\begin{align}
 F_{\psi\psi}&=\frac{H_yV_{\psi\psi}+\omega_\psi^2}{H_x^2},&
 F_{\psi\phi}&=\frac{H_yV_{\psi\phi}+\omega_\psi\omega_\phi}{H_x^2},&
 F_{\phi\phi}&=\frac{H_yV_{\phi\phi}+\omega_\phi^2}{H_x^2}.
 \label{eq:scalarangular}
\end{align}
The constants in~\eqref{eq:basis} are $q_A=A^t{}_A$ and
$b_{iA}=A^i{}_A$, with $i=\psi,\phi$, $A=\Psi,\Phi$.
They implement $\dd t=\dd T+q_A\dd\varphi^A$ and
$\dd x^i=b_{iA}\dd\varphi^A$. Defining
\begin{equation}\begin{aligned}
 \omega_A={}&b_{\psi A}\omega_\psi+b_{\phi A}\omega_\phi,\\
 \mathcal F_{AB}={}&b_{\psi A}b_{\psi B}F_{\psi\psi}
 +(b_{\psi A}b_{\phi B}+b_{\phi A}b_{\psi B})F_{\psi\phi}
 +b_{\phi A}b_{\phi B}F_{\phi\phi},
\end{aligned}\label{eq:scalarphysicalfunctions}\end{equation}
the physically normalised metric becomes
\begin{equation}\begin{aligned}
 ds_A^2={}&-\frac{H_y}{H_x}
 \left[\dd T+\left(\frac{\omega_\Psi}{H_y}+q_\Psi\right)\dd\Psi
             +\left(\frac{\omega_\Phi}{H_y}+q_\Phi\right)\dd\Phi\right]^2\\
 &+H_x\left[P(\dd\rho^2+\dd z^2)
 +\frac{\mathcal F_{\Psi\Psi}\dd\Psi^2
       +2\mathcal F_{\Psi\Phi}\dd\Psi\dd\Phi
       +\mathcal F_{\Phi\Phi}\dd\Phi^2}{H_y}\right].
\end{aligned}\label{eq:scalarphysical}\end{equation}
The physical angles $\Psi$ and $\Phi$ both have period $2\pi$.
The constants $q_\Psi,q_\Phi$
are time shifts, distinct from the cone factors $q_L,q_R$.
Subsection~\ref{app:scalar} also gives scalar algebraic formulas for
these constants and $b_{iA}$. To obtain a balanced Saturn, we impose
both the auxiliary-point conditions~\eqref{eq:phantoms} and the axis
conditions~\eqref{eq:balance}.

The stationary Killing field becomes null at an ergosurface:
$g_{TT}=-H_y/H_x=0$, so $H_y=0$ when $H_x$ is finite and nonzero.
The factors $1/H_y$ in the completed-square form
\eqref{eq:scalarphysical} might then appear singular. They arise only
from completing the square: expanding it cancels these denominators
and recovers~\eqref{eq:scalarraw}, followed by the constant change of
coordinates. The latter form is therefore the appropriate expression
at an ergosurface. One useful algebraic consistency identity is

\begin{equation}
 F_{\psi\psi}F_{\phi\phi}-F_{\psi\phi}^2
             =\frac{\rho^2H_y}{H_x}.
 \label{eq:scalarangularidentity}
\end{equation}
Indeed, the completed-square form gives
$\det G=-(H_x/H_y)\det(F_{ij})$, so
\eqref{eq:scalarangularidentity} is precisely the Weyl constraint
$\det G=-\rho^2$ expressed in scalar functions. It provides a check
on their relative normalisation. Smoothness at $H_y=0$ must still
be checked in the component form, where no division by $H_y$ occurs.

The choice~\eqref{eq:scalarP} makes the comparison with the singly
spinning metric particularly direct. Under the degeneration and
endpoint map of appendix~\ref{app:ef}, the scalar identities are
\begin{equation}\begin{aligned}
 H_x&\longrightarrow H_x^{\rm EF},&
 H_y&\longrightarrow H_y^{\rm EF},&P&\longrightarrow P^{\rm EF},\\
 \omega_\psi&\longrightarrow-\omega_\psi^{\rm EF},&
 \omega_\phi&\longrightarrow0,&V_{\psi\phi}&\longrightarrow0,\\
 F_{\psi\psi}&\longrightarrow G_y^{\rm EF},&
 F_{\psi\phi}&\longrightarrow0,&
 F_{\phi\phi}&\longrightarrow
                 \frac{H_y^{\rm EF}}{H_x^{\rm EF}}G_x^{\rm EF}.
\end{aligned}\label{eq:scalarEF}\end{equation}
The common angular reflection and canonical normalisation already
specified in that appendix then give precisely the
singly spinning Black Saturn line element~\cite{EF}.

\subsection{Independent parameters and scale conventions}
\label{sec:parameters}

After setting $a_1=0$, the construction has five endpoint positions and
six polarisation constants. The preceding conditions reduce the count:
\begin{center}
\begin{tabular}{lr}\toprule
Data or constraint & Remaining parameters\\\midrule
Five endpoints and six polarisations & 11\\
Removal of the two auxiliary points, \eqref{eq:phantoms} & 9\\
Three axis and balance equations, \eqref{eq:balance} & 6\\\bottomrule
\end{tabular}
\end{center}
Therefore, a convenient local choice of the 6 free parameters is
\begin{equation}
 p=(g_1,g_2,g_3,g_4,g_5,C_2),\qquad g_i=a_{i+1}-a_i>0.
 \label{eq:controls}
\end{equation}
The gaps $g_2$ and $g_4$ are the ring and central-hole rod lengths,
and $g_3$ is the intervening axis length in Weyl coordinates.
The remaining gaps $g_1,g_5$ locate the auxiliary points and affect
the final metric even after those points have been made smooth.

For given $p$,~\eqref{eq:phantoms} fixes $C_1,C_6$, and we solve
\eqref{eq:balance} for $(C_3,B_4,B_5)$. These three quantities are
locally smooth functions of $p$ when the derivative of the three
balance equations with respect to them is invertible. This condition
is verified near the reference configuration in
subsection~\ref{app:nonextbounds}. Section~\ref{sec:independence} relates
the six parameters $p$ to the two horizon masses and four horizon
angular momenta. One parameter sets the overall length scale of the
entire hole-ring configuration and fixing it leaves five dimensionless
parameters specifying shape and rotation.

We introduce two scale conventions. The metric $ds_A^2$ has the chosen
$2\pi$ angular periods. Replacing $(\rho,z)$ by
$(\widehat\rho,\widehat z)$ is a coordinate change of the same geometry.
For comparison with limiting metrics we also use a constant rescaling
of the entire solution. We define
\begin{equation}
 B=A\diag(1,\delta^{-1/2},\delta^{-1/2}),\qquad
 \bar G=B^{\mathsf T}GB,\quad \bar f=f/\delta,
 \qquad\det\bar G=-\rho^2,
 \label{eq:canonical}
\end{equation}
so that it gives $ds_{\rm can}^2=\delta^{-1}ds_A^2$ after
$T_A=\sqrt\delta\,T_{\rm can}$, with the angular periods unchanged.
We use $ds_A^2$ for the nonextremal geometry and the reductions to
Myers-Perry and Pomeransky-Sen'kov described in
appendices~\ref{app:mp} and~\ref{app:ps}. We use $ds_{\rm can}^2$ for
the extremal analysis and for the analytic reduction to the singly
spinning Saturn of~\cite{EF} in appendix~\ref{app:ef}.
The corresponding thermodynamic conversion is given after the charges
have been defined, in section~\ref{sec:asymptotic}.

\section{Geometry, thermodynamics and nonextremal regularity}
\label{sec:properties}

We now compute the temperatures, entropies and angular velocities of
both horizons, the ADM charges at infinity and the individual horizon
Komar charges. We explain the resulting Smarr relations and first law,
construct coordinates through the horizons, and state sufficient
conditions for a regular causal exterior. All formulas in this section
use $ds_A^2$ and the asymptotic normalisation of section~\ref{sec:balance}.

\subsection{Horizon geometry, temperature and entropy}
\label{sec:nonexthorizons}

The rods $[a_2,a_3]$ and $[a_4,a_5]$ represent the ring ($h=\R$)
and central hole ($h=\Hh$). Their generators are
\begin{equation}
 \chi_h=\partial_T+\Omega_h^A\partial_{\varphi^A},\qquad
 (\varphi^A)=(\Psi,\Phi),
 \label{eq:horizongenerator}
\end{equation}
where the repeated angular indices are summed. Both sets of angular velocities
refer to the common asymptotic frame fixed in section~\ref{sec:balance}.

The generator lies in the kernel of the Killing block, which
fixes its angular velocities. We can evaluate the leading rod coefficients
$(G^{[s]},f^{[s]})$ defined in~\eqref{eq:rodblockformulas} and
\eqref{eq:rodconformal}, at an interior point of the horizon interval,
with $s=2$ for the ring and $s=4$ for the hole. Their parameter-only
expressions are given in subsection~\ref{app:analyticthermo}. Setting
$I=(\psi,\phi)$, the kernel and the scalar $q_h$ are
\begin{equation}
 v_h=\begin{pmatrix}1\\-(G^{[s]}_{II})^{-1}G^{[s]}_{It}\end{pmatrix},
 \quad q_h=-\frac{1}{f^{[s]}\det G^{[s]}_{II}},\quad
 w_h=A^{-1}v_h.
 \label{eq:algebraichorizon}
\end{equation}
Here $s=2$ labels the ring and $s=4$ the central hole. The raw kernel
$v_h$ is chosen to have $t$ component one. Applying $A^{-1}$ expresses
it in the physical coordinates. Dividing the result by $w_h^T$ gives
$\chi_h$, whose $T$ component is one. The invariant $q_h$ is the
horizon analogue of the axis cone factor: it is negative, and its
magnitude determines the surface gravity. In corotating coordinates the leading time--radial metric has the Rindler form:
$-\kappa_h^2R_\perp^2\dd T^2+\dd R_\perp^2$, with proper radial
distance $R_\perp=\sqrt{f_0}\rho$ at fixed $z$. Hence
$\cG(\chi_h,\chi_h)=-\kappa_h^2f_0\rho^2+O(\rho^4)$.
The exact rod data make the surface gravity $\kappa_h$ and angular velocities $\Omega_h^A$ constant on each
component. The Hawking temperature is $T_h=\kappa_h/(2\pi)$:
\begin{equation}
\kappa_h=\frac{\sqrt{-q_h}}{|w_h^T|},\qquad
 T_h=\frac{\sqrt{-q_h}}{2\pi|w_h^T|},\qquad
 \Omega_h^A=\frac{w_h^A}{w_h^T}.
 \label{eq:algebraictemperature}
\end{equation}
To obtain the entropy, we set $\lambda_{AB}=\cG_{AB}$ and use a subscript
zero for the horizon limit. A spatial horizon section has area
\begin{equation}
 \cA_h=(2\pi)^2\int_{a_-}^{a_+}\sqrt{f_0\det\lambda_0}\,\dd z.
 \label{eq:areaintegral}
\end{equation}
The determinant condition gives
$\kappa_h^2f_0\det\lambda_0=\delta^2$. Since $\kappa_h$ is constant,
the area density is simply $\delta/\kappa_h$. Writing
$\ell_\R=a_3-a_2$, $\ell_\Hh=a_5-a_4$ and using $S_h=\cA_h/(4G_5)$,
we obtain
\begin{equation}
\cA_h=\frac{4\pi^2\delta\ell_h}{\kappa_h},\qquad
 S_h=\frac{\pi^2\delta\ell_h|w_h^T|}{G_5\sqrt{-q_h}},\qquad
 T_hS_h=\frac{\pi\delta\ell_h}{2G_5}.
 \label{eq:area}
\end{equation}
The total entropy is the sum of the two horizon entropies,
$S_{\rm tot}=S_\R+S_\Hh$. Hence
\begin{equation}
S_{\rm tot}=\frac{\pi^2\delta}{G_5}
       \left(\frac{a_3-a_2}{\kappa_\R}+\frac{a_5-a_4}{\kappa_\Hh}\right).
 \label{eq:entropy}
\end{equation}
These formulas apply to nonextremal rods. At extremality both
$\ell_h$ and $\kappa_h$ vanish and their ratio is replaced by data from
the limiting horizon geometry in section~\ref{sec:extthermo}.

\subsection{Smooth coordinates across each nonextremal horizon}
\label{sec:nonextcontinuation}

The divergence of the radial metric in Weyl coordinates at a horizon
is a coordinate effect if it can be removed by a nonsingular extension.
We construct that extension separately for the ring and central black hole.
For either horizon, we suppress its label $h$, set $x=\rho^2$ and
$\theta^A=\varphi^A-\Omega^AT$, and define the actual metric coefficients
\begin{equation}
 \lambda_{AB}=\cG_{AB},\quad
 \beta_A=\frac{\cG(\chi,\partial_{\varphi^A})}{x},\quad
 \alpha=-\frac{\cG(\chi,\chi)}x,\quad
 a=\frac{\alpha-\kappa^2f}{x}.
\end{equation}
On the certified branch these are analytic in $(x,z)$ along each open
horizon. Bounded Laurent expressions and the exact kernel remove the
first apparent poles. The horizon identity
$\alpha(0,z)=\kappa^2f(0,z)$ removes the last. The metric becomes
\begin{equation}
 ds^2=-x\alpha\dd T^2+2x\beta_A\dd T\dd\theta^A
       +\lambda_{AB}\dd\theta^A\dd\theta^B
       +f\left(\frac{\dd x^2}{4x}+\dd z^2\right).
\end{equation}
With an arbitrary constant $x_*>0$, the future coordinate
$v=T+(2\kappa)^{-1}\log(x/x_*)$ gives
\begin{align}
 ds^2={}&-x\alpha\dd v^2+\frac\alpha\kappa\dd v\dd x
 -\frac a{4\kappa^2}\dd x^2+f\dd z^2
 +2x\beta_A\dd v\dd\theta^A\nonumber\\
 &-\frac{\beta_A}\kappa\dd x\dd\theta^A
       +\lambda_{AB}\dd\theta^A\dd\theta^B,
 \quad\det g=-\frac{\delta^2f^2}{4}.
 \label{eq:nonextingoing}
\end{align}
Thus the extension through $x=0$ is analytic and nondegenerate.
The outgoing chart reverses the logarithm and the two mixed radial
signs. The cancellation is explicit: substituting
$\dd T=\dd v-\dd x/(2\kappa x)$ combines the singular radial terms
into $-(\alpha-\kappa^2f)\dd x^2/(4\kappa^2x)$, which is finite by
the horizon identity. When $f$ is finite and positive, the determinant
is nonzero and the chart extends across $x=0$ to negative $x$.

The horizons can be extended separately even when their $\kappa_h$
and $\Omega_h^A$ differ. Subsection~\ref{app:nonextcharts} gives the
two-sided Kruskal coordinates and joint horizon-axis coordinates at all four poles.

\subsection{Asymptotic flatness and thermodynamic charges}
\label{sec:asymptotic}
Here we exhibit the asymptotically Minkowskian coordinates and then
read off the total mass and angular momenta. Use the physical Killing
coordinates $(T,\Psi,\Phi)$ of~\eqref{eq:physical}, with both angles
of period $2\pi$ and $\delta=\det A>0$. We set the irrelevant origin
shift $z_0$ to zero and define
\begin{equation}\begin{gathered}
 \rho=\frac{r^2}{2\delta}\sin2\theta,\qquad
 z=\frac{r^2}{2\delta}\cos2\theta,\qquad 0\leq\theta\leq\frac\pi2,\\
 R_W=\sqrt{\rho^2+z^2}=\frac{r^2}{2\delta},\qquad
 c=\frac{z}{R_W}=\cos2\theta.
\end{gathered}\label{eq:asymptoticcoordinates}\end{equation}
In the canonical Weyl coordinates $\widehat\rho=\delta\rho$ and
$\widehat z=\delta z$, this is the usual five-dimensional
spherical-coordinate transformation~\cite{Harmark}. The factor $\delta$
is essential when using the unrescaled metric $ds_A^2$.

To see the limit directly, we expand the pole functions as
$u_i=(1-c)(R_W+a_i)+O(R_W^{-1})$ and substitute in
\eqref{eq:G}--\eqref{eq:f}, or equivalently in the scalar expressions.
After the constant change of Killing coordinates by $A$, the leading
coefficients are
\begin{equation}\begin{aligned}
 f&=\frac{\delta}{2R_W}\bigl[1+O(R_W^{-1})\bigr],&
 \cG_{TT}&=-1+O(R_W^{-1}),\\
 \cG_{\Psi\Psi}&=\delta R_W(1-c)+O(1),&
 \cG_{\Phi\Phi}&=\delta R_W(1+c)+O(1),\\
 \cG_{T\Psi},\cG_{T\Phi}&=O(R_W^{-1}),&
 \cG_{\Psi\Phi}&=O(1).
\end{aligned}\label{eq:asymptoticleading}\end{equation}
Here the time normalisation and the flat spatial normalisation are
both consequences of the metric. In particular, the leading coefficient $f_1$ in
$f=f_1/R_W+\cdots$ is angle independent. Normalising each outer axis
to period $2\pi$ makes the two leading angular coefficients
$2f_1R_W(1\mp c)$. The determinant condition
$\det\cG=-\delta^2\rho^2$ then gives $2f_1=\delta$ on the branch
$f_1>0$. The leading time-angular constants vanish in this same
outer-axis basis. We give the algebraic coefficient check
in~\ref{app:asymptoticchecks}.

The orbit-space part transforms exactly according to
\begin{equation}
 \dd\rho^2+\dd z^2=\frac{r^2}{\delta^2}
                 (\dd r^2+r^2\dd\theta^2).
 \label{eq:asymptoticbase}
\end{equation}
Combining this identity with~\eqref{eq:asymptoticleading} gives
\begin{equation}
 ds_A^2=ds_{\rm flat}^2+h,\qquad
 ds_{\rm flat}^2=-\dd T^2+\dd r^2+
 r^2\bigl(\dd\theta^2+\sin^2\theta\,\dd\Psi^2+
                       \cos^2\theta\,\dd\Phi^2\bigr).
 \label{eq:asymptoticflat}
\end{equation}
The correction has components $h_{\mu\nu}=O(r^{-2})$ in the
asymptotically Cartesian coordinates
\begin{equation}\begin{aligned}
 (X_1,X_2)&=r\sin\theta(\cos\Psi,\sin\Psi),&
 (X_3,X_4)&=r\cos\theta(\cos\Phi,\sin\Phi).
\end{aligned}\label{eq:asymptoticcartesian}\end{equation}
Their first spatial derivatives are $O(r^{-3})$. These coordinates
also remove the polar-coordinate degeneracies at $\theta=0,\pi/2$ and
the outer-axis conditions supply the smooth matching there. Together
with the independent $2\pi$ identifications, this gives the standard
$S^3$ asymptotic metric.

\subsection{Thermodynamic charges}
\label{sec:charges}
The leading time components of the metric beyond flat space define the ADM charges:
\begin{align}
 \cG_{TT}&=-1+\frac{8G_5M}{3\pi r^2}+O(r^{-4}),\nonumber\\
 \cG_{T\Psi}&=-\frac{4G_5J_\Psi}{\pi r^2}\sin^2\theta+O(r^{-4}),&
 \cG_{T\Phi}&=-\frac{4G_5J_\Phi}{\pi r^2}\cos^2\theta+O(r^{-4}).
 \label{eq:adm}
\end{align}
Equivalently,
\begin{equation}
 M=\lim_{R_W\to\infty}\frac{3\pi\delta R_W}{4G_5}(\cG_{TT}+1),
 \quad J_{\Psi,\Phi}=-\lim_{R_W\to\infty}
       \frac{\pi\delta R_W}{G_5(1\mp c)}\cG_{T\Psi,T\Phi}.
\end{equation}
The ADM quantities describe the entire spacetime. To distinguish the
ring and central hole, we integrate the same stationary and rotational
Killing fields over a section $\mathcal H_h$ of each horizon. With
outward orientation, their Komar charges are
\begin{equation}
 M_h=-\frac3{32\pi G_5}\int_{\mathcal H_h}\star\dd\xi^\flat,
 \qquad J_{hA}=\frac1{16\pi G_5}\int_{\mathcal H_h}\star\dd m_A^\flat,
 \quad \xi=\partial_T,\quad m_A=\partial_{\varphi^A}.
 \label{eq:komar}
\end{equation}
These reduce to one-dimensional integrals of the Weyl matrix current:
\begin{align}
 Q_a(z)&=\lim_{\rho\to0}\delta\rho
                 (\cG^{-1}\partial_\rho\cG)^T{}_a,\nonumber\\
 M_h&=\frac{3\pi}{8G_5}\int_{a_-}^{a_+}Q_T\dd z,&
 J_{hA}&=-\frac\pi{4G_5}\int_{a_-}^{a_+}Q_A\dd z.
 \label{eq:currents}
\end{align}
The reduction to one dimension integrates out the two periodic angles.
More usefully, the vacuum matrix equation~\eqref{eq:reducedvacuum}
makes the current locally integrable: there is a matrix potential
$\mathcal P$ with
\begin{equation}
 \partial_z\mathcal P=\rho G^{-1}\partial_\rho G,\qquad
 \partial_\rho\mathcal P=-\rho G^{-1}\partial_zG.
 \label{eq:currentpotential}
\end{equation}
A horizon integral is consequently an endpoint difference. For this
solution $\mathcal P$ is algebraic in the pole functions, endpoints and
polarisation constants, and its
explicit expression is~\eqref{eq:analyticprimitive} in
subsection~\ref{app:currentprimitive}.
Under the change from $(t,\psi,\phi)$ to the normalised coordinates
$(T,\Psi,\Phi)$ in~\eqref{eq:basis}, it becomes $A^{-1}\mathcal PA$.
Its $T$ row is constant on each smooth spacelike axis, so only three
axis values enter the charges.

Let $\mathcal P^{[0]},\mathcal P^{[3]},\mathcal P^{[6]}$ denote its
algebraic values on the lower, middle and upper axes, respectively.
Subsection~\ref{app:analyticthermo} gives these values explicitly as
ratios of finite determinant coefficients, with no unevaluated axis
limit. Define the three-component row
\begin{equation}
 p_a^{[s]}=(A^{-1}\mathcal P^{[s]}A)^T{}_a,\qquad
 \Delta_\R p=p^{[3]}-p^{[0]},\quad
 \Delta_\Hh p=p^{[6]}-p^{[3]},\quad
 \Delta_\infty p=p^{[6]}-p^{[0]}.
 \label{eq:potentialjumps}
\end{equation}
All ADM and individual Komar charges are consequently
\begin{equation}
\begin{aligned}
 M_h&=\frac{3\pi\delta}{8G_5}\Delta_h p_T,&
 J_{hA}&=-\frac{\pi\delta}{4G_5}\Delta_h p_A,\\
 M&=\frac{3\pi\delta}{8G_5}\Delta_\infty p_T,&
 J_A&=-\frac{\pi\delta}{4G_5}\Delta_\infty p_A,
 \qquad A=\Psi,\Phi.
 \end{aligned}
 \label{eq:algebraiccharges}
\end{equation}
Adding the two differences cancels the common middle-axis value
$p^{[3]}$, leaving
only the difference between the outer axes. This explains additivity
directly in the algebraic formulas. It also follows geometrically. In vacuum,
$\dd\star\dd k^\flat=0$ for any Killing field $k$. Stokes' theorem on
a spatial exterior bounded by infinity and the two horizons gives
\begin{equation}
 M=M_\R+M_\Hh,\qquad J_A=J_{\R A}+J_{\Hh A}.
 \label{eq:sums}
\end{equation}
This requires a regular exterior with no additional boundary sources. The potential jumps
make the same conservation law explicit. We also checked this conservation law by extracting the charges
independently at infinity and on the horizons
(footnote~\ref{app:repro}).

\paragraph{Scale dependence of the thermodynamic quantities.}
The change from $ds_A^2$ to $ds_{\rm can}^2$ in~\eqref{eq:canonical} gives
\begin{equation}
 M_{\rm can}=M_A/\delta,\quad
 (J_{\rm can},\cA_{\rm can})=(J_A,\cA_A)/\delta^{3/2},\quad
 (\kappa_{\rm can},\Omega_{\rm can})=\sqrt\delta(\kappa_A,\Omega_A).
 \label{eq:scales}
\end{equation}
This operation changes the scale of all proper lengths by
$L=\delta^{-1/2}$. More generally, multiplying a vacuum line element
by $L^2$, and rescaling time so that its norm at infinity remains $-1$,
produces a geometrically similar configuration. At fixed $G_5$, its
mass is $L^2M$, its spins and horizon areas are $L^3J_A$ and
$L^3\cA_h$, and its temperatures and angular velocities are
$T_h/L$ and $\Omega_h^A/L$. Setting $L=\delta^{-1/2}$ gives
\eqref{eq:scales} and the entropy changes by the same factor as area. The same factors apply separately to each horizon. The dimensionless
spins $J_A/M^{3/2}$ and areas $\cA_h/M^{3/2}$, with the fixed $G_5$
factors in~\eqref{eq:reduced}, are unchanged by this conversion.

\subsection{Smarr relations, the first law and independent charges}
\label{sec:independence}

The individual Komar charges satisfy
\begin{equation}
 \frac23M_h=T_hS_h+\Omega_h^A J_{hA}.
 \label{eq:smarr}
\end{equation}
To derive this relation, insert $\xi=\chi_h-\Omega_h^A m_A$ into
the mass integral~\eqref{eq:komar}. With our orientation,
$\int_{\mathcal H_h}\star\dd\chi_h^\flat=-2\kappa_h\cA_h$, while
the remaining terms are $\Omega_h^A$ times the angular-momentum
integrals. The coefficients in~\eqref{eq:komar} then give
$M_h=3\kappa_h\cA_h/(16\pi G_5)+(3/2)\Omega_h^AJ_{hA}$,
which is~\eqref{eq:smarr}. This derives the identity without defining
the mass from the Smarr relation. Summing the two identities and using
\eqref{eq:sums} gives the total Smarr relation.

For variations between balanced stationary solutions with the same
asymptotic normalisation, the boundary-term derivation of black hole
mechanics~\cite{IyerWald,EEF} gives
\begin{equation}
 \dd M=\sum_{h=\R,\Hh}
       \left(T_h\dd S_h+\Omega_h^A\dd J_{hA}\right).
 \label{eq:firstlaw}
\end{equation}
Each horizon contributes its own boundary term, with its own intensive
variables. This is the two-spin form of disconnected horizon mechanics
in~\cite{EEF}. The varied parameters must continue to satisfy the axis
conditions~\eqref{eq:balance} otherwise a conical defect may appear
and its stress would contribute an additional work term. As a check,
a uniform change of length gives $\dd M=2M\,\dd\log L$ and
$(\dd S_h,\dd J_{hA})=3(S_h,J_{hA})\,\dd\log L$.
Equation~\eqref{eq:firstlaw} then reproduces the summed Smarr relation.

The algebraic Smarr identity for a component is not an isolated hole
differential law for its Komar mass. Changing a Saturn parameter changes
the interaction as well as the horizons. In general,
$\dd M_h-T_h\dd S_h-\Omega_h^A\dd J_{hA}$ is nonzero and only the sum
vanishes. Mutual thermal equilibrium imposes the further requirements
$T_\R=T_\Hh$ and $\Omega_\R^A=\Omega_\Hh^A$, studied in
section~\ref{sec:thermal}.

The physical content of the parameter count can be tested with the six
horizon quantities
\begin{equation}
 Q_h=(M_\R,J_{\R\Psi},J_{\R\Phi},M_\Hh,J_{\Hh\Psi},J_{\Hh\Phi}).
 \label{eq:horizonchargevector}
\end{equation}
At the reference solution, we compute the $6\times6$ derivative matrix
of these quantities with respect to $p$ in~\eqref{eq:controls}.
The computed matrix is invertible and remains so under refinement of
the derivatives. Thus, numerically, small changes of the six parameters
can independently change the six horizon charges. Only their three
sums are measured at infinity. We checked these derivatives under step refinement, together with
the first law (footnote~\ref{app:repro}).

\subsection{Axes, causality and the tested parameter domain}
\label{sec:nonextregularity}

Three kinds of points require separate treatment: the open exterior,
the rotation axes, and the horizons including their poles. In the open
exterior the metric coefficients must be smooth and nondegenerate.
On an axis the shrinking angular circle must close with circumference
$2\pi$ times its proper radius. At a horizon the coordinates of
section~\ref{sec:nonextcontinuation} must give a smooth extension.
At the auxiliary points $a_1,a_6$, the square roots introduced by the
construction cancel by~\eqref{eq:phantoms} and the remaining denominators
must also stay nonzero so that this cancellation leaves finite metric
components.

For example, on a $\Phi$ axis the no-conical-defect condition is
$\cG_{\Phi\Phi}=\rho^2f_0+O(\rho^4)$. The part of the metric
perpendicular to the axis is then
$f_0(\dd\rho^2+\rho^2\dd\Phi^2)+\cdots$, which is smooth in
$(\rho\cos\Phi,\rho\sin\Phi)$ for $f_0>0$ and a $2\pi$-periodic
$\Phi$. The axis matching of section~\ref{sec:balance} ensures that
this angle runs once around the shrinking circle. Where a horizon
meets an axis, subsection~\ref{app:nonextcharts} uses coordinates regular
in both directions at once.

To exclude closed timelike curves, it is insufficient to check the
$\Psi$ and $\Phi$ circles separately: a linear combination could still
be timelike. Instead, we consider their full $2\times2$ metric
$\Lambda=(\cG_{AB})_{A,B=\Psi,\Phi}$. At every finite off-axis
exterior point, sufficient conditions are finite smooth coefficients and

\begin{equation}
 f>0,\qquad \cG_{\Psi\Psi}>0,\qquad \det\Lambda>0.
 \label{eq:positivity}
\end{equation}
The last two inequalities make the angular block positive definite,
so no angular combination is timelike. The base is positive and the
Killing determinant negative, giving Lorentzian signature. More strongly,
the cofactor formula for the inverse metric gives
\begin{equation}
 g^{TT}=-\frac{\det\Lambda}{\delta^2\rho^2}<0.
 \label{eq:temporal}
\end{equation}
The angular identifications leave $T$ single-valued, and
\eqref{eq:temporal} says that its gradient is timelike. Thus $T$
increases strictly along every future causal curve, which rules out
closed causal curves and establishes stable causality. An ergoregion
has $g_{TT}>0$, but this concerns the stationary vector rather than
the gradient of $T$ and does not contradict $g^{TT}<0$.
At an axis a circle has zero size, so the strict angular inequalities
are replaced by the smooth Cartesian-axis conditions above. This
combination of metric extension and a global time function follows the
strategy used for singly spinning Saturns~\cite{CESSmooth,SzybkaCausal}.

Subsection~\ref{app:regularity} writes~\eqref{eq:positivity} as explicit
algebraic inequalities for the six-parameter Saturn metric. We check that they are obeyed at least in a certain region of parameter space, together with regularity of all finite
boundary components. A broader numerical scan varies every parameter
by up to $30\%$ around the reference configuration and finds no
regularity or causality violation. These results support a regular nonextremal
regime but a complete proof of causality and regularity in the entire exterior is beyond the scope of this work. These checks are given in
appendices~\ref{app:nonextbounds} and~\ref{app:broadscan}.
For the double-extremal examples in the next section, the bounds do
cover the entire exterior.

\section{Two extremal horizons}
\label{sec:extremal}
The black Saturn metric admits a limit in which both horizons are
extremal. We construct the limiting metric, give coordinates that extend
through its horizons and compute its thermodynamic quantities.
Regularity and causality are established for the parameter interval
and examples specified in subsection~\ref{app:extbalanced}.
Appendix~\ref{app:mixedextremal} treats the two other possibilities:
an extremal central hole surrounded by a nonextremal ring, and an
extremal ring surrounding a nonextremal hole.

\subsection{A correlated degeneration of the metric}

For a nonextremal horizon,~\eqref{eq:area} relates its Weyl rod length
to $\kappa_h\cA_h$. A finite-area extremal horizon therefore becomes a
point in the orbit space. The geometry near that point can nevertheless
contain an entire compact horizon, at the end of an infinite spatial
throat. To retain this geometry we must scale the rod lengths,
polarisations and normalisation together. We fix $\lambda>0$ and introduce finite constants
$(Z,c,d,h,v_*,w,r_0,\beta)$, with $Z,d,h,r_0>0$ and $\beta\ne0$. It is useful to define
\begin{align}
 s&=\sqrt{r_0^4+\lambda^2},&r_+&=\sqrt{s+\lambda},&
 z_c&=Z-\tfrac12(s-r_0^2),\nonumber\\
 a_2&=c\lambda,&a_3&=c\lambda+d\lambda^2,&
 a_4&=z_c-\lambda/2,\nonumber\\
 a_5&=z_c+\lambda/2,&a_6&=a_5+h/\lambda,&a_1&=0.
 \label{eq:extrods}
\end{align}
As $\lambda\to0^+$, the ring and hole rod lengths vanish as
$d\lambda^2$ and $\lambda$, while their positions approach $0$ and
$Z$. Thus $Z$ is the limiting separation in Weyl coordinates and $d$
controls the relative collapse rate. The auxiliary endpoint $a_6$
moves to infinity as $h/\lambda$. The remaining constants specify
how the rotation parameters approach this boundary of the family.
We keep the auxiliary-point conditions~\eqref{eq:phantoms} and set
\begin{align}
 \alpha&=-\frac{r_0}{\beta r_+},& b_0&=\frac{\beta r_0}{r_+},\nonumber\\
 B_4&=\frac{\sqrt{2\lambda}-2C_6\lambda b_0}
                 {C_6\sqrt{2\lambda}+b_0},&
 B_5&=\frac{\sqrt{2\lambda}(a_6-a_5)C_6\alpha}{a_5+1},\nonumber\\
 \eta&=1-C_1C_2C_3=c(1+Z^{-1})\lambda+w\lambda^2,&
 C_2&=\frac{1-\eta}{C_1C_3}.
 \label{eq:extpolar}
\end{align}
It is convenient to write the limiting parameter as a square,
so that the expressions can be expanded in integer powers of $\tau$:
\begin{equation}
 \lambda=\tau^2/2,\qquad h=H_*^2/4,\qquad C_3=v_*\tau.
 \label{eq:extconvention}
\end{equation}
The $+1$ in~\eqref{eq:extpolar} comes from the choice $b=-1$ in
$P_b$ of section~\ref{sec:metric}. Equations~\eqref{eq:extrods}-\eqref{eq:extconvention} specify how both
the rod endpoints and the Belinski-Zakharov polarisation constants
$(C_1,C_2,C_3,B_4,B_5,C_6)$ vary as $\lambda$ tends to zero.
The latter constants are the coefficients in the dressing vectors
of section~\ref{sec:metric} and they control the mixing of time and the
two rotational Killing directions. Thus the prescription fixes the
rotation data as well as the collapsing horizon rods.

The relation for $\eta$ cancels the leading pole in the dressed
Killing block. This is why the polarisation constants must be tuned
together with the rod lengths. The algebraic cancellation is displayed
in subsection~\ref{app:extscalar}.

The original Killing coordinates also become poorly normalised in
this limit. Before extracting the finite metric we apply a
$\tau$-dependent linear change of those coordinates, with matrix
\begin{equation}
 u_b=\tfrac12(\tau^{-1}+\tau),\quad
 v_b=\tfrac12(\tau^{-1}-\tau),\qquad
 \mathcal C=\begin{pmatrix}
 1&-v_b/C_6&-u_b/C_6\\0&u_b&v_b\\0&v_b&u_b
 \end{pmatrix},\quad \det\mathcal C=1.
\end{equation}
We substitute~\eqref{eq:extrods}--\eqref{eq:extconvention} in the metric,
apply this matrix, and expand in $\tau$. The notation $[\tau^0]$
means the coefficient independent of $\tau$ after cancelling common
powers. The limiting metric is
\begin{equation}
 G_e=[\tau^0](\mathcal C^{\mathsf T}G\mathcal C),\qquad
 f_e=[\tau^0](f/\lambda^3).
 \label{eq:extmetric}
\end{equation}
The multiplication of $f$ by a parameter-dependent constant is an
allowed integration-constant choice before normalisation. We then recompute
the outer kernels and cone factors of $(G_e,f_e)$, and apply the
canonical normalisation~\eqref{eq:canonical}. We denote the resulting
metric $(\bar G_e,\bar f_e)$ and have $\det\bar G_e=-\rho^2$.
Equations~\eqref{eq:rows}--\eqref{eq:f} and
\eqref{eq:extrods}--\eqref{eq:extmetric} are a finite algebraic
specification of the exact extremal spacetime.

Balance is imposed on this limiting metric by the same axis conditions
as in section~\ref{sec:balance}. The auxiliary path used to calculate
its coefficients need not itself be balanced at every positive
$\lambda$. Subsection~\ref{app:extscalar} evaluates~\eqref{eq:extmetric}
algebraically: it gives the finite scalar polynomials whose ratios are
the metric components at $\lambda=0$. This provides a formula that can
be used directly.

For the regularity analysis it is useful to replace the unbounded
half-plane $(\rho,z)$ by a square. Define the distances to the two
horizon points and the coordinates

\begin{equation}
 u=\frac{R_\R+R_\Hh-Z}{R_\R+R_\Hh+Z},\qquad
 y=\frac{R_\R-R_\Hh+Z}{2Z},\qquad
 R_\R=\sqrt{\rho^2+z^2},\quad R_\Hh=\sqrt{\rho^2+(z-Z)^2}.
 \label{eq:compact}
\end{equation}
The finite off-axis exterior is $0<u,y<1$. The three axes form the
edges $u=0$ and $y=0,1$ while the horizons are the corners $(0,0)$ and
$(0,1)$, and spatial infinity is $u=1$. On the subfamily studied below
the metric is rational in $u,y$, allowing uniform polynomial sign
bounds over the whole exterior.

\subsection{Explicit scalar form at zero temperature}
\label{sec:extscalar}

The correlated limit also has a scalar-function presentation. In the
Killing coordinates obtained after the transformation $\mathcal C$,
and before normalising the angular periods, define seven functions
$K,Y,W_\psi,W_\phi,V_{\psi\psi},V_{\psi\phi},V_{\phi\phi}$ by the
finite polynomial formulas of subsection~\ref{app:extscalar}. Then
\begin{equation}\begin{aligned}
 ds_e^2={}&\frac{1}{K}\bigl[-Y\,dt^2-2W_\psi\,dt\,d\psi
 -2W_\phi\,dt\,d\phi+V_{\psi\psi}\,d\psi^2\\
 &\hspace{20mm}+2V_{\psi\phi}\,d\psi\,d\phi
 +V_{\phi\phi}\,d\phi^2\bigr]+K(d\rho^2+dz^2).
\end{aligned}\label{eq:extscalarraw}\end{equation}
Here the scalar convention sets the analogue of $P$ to one:
$K=f_e$, $Y=-f_e(G_e)_{tt}$, $W_i=-f_e(G_e)_{ti}$ and
$V_{ij}=f_e(G_e)_{ij}$. These identities identify the functions and
their explicit evaluation does not require constructing $G_e$.
With $i,j=\psi,\phi$, put
\begin{equation}
 J_{ij}=\frac{YV_{ij}+W_iW_j}{K^2}.
 \label{eq:extscalarJ}
\end{equation}
The doubly spinning extremal metric can thus be written as
\begin{equation}\begin{aligned}
 ds_e^2={}&-\frac{Y}{K}
 \left(dt+\frac{W_\psi}{Y}d\psi+\frac{W_\phi}{Y}d\phi\right)^2\\
 &+K\left[d\rho^2+dz^2+
 \frac{J_{\psi\psi}d\psi^2+2J_{\psi\phi}d\psi\,d\phi
       +J_{\phi\phi}d\phi^2}{Y}\right].
\end{aligned}\label{eq:extscalarcompleted}\end{equation}
As in the nonextremal case, $Y=0$ is an ergosurface rather than a
singularity caused by completing the square. At such a surface use
\eqref{eq:extscalarraw}, whose coefficients do not divide by $Y$. In particular, $J_{\psi\psi}J_{\phi\phi}-J_{\psi\phi}^2
=\rho^2Y/K$.

For the physical canonical metric, let $B_e$ be the constant basis
obtained by the outer-axis normalisation and canonical rescaling
specified after~\eqref{eq:extmetric}. Write
$dt=dT+q_A d\varphi^A$, $dx^i=b_{iA}d\varphi^A$,
where $(\varphi^1,\varphi^2)=(\Psi,\Phi)$ have period $2\pi$,
and let $\delta_e$ be the determinant of the outer-axis basis before
canonical rescaling. The physical scalar functions are explicitly
\begin{equation}\begin{aligned}
 \bar K&=K/\delta_e,\qquad \bar Y=Y/\delta_e,\qquad
 \bar W_A=(Yq_A+b_{iA}W_i)/\delta_e,\\
 \bar V_{AB}&=\frac{b_{iA}b_{jB}V_{ij}
   -q_A b_{iB}W_i-q_B b_{iA}W_i-Yq_Aq_B}{\delta_e}.
\end{aligned}\label{eq:extscalarphysical}\end{equation}
Replacing all functions in~\eqref{eq:extscalarraw}--\eqref{eq:extscalarcompleted}
by their barred versions gives $(\bar G_e,\bar f_e)$.

Subsection~\ref{app:extscalar} retains every parameter of the correlated
limit before balance is imposed. It expresses the seven functions by
finite polynomial arithmetic, so the metric can be evaluated directly
at zero temperature. The \ancillarylink{ancillary Mathematica file} implements these components.
Subsection~\ref{app:extbalanced} specifies a balanced parameter interval
and further examples of regular doubly extremal black Saturns, together
with the bounds used to establish their regularity.

\subsection{Coordinates through the degenerate horizons}
\label{sec:extcharts}

Although both horizon rods have collapsed, their neighbourhoods can
be resolved by polar coordinates about the points $z_H=0,Z$.
We first work with $R>0$ in the exterior, then continue $R$ through
zero to obtain an interior coordinate:
\begin{equation}
 \rho=R\sin\theta,\qquad z=z_H+R\cos\theta.
 \label{eq:signed}
\end{equation}
The choice of square-root branch is essential. The root centred at
the chosen horizon must continue as the signed coordinate $R$.
Replacing it by $|R|$ would create a cusp at $R=0$.
At the ring the two roots in~\eqref{eq:compact} are
$(R,\sqrt{Z^2-2ZR\cos\theta+R^2})$, and at the hole they are
$(\sqrt{Z^2+2ZR\cos\theta+R^2},R)$. The other root is analytic
for $|R|<Z$.

In canonical coordinates set
\begin{equation}
 F=R^2\bar f_e,\quad a=\frac{\sin^2\theta}{\det\Lambda},\quad
 \beta^A=(\Lambda^{-1})^{AB}(\bar G_e)_{TB},\quad
 \varphi^A=\Phi^A-\Omega_H^AT,\quad
 k^A=\frac{\beta^A+\Omega_H^A}{R}.
\end{equation}
Here $\Phi^A=(\Psi,\Phi)$ and $\Lambda$ is the canonical angular
block. Completing the angular square yields
\begin{equation}
 ds^2=-R^2a\dd T^2+F\left(\frac{\dd R^2}{R^2}+\dd\theta^2\right)
 +\Lambda_{AB}(\dd\varphi^A+Rk^A\dd T)
                       (\dd\varphi^B+Rk^B\dd T).
 \label{eq:extlocal}
\end{equation}
Write each coefficient as a Taylor expansion in signed $R$, for
example $F=F_0+RF_1+O(R^2)$. Direct expansion of the metric shows that $\Omega_H^A,k_0^A$ and
\begin{equation}
 C^2=\frac{a_0}{F_0}>0,\qquad
 T_1=\frac{a_1/C^2-F_1}{2CF_0}
 \label{eq:extconstants}
\end{equation}
are independent of $\theta$ on each horizon, with $C>0$.
These constants determine the coordinate shifts needed to remove the
$1/R^2$ and $1/R$ radial poles. Since they do not depend on angle, the
shifts define actual coordinates:
\begin{equation}
 T=v+\frac1{CR}+T_1\log|R|,\qquad
 \varphi^A=\chi^A+\frac{k_0^A}{C}\log|R|.
 \label{eq:extingoing}
\end{equation}
This pole-cancellation construction is related to the extremal
extensions in~\cite{ArmasDipole,DurkeePS}. With
\begin{equation}
 U_R=-\frac1{CR^2}+\frac{T_1}{R},\qquad
 Q^A=\frac{k_0^A-k^A}{CR}+T_1k^A,
\end{equation}
the transformed metric is
\begin{align}
 ds^2={}&-R^2a(\dd v+U_R\dd R)^2+\frac F{R^2}\dd R^2
             +F\dd\theta^2\nonumber\\
 &+\Lambda_{AB}(\dd\chi^A+Rk^A\dd v+Q^A\dd R)
                    (\dd\chi^B+Rk^B\dd v+Q^B\dd R).
 \label{eq:extregular}
\end{align}
Since $k-k_0=O(R)$, $Q$ is analytic. The apparently singular radial
coefficient reduces to
\begin{equation}
 \frac{F-a/C^2}{R^2}+\frac{2aT_1}{CR}-aT_1^2,
\end{equation}
whose $R^{-2}$ and $R^{-1}$ terms vanish by~\eqref{eq:extconstants}.
The resulting analytic metric obeys
\begin{equation}
 g_{vR}|_{R=0}=CF_0\ne0,\qquad
 \det g_{(v,R,\theta,\chi^1,\chi^2)}=-F^2\sin^2\theta.
 \label{eq:extdet}
\end{equation}
Thus $R=0$ is a regular null hypersurface in $(v,R,\theta,\chi^A)$.
Reversing the radial shifts gives an outgoing chart. The generator
$\partial_v$ has vanishing surface gravity, as required for an
extremal horizon.

The coordinates above still have the usual polar-coordinate
singularity at $\theta=0,\pi$. To include the poles of the compact
horizon, replace the shrinking angle by Cartesian coordinates. Put
$x_p=\sqrt{2(1-|\cos\theta|)}$,
$X=x_p\cos\chi_{\rm deg}$, $Y=x_p\sin\chi_{\rm deg}$.
The shrinking cycle is $\Phi$ at both ring poles and the lower hole
pole, and $\Psi$ at the upper hole pole. Exact kernel and unit-cone
identities, together with joint analyticity in $(R,\cos\theta)$,
remove the polar singularities even at $R=0$. The Cartesian
determinant is $-F^2$, and the remaining circle has positive norm.
The primitive angular lattice excludes an orbifold quotient.

\subsection{Global exterior and interpretation of the horizons}

For the balanced parameter interval of proposition~\ref{prop:ext}
and the seven roots in table~\ref{tab:extareas}
(subsection~\ref{app:extbalanced}), the denominators of the exact rational
metric stay nonzero throughout $0<u,y<1$, and the canonical versions
of~\eqref{eq:positivity} hold there. After removing the known factors associated with a shrinking circle,
the inequalities also hold on each open axis. The
charts above cover the two focal corners. At infinity we use
\eqref{eq:asymptoticcoordinates} with $\delta=1$, because this metric
is in the canonical convention. Expanding the limiting metric itself
gives~\eqref{eq:asymptoticleading} and hence the flat metric
\eqref{eq:asymptoticflat}. The expansion must be made after taking
the extremal limit: the auxiliary endpoint $a_6$ recedes to infinity,
so an interchange of these limits cannot be assumed. The all-angle
bounds described in subsection~\ref{app:regularity} control the
asymptotic end of this limiting metric. These statements establish a complete exterior cover
for that interval and those individual roots. They do not establish
regularity for arbitrary constants in the extremal ansatz.

The resulting global time has timelike gradient. More explicitly,
the lapse after completing the angular square is
$N_T^2=\rho^2/\det\Lambda>0$. Along an orbit-space path to the
asymptotic end, choose $\dd\Phi^A=-\beta^A\dd T$ and increase
$\dd T$ sufficiently to obtain a future timelike escaping curve.
At a future horizon a future causal vector $V$ satisfies
$g(\partial_v,V)=CF_0V^R\leq0$ and cannot cross outward from the
locally attached interior. This identifies the outer null boundaries
in the chosen future extension. It would be interesting to pursue a maximal analytic extension.

\subsection{Thermodynamics at double extremality}
\label{sec:extthermo}

The constants in the regular canonical chart give the entropies
without integrating a horizon section. Indeed,
$F_0\det\Lambda_0=\sin^2\theta/C_h^2$ by
\eqref{eq:extconstants}, and the two angles have period $2\pi$.
Consequently
\begin{equation}
 T_h=0,\quad\cA_h=\frac{8\pi^2}{C_h},\quad
 S_h=\frac{2\pi^2}{G_5C_h},\quad
 S_{\rm tot}=\frac{2\pi^2}{G_5}
            \left(C_\R^{-1}+C_\Hh^{-1}\right).
 \label{eq:extanalyticentropy}
\end{equation}
Here $C_h$ and $\Omega_h^A$ are constants determined by the exact
metric near the horizon. To extract them, expand a component at fixed
$\theta$ in the signed radial coordinate $R$ of
section~\ref{sec:extcharts}. We write $[R^n]Q$ for the coefficient of
$R^n$ in this expansion: if $Q=Q_0+RQ_1+O(R^2)$, then
$[R^0]Q=Q_0$. In particular, $\bar f_e$ has a second-order pole,
but $R^2\bar f_e$ has a finite constant coefficient.
Since $C_h$ and $\Omega_h^A$ are independent of $\theta$, they may be
computed at the equator $\theta=\pi/2$, where $\sin^2\theta=1$:
\begin{equation}\begin{gathered}
 L_h=[R^0]\Lambda,\quad F_h=[R^0](R^2\bar f_e),\quad
 (b_h)_A=[R^0](\bar G_e)_{AT},\\
 C_h=(F_h\det L_h)^{-1/2},\qquad
 \Omega_h=-L_h^{-1}b_h.
\end{gathered}
 \label{eq:extalgebraicconstants}
\end{equation}
Here $L_h$ is the $2\times2$ angular metric, $b_h$ is a two-component
column with $A\in\{\Psi,\Phi\}$, and the positive root is chosen for
$C_h$. 

The central and ring extremal area-spin relations provide useful
additional consistency checks:
\begin{equation}
 \cA_\Hh=8\pi G_5\sqrt{|J_{\Hh\Psi}J_{\Hh\Phi}|},\qquad
 \cA_\R=8\pi G_5|J_{\R\Phi}|.
 \label{eq:extareaspin}
\end{equation}
They agree with the spherical and ring extremal near-horizon
relations~\cite{HollandsExt}. They do not identify the whole
interacting metric with a pair of isolated solutions. The horizon
angular velocities, redshifts and Komar masses retain the interaction.

There is a short way to obtain the individual extremal charges.
The area--spin relations~\eqref{eq:extareaspin} fix one ring spin and the product of the
central spins, while conservation fixes the remaining combinations.
Together with the independently determined ADM spins this determines
all four horizon spins. The horizon Smarr relations then give the two
Komar masses. On a connected regular branch fix
$\sigma_\R=\operatorname{sgn}J_{\R\Phi}$ and
$\sigma_\Hh=\operatorname{sgn}(J_{\Hh\Psi}J_{\Hh\Phi})$.
The near-horizon identities~\eqref{eq:extareaspin} and charge
conservation give
\begin{equation}
\begin{aligned}
 J_{\R\Phi}&=\frac{\sigma_\R\pi}{G_5 C_\R},&
 J_{\Hh\Phi}&=J_\Phi-\frac{\sigma_\R\pi}{G_5 C_\R},\\
 J_{\Hh\Psi}&=\frac{\sigma_\Hh\pi^2}
                         {G_5^2 C_\Hh^2 J_{\Hh\Phi}},&
 J_{\R\Psi}&=J_\Psi-J_{\Hh\Psi},\\
 M_h&=\frac32\left(\Omega_h^\Psi J_{h\Psi}
                         +\Omega_h^\Phi J_{h\Phi}\right).
 \end{aligned}
 \label{eq:extanalyticcharges}
\end{equation}
A finite nonzero central area excludes $J_{\Hh\Phi}=0$. The signs
are discrete branch data, not freely adjustable continuous parameters.
In the angular orientation of the extremal phase plots they are
$(\sigma_\R,\sigma_\Hh)=(+1,-1)$. Here the final equality uses the
extremal horizon Smarr identity. Direct Komar surfaces provide an
independent check of every local charge, including the masses.
The ADM mass is obtained separately at infinity and agrees with
$M_\R+M_\Hh$.

Explicitly, in the canonical convention put $\xi=1/R_W$, $z=0$ and $\rho=1/\xi$.
The notation $g_{Ta}^{(1)}=[\xi^1](\bar G_e)_{Ta}$ denotes the
coefficient linear in inverse radius in the asymptotic expansion.
In the asymptotic coordinates of~\eqref{eq:asymptoticcoordinates},
$z=0$ corresponds to $\theta=\pi/4$ (distinct from the local horizon
polar angle used above). The ADM charges are
\begin{equation}
 M=\frac{3\pi}{4G_5}g_{TT}^{(1)},\qquad
 J_\Psi=-\frac\pi{G_5}g_{T\Psi}^{(1)},\qquad
 J_\Phi=-\frac\pi{G_5}g_{T\Phi}^{(1)}.
 \label{eq:extanalyticadm}
\end{equation}
The rational functions in~\eqref{eq:compact} make this a finite
polynomial calculation: at $z=0$ we substitute
$u=1-Z\xi+O(\xi^2)$, $y=\tfrac12-Z\xi/4+O(\xi^2)$. After cancelling the
common leading power, the linear coefficient of
$(n_0+n_1\xi)/(d_0+d_1\xi)$ is $(n_1d_0-n_0d_1)/d_0^2$.
Equations~\eqref{eq:extalgebraicconstants}--\eqref{eq:extanalyticadm}
therefore specify all extremal thermodynamic quantities.

\section{Phase diagrams, blackfold predictions and frame dragging}
\label{sec:phases}

The thermodynamic data allow three complementary comparisons. First,
we compare solutions in mutual thermal and rotational equilibrium at
fixed angular momentum in the second rotation plane. Next, we follow a double-extremal
Saturn curve, whose two angular momenta both vary, and compare it with
isolated black holes, rings and the blackfold approximation. Finally,
we hold conserved charges fixed to study non-uniqueness and isolate
frame dragging in the second rotation plane.

\subsection{Reduced variables}
At fixed ADM mass, entropy is proportional to total horizon area.
Following the five-dimensional phase-diagram conventions
of~\cite{ER,EF,EEF}, we remove the overall length scale by defining
\begin{equation}\begin{gathered}
 j_A=\sqrt{\frac{27\pi}{32G_5}}\frac{J_A}{M^{3/2}},\qquad
 a_h=\frac3{16}\sqrt{\frac3\pi}\frac{\cA_h}{(G_5M)^{3/2}},\\
 a_{\rm tot}=a_\R+a_\Hh,\qquad
 \omega_{hA}=\sqrt{\frac{8G_5M}{3\pi}}\Omega_{hA}.
\end{gathered}
 \label{eq:reduced}
\end{equation}
Individual reduced Komar spins use the same total $M$. Since there
are two ADM spins, a phase comparison requires a surface or fixed-spin
slices. For a Myers-Perry (MP) black hole with parameters $(\mu,a,b)$,
$j_\Psi=a/\sqrt\mu$, $j_\Phi=b/\sqrt\mu$, we have that
\begin{equation}
 a_{\rm MP}=2\sqrt{1-j_\Psi^2-j_\Phi^2+
 \sqrt{[1-(j_\Psi+j_\Phi)^2][1-(j_\Psi-j_\Phi)^2]}},\qquad
 |j_\Psi|+|j_\Phi|\leq1.
 \label{eq:MPphase}
\end{equation}
For a Pomeransky-Sen'kov (PS) ring, we take $\Psi$ as the ring plane and
$0\leq\nu<1$, $2\sqrt\nu\leq\lambda<1+\nu$. The positive-spin sheet is
\begin{align}
 j_\Psi&=\frac{(1+\lambda-6\nu+\lambda\nu+\nu^2)
                  \sqrt{1+\lambda+\nu}}{2\sqrt{2\lambda}(1-\nu)^2},\nonumber\\
 j_\Phi&=\frac{(1-\lambda+\nu)\sqrt{\nu(1+\lambda+\nu)}}
                   {\sqrt{2\lambda}(1-\nu)^2},\nonumber\\
 a_{\rm PS}&=\frac{2(1+\lambda+\nu)(1-\lambda+\nu)^{3/2}}
 {\sqrt\lambda(1-\nu)^2(y_h^{-1}-y_h)},\qquad
 y_h=-\frac2{\lambda+\sqrt{\lambda^2-4\nu}}.
 \label{eq:PSphase}
\end{align}
Both PS branches are included wherever the same pair of spins admits
two rings. Reflection of either angular coordinate supplies other
sign sectors.

\subsection{Configurations in mutual thermal equilibrium}
\label{sec:thermal}
We now restrict the balanced black Saturn family to configurations in
mutual thermodynamic equilibrium.
Mutual equilibrium imposes three conditions in addition to balance:
\begin{equation}
 \kappa_\R=\kappa_\Hh,\qquad
 \Omega_\R^\Psi=\Omega_\Hh^\Psi,\qquad
 \Omega_\R^\Phi=\Omega_\Hh^\Phi.
 \label{eq:thermal}
\end{equation}
The temperature equality prevents heat exchange, while the two
angular-velocity equalities prevent rotational exchange in either
plane. At zero temperature the first condition is automatic, but the
two angular conditions remain nontrivial. The double-extremal solution
of section~\ref{sec:extremal} cannot satisfy both while retaining a
regular axis between the horizons: appendix~\ref{app:extremequilibrium}
proves this for the full finite parameter chart and every real sign
sector. We provide strong numerical evidence that doubly spinning extremal black saturns cannot be in thermal equilibrium, which was already predicted in the blackfold regime \cite{AHOExt}.

\begin{figure}[htbp]\centering
 \includegraphics[width=\textwidth]{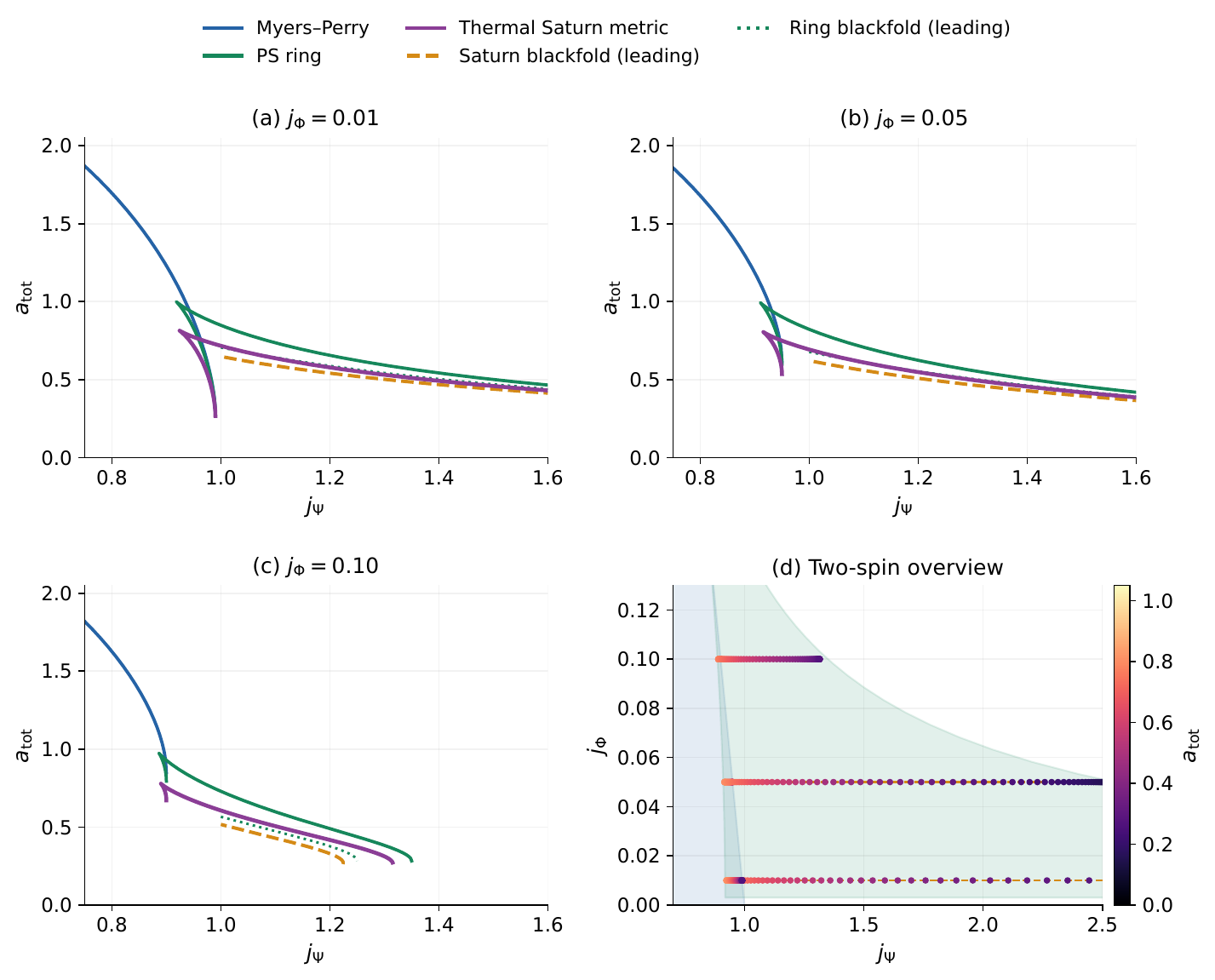}
 \caption{Mutual thermal equilibrium. Panels (a)--(c) compare total
 reduced area at fixed $j_\Phi=0.01,0.05,0.10$ while (d) locates these
 sections in the two-spin plane, with colour denoting Saturn area.
 Purple: the auxiliary-boundary metric~\eqref{eq:thermalmetric},
 blue and green: Myers-Perry and Pomeransky-Sen'kov.
 Dashed orange and dotted green: leading blackfold Saturn and
 isolated-ring predictions, displayed only for $j_\Psi\geq1$.
 Panel (c) is a qualitative comparison as the blackfold approximation is not strictly valid there. The hierarchy
 of scales and the larger mismatch at $j_\Phi=0.10$ are discussed in
 appendix~\ref{app:blackfold}. The Saturn sections are numerical
 continuations, not a complete equilibrium surface.}
 \label{fig:thermal}\label{fig:thermaloverview}
\end{figure}

At nonzero temperature, solving these conditions within the balanced Saturn family
leads to a branch approaching the auxiliary boundary $a_6=a_5$ while both
physical horizon rods retain nonzero length. Thus neither horizon
becomes extremal at this boundary. A finite expression for the
limiting metric follows by setting
\begin{equation}
 a_6=a_5+\epsilon^2,\quad B_5=\epsilon\bar B_5,\quad
 C_6=\frac{\bar C_6}{\epsilon}+O(\epsilon),\quad
 \bar C_6^2=\frac{a_5}{2(a_5-a_2)}.
\end{equation}
The other polarisations are finite. We collect the exact monomials of
$f$ and $fG$ before taking the coefficient:
\begin{equation}
 f_*=[f]_{\epsilon^0},\qquad
 G_*=[fG]_{\epsilon^0}/f_*.
 \label{eq:thermalmetric}
\end{equation}
The retained powers have nonnegative degrees, and $u_6=u_5$ in the
coefficient. This defines the complete limiting vacuum metric on
its regular off-axis domain, including its conformal factor. For each fixed $j_\Phi$, we trace a one-parameter family using
\[
 q=\log\frac{a_4-a_3}{a_2-a_1}.
\]
Here $a_4-a_3>0$ is the Weyl-coordinate gap between the ring and
central-hole horizon rods. We use $a_2-a_1=1$ to fix the scale in the
continuation. This gap is a coordinate parameter, not the proper
distance between the horizons. At every step we adjust the remaining
parameters to solve balance, the three equilibrium conditions and the
chosen value of $j_\Phi$. Figure~\ref{fig:thermal} shows the resulting
slices at $j_\Phi=0.01,0.05,0.10$. Points are connected in increasing
$q$, so the curve continues smoothly even where $j_\Psi(q)$ reverses
direction. Near such a reversal, two distinct configurations can have
the same $j_\Psi$ and different areas. We checked the exterior and rod signs along this branch, providing
numerical evidence for regularity (footnote~\ref{app:repro}).

In mutual equilibrium the first law simplifies to
\begin{equation}
 \dd M=\frac\kappa{8\pi G_5}\dd(\cA_\R+\cA_\Hh)
                      +\Omega_A\dd J_A.
\end{equation}
At fixed mass, this gives
\begin{equation}
 \dd a_{\rm tot}=-2\sqrt2\sum_A\frac{\Omega_A}\kappa\dd j_A.
\end{equation}
Equivalently, along a fixed-$j_\Phi$ curve,
\[
 \frac{\dd a_{\rm tot}}{\dd q}
 =-2\sqrt2\,\frac{\Omega_\Psi}{\kappa}
                   \frac{\dd j_\Psi}{\dd q}.
\]
At a turning point of $j_\Psi(q)$ the right-hand side vanishes, so
$a_{\rm tot}(q)$ is stationary as well, provided
$\Omega_\Psi/\kappa$ is finite. If this ratio is also nonzero, the two
stationary-point conditions are equivalent. This explains the paired
area and spin extrema seen when the branch doubles back in the plot.

On the plotted equilibrium branches, the Saturn has a smaller total
reduced area than the competing PS ring at the same two ADM spins,
where both exist. This is consistent with the singly spinning
equilibrium phase diagram of~\cite{EF,EEF}. The leading nonextremal
blackfold curves are derived in appendix~\ref{app:blackfold}.
They reproduce part of the thin-ring behaviour, while agreement
deteriorates on the $j_\Phi=0.10$ slice. That appendix explains the
scale hierarchy and endpoint sensitivity responsible for treating
this slice as a qualitative comparison.

\subsection{A double-extremal phase curve}
The extremal comparison that we provide here is outside mutual rotational equilibrium.
Both temperatures vanish, but the regular double-extremal family
cannot have both angular velocities equal, as proved in
appendix~\ref{app:extremequilibrium}. The blackfold curves are likewise
non-equilibrium configurations in their controlled regime. Fix~\eqref{eq:extfixed} and
continue the exact coefficient metric by varying $d$ and solving
balance for $(c,H_*,v_*)$. The sampled curve spans $0.00014\leq d\leq14$, with two positive
finite areas at each point. Increasing $d$ grows the ring from a
small perturbation to a component with an area larger than the centre.
Some representative numerical examples used to plot these curves are
\begin{center}
\begin{tabular}{rrrr}\toprule
 $d$&$j_\Psi$&$j_\Phi$&$a_{\rm tot}$\\\midrule
 $0.00014$&$0.597991542$&$-0.402030258$&$1.386724266$\\
 $0.014$&$0.602759415$&$-0.399402423$&$1.377742570$\\
 $14$&$0.895587912$&$0.123039029$&$0.438320709$\\\bottomrule
\end{tabular}
\end{center}
ADM charges are obtained from the exact asymptotic coefficients,
and areas from exact horizon coefficients. All these examples are globally regular.

For a meaningful blackfold comparison we must specify what is held
fixed. At each exact Saturn point, we choose the isolated reference
centre to have the same
$(\cA_\Hh,J_{\Hh\Psi},J_{\Hh\Phi})$.
The central-horizon identity~\eqref{eq:extareaspin} is precisely the
extremal MP area--spin relation, so the isolated reference centre is
extremal. Matching the Weyl separation fixes
$R^2=2Z+(\mu-a^2-b^2)/2$, and the transverse potential fixes
$m=k/(2|\Omega_\R^\Phi|)$ and
$\alpha=\operatorname{sign}\Omega_\R^\Phi$. Radial balance
determines the orbital velocity providing a prediction for the curve of doubly-spinning extremal Saturns.

At the thin end $m/\mathcal R\simeq1.02\times10^{-6}$, the
relative discrepancies are $3.74\times10^{-6}$ in ring orbital
spin, $2.05\times10^{-6}$ in transverse spin and area, and
$2.49\times10^{-7}$ in the ADM mass increment. At $d=0.014$ they
are approximately one hundred times larger. Agreement improves
with decreasing thickness, as expected from a leading blackfold
expansion. At the thick end $m/\mathcal R\simeq0.182$, the orbital
spin discrepancy is about $40\%$ and is not surprising since this part of the blackfold line
is an extrapolation beyond its regime of validity.

\begin{figure}[htbp]\centering
 \includegraphics[width=\textwidth]{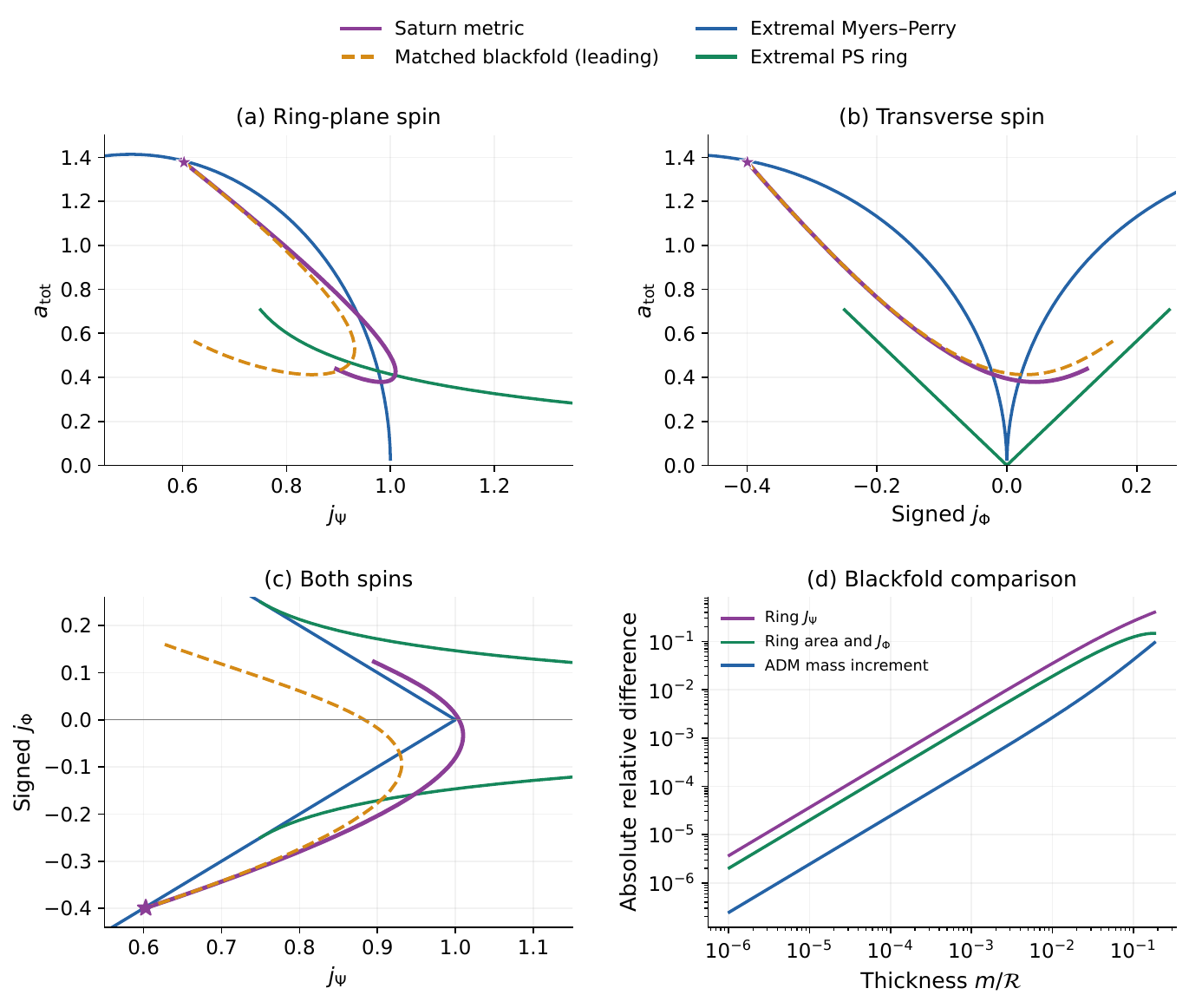}
 \caption{Double-extremal configurations. Panels (a) and (b) show
 area against each ADM spin, (c) shows the same configurations in
 the two-spin plane, (d) compares the blackfold prediction with
 the exact ring charges, area and ADM mass.
 Solid purple: the limiting Saturn metric, dashed orange:
 matched leading blackfold, blue and green: extremal MP and PS.
 The star marks $d=0.014$. Both spins vary along the Saturn curve,
 and the two horizons need not have equal angular velocities.}
 \label{fig:extphase}\label{fig:extdetail}
\end{figure}

The isolated extremal boundaries are
\begin{align}
 \text{MP: }&|j_\Psi|+|j_\Phi|=1,&
 a_{\rm MP}&=2\sqrt{2|j_\Psi j_\Phi|},\nonumber\\
 \text{PS: }&j_\Psi=\frac1{8|j_\Phi|}+|j_\Phi|,&
 a_{\rm PS}&=2\sqrt2|j_\Phi|,\quad0<|j_\Phi|<\tfrac14.
\end{align}
They are shown with the solid Saturn curve and the
matched dashed blackfold curve in figure~\ref{fig:extphase}. The
two-spin view in panel (c) shows which configurations can be
compared at the same conserved charges. Panel (a) suppresses $j_\Phi$:
two points at the same horizontal position can therefore have
different transverse spins, even when their areas coincide.
Likewise, panel (b) suppresses $j_\Psi$. In panel (c), two
configurations have the same pair $(j_\Psi,j_\Phi)$ only if they
occupy the same point. After fixing the common mass, an entropy
comparison is meaningful only at such matching spin pairs.
The Saturn curve crosses total
$j_\Phi=0$ by cancellation between the two horizon contributions,
without either horizon losing extremality. At $d=0.014$,
$|j_\Psi|+|j_\Phi|>1$ and $|j_\Phi|>1/4$, so this regular example from
subsection~\ref{app:extbalanced} has no regular isolated MP hole or balanced
PS ring with the same ADM mass and both angular momenta. The MP bound
is $|j_\Psi|+|j_\Phi|\leq1$, whereas PS requires $|j_\Phi|\leq1/4$.

\FloatBarrier

\subsection{Non-uniqueness and dragging in the second plane}

With six independent horizon charges but only three sums measured
at infinity, one expects continuous non-uniqueness. We exhibit this
by following a numerical family with fixed ADM charges ($G_5=1$):
\begin{equation}\begin{gathered}
 (M,J_\Psi,J_\Phi)=(16.655554840,38.214561224,2.087501673),\\
 (j_\Psi,j_\Phi)=(0.915318098,0.05),
\end{gathered}
\end{equation}
while $a_{\rm tot}$ ranges from $0.696024735$ to $0.746032225$.
At these charges the MP area is $1.078054506$ and the two PS areas
are $0.978641732$ and $0.972337169$. This particular Saturn path
has smaller entropy than MP and PS black holes. The
varying area gives direct numerical non-uniqueness at fixed
conserved charges.

\begin{figure}[htbp]\centering
 \includegraphics[width=.48\textwidth]{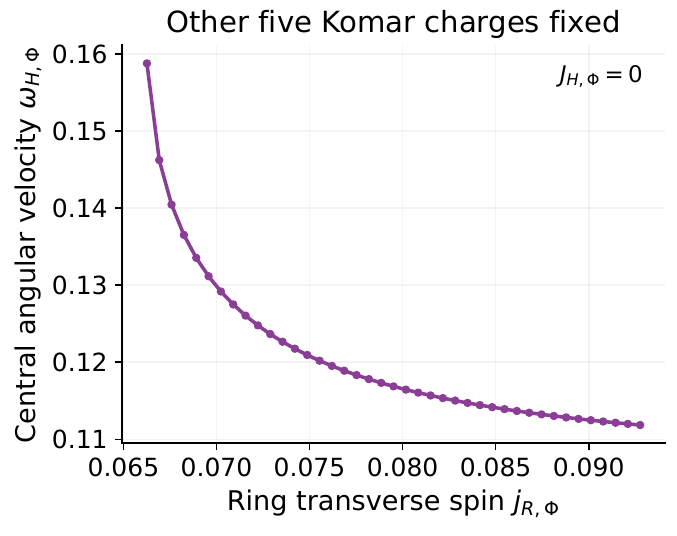}\hfill
 \includegraphics[width=.48\textwidth]{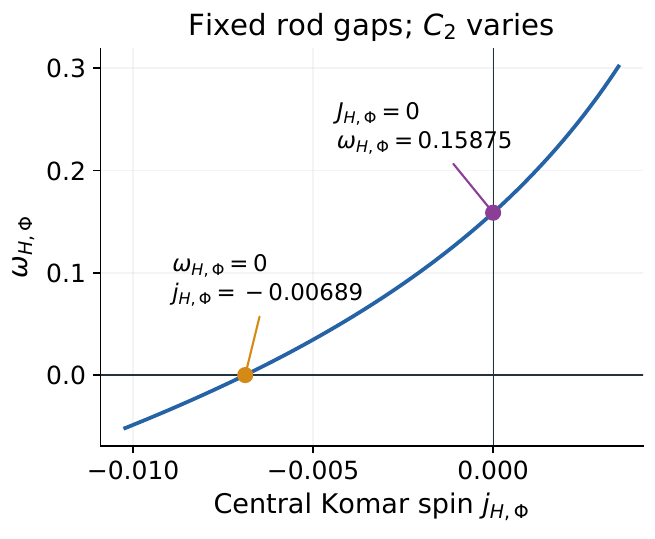}\\[6pt]
 \includegraphics[width=.62\textwidth]{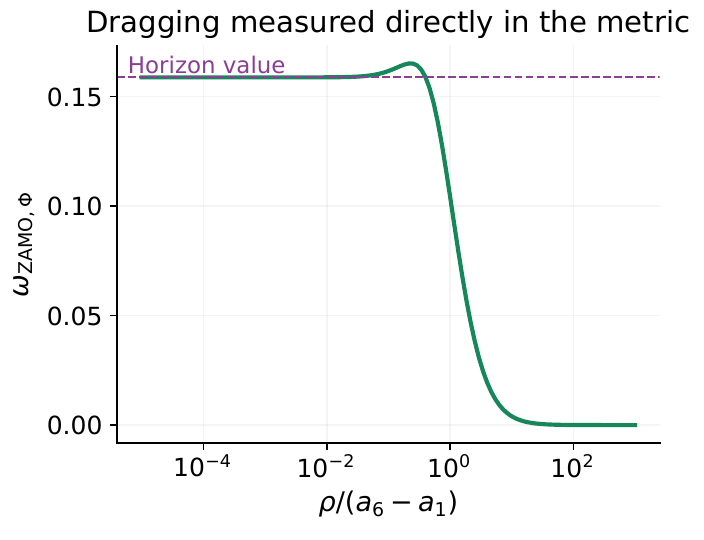}
 \caption{Transverse frame dragging. Upper left: the central
 angular velocity changes as the ring transverse spin varies, with
 the other five Komar charges fixed and $J_{\Hh\Phi}=0$. Upper right:
 on a separate fixed-rod continuation, a central counter-spin is
 needed to make $\omega_{\Hh\Phi}=0$. Lower panel: the local
 zero-angular-momentum angular velocity approaches the central
 horizon value and decays to zero at infinity.}
 \label{fig:dragging}
\end{figure}

For the second-plane analogue of the original Saturn dragging
experiment in~\cite{EF}, we impose $J_{\Hh\Phi}=0$ while retaining the central
orbital spin. Solving balance together with this zero-spin condition gives a
configuration whose central horizon still rotates in the transverse
plane relative to the nonrotating frame at infinity.
For this example and figure~\ref{fig:dragging}, we reverse both
angular coordinates relative to the $A$-normalised basis of
section~\ref{sec:metric}: $(\Psi,\Phi)\mapsto(-\Psi,-\Phi)$.
This reverses both $J_{hA}$ and $\Omega_h^A$ for each plane and
leaves products $\Omega_h^A J_{hA}$ unchanged.
In this orientation the horizon charges are
\begin{equation}\begin{aligned}
 (M_\R,J_{\R\Psi},J_{\R\Phi})&=(13.676724457,33.413964321,2.706293961),\\
 (M_\Hh,J_{\Hh\Psi},J_{\Hh\Phi})&=(2.736435333,3.792434226,0).
\end{aligned}\end{equation}
Direct integration over the central horizon gives
$|J_{\Hh\Phi}|<10^{-24}$, whereas
\begin{equation}
 \omega_{\Hh\Phi}=0.158748079001\ne0.
 \label{eq:dragging}
\end{equation}
For comparison, an isolated five-dimensional MP hole obeys
\[
 J_\Phi^{\rm MP}=\frac{\pi\mu b}{4G_5},
 \qquad \Omega_\Phi^{\rm MP}=\frac{b}{r_+^2+b^2}.
\]
For positive mass and a regular horizon, $J_\Phi^{\rm MP}=0$
therefore implies $b=0$ and $\Omega_\Phi^{\rm MP}=0$, even if the
other spin is nonzero. For the central Saturn horizon, by contrast, $J_{\Hh\Phi}=0$
while $\Omega_\Hh^\Phi\ne0$. The surrounding rotating ring therefore
induces transverse rotation relative to infinity even when the central
horizon carries no transverse Komar angular momentum. This rotation is induced by the ring's gravitational field.

To separate this interaction from changing the other horizon charges,
we hold the remaining five Komar charges fixed and increase
$J_{\R\Phi}$ by up to $40\%$.
The central transverse spin stays zero, while
$\omega_{\Hh\Phi}$ decreases from $0.158748079001$ to
$0.111864934012$. Both masses, both ring-plane spins and $J_{\Hh\Phi}=0$ are
held fixed, but balance requires the shape and separation to adjust.
The decreasing angular velocity therefore measures the response along
this constrained family, it does not assert that increasing ring spin
at fixed geometry weakens dragging. Conversely, a separate
continuation at fixed rod gaps, varying $C_2$ and readjusting the
balance parameters, reaches
$\omega_{\Hh\Phi}=0$ with
$j_{\Hh\Phi}=-0.006885711854\ne0$. Here a central counter-spin
cancels the induced horizon rotation.

The local shift seen by a zero-angular-momentum observer is
\begin{equation}
 \Omega_{\rm ZAMO}^A=-(\Lambda^{-1})^{AB}\cG_{TB}.
\end{equation}
This follows by setting the two angular covariant momenta
$p_A=\Lambda_{AB}\dot\varphi^B+\cG_{AT}\dot T$ to zero and
solving for $\dot\varphi^A/\dot T$. Along the central equatorial
Weyl profile, the transverse component approaches the horizon
angular velocity~\eqref{eq:dragging} near the centre and vanishes
at infinity. 

Figure~\ref{fig:dragging} shows three complementary tests. The upper
left panel measures how the zero-spin central horizon responds as
the ring spin varies with the other five Komar charges fixed. The
upper right panel shows that zero central spin and zero central
angular velocity occur at different points on the fixed-rod
continuation. The lower panel exhibits the corresponding angular
shift directly in the metric, interpolating between the rotating
horizon and the nonrotating asymptotic frame. The plotted quantity is
the dimensionless shift
\begin{equation}
 \omega_{\rm ZAMO}^A=
 \sqrt{\frac{8G_5M}{3\pi}}\,\Omega_{\rm ZAMO}^A,
\end{equation}
using the total ADM mass $M$ of each configuration. Thus uppercase
$\Omega_{\rm ZAMO}^A$ denotes the angular velocity in inverse-length
units, while lowercase $\omega_{\rm ZAMO}^A$ denotes its reduced value.
Together these
diagnostics distinguish the horizon's intrinsic angular momentum
from the angular velocity induced by its environment.
\FloatBarrier

\section{Discussion}
\label{sec:discussion}

We have constructed a six-parameter family of doubly spinning black
Saturns in five-dimensional vacuum gravity, including balanced
configurations with two finite-area extremal horizons and a regular,
causal exterior in the domains established here. To our knowledge, these
provide the first regular exact solutions with standard asymptotic
flatness and disconnected extremal horizons of topologies $S^3$ and
$S^1\times S^2$. Despite the existent inverse scattering method, without AI tools, this would have been a Herculean task: the seven scalar numerators in appendix~\ref{app:scalar} contain
1643 monomials! Nevertheless, their compact representation makes it possible to
compute thermodynamic charges, check regularity and causality, and
study phase diagrams and transverse frame dragging.

The nonextremal family admits configurations in mutual thermodynamic
equilibrium. For the doubly extremal metric of
section~\ref{sec:extremal}, however, equality of both horizon angular
velocities is incompatible with a regular intervening axis. This exact
obstruction agrees with the absence of mutual equilibrium in the
thin-ring blackfold regime~\cite{AHOExt}. Appendix~\ref{app:extremequilibrium}
extends the obstruction to specified boundary limits and gives further
numerical evidence from independently retuned parameters.

The blackfold approach also predicts extremal di-rings and bi-rings,
with two disconnected ring horizons~\cite{AHOExt}. Within its regime
of validity, extremal di-rings can attain mutual thermodynamic
equilibrium. For extremal bi-rings, the matching conditions instead
fall outside that regime. Exact inverse-scattering constructions would test these predictions
beyond the thin-ring approximation and clarify the role of topology
in the extremal phase diagram.

Appendix~\ref{app:regularity} combines algebraic regularity conditions
with interval bounds and numerical sampling. For the specified doubly extremal
subfamily, the interval bounds cover the complete exterior and both
horizon extensions. The wider nonextremal scans found no violations at tested points, but
do not establish global regularity throughout the sampled parameter range. A natural next step is to
extend the bounds to a full exterior cover on an open nonextremal
parameter region and seek analytic proofs analogous to those for
smoothness and stable causality of the singly spinning
Saturn~\cite{CESSmooth,SzybkaCausal}.

Our local horizon coordinates provide a starting point for studying
geodesics and the interior spacetime. Geodesic
methods for doubly spinning rings~\cite{DurkeePS} and the extension
techniques developed for black
rings~\cite{CCRingExtension,CCGPSGlobal,ArmasDipole} suggest useful
approaches. A maximal analytic extension would determine the global causal
structure beyond the locally extended horizons.

These solutions are also relevant to the semiclassical gravitational
path integral, in which distinct geometries with the same boundary
data can compete~\cite{GH}. Disconnected-horizon configurations should
be included when compatible with those data. A generic stationary
Saturn has different temperatures and angular velocities on its two
components, so it does not automatically admit a smooth Euclidean
continuation with a single thermal identification. The subfamilies in
mutual equilibrium are therefore especially relevant to comparisons
at fixed temperature and angular potentials, while entropy comparisons
at fixed conserved charges can be made more generally.

\subsubsection*{Acknowledgements}
J. Armas is partly funded by the Dutch Institute for Emergent Phenomena (DIEP) cluster
at the University of Amsterdam via the DIEP programme Foundations and Applications of
Emergence (FAEME) and the national NWA-ORC 2023 consortium EAAS. The author used
OpenAI Codex Astra 6 Pro, and Anthropic Claude Fable 5.1 to assist with manuscript revision, literature cross-checks, and
review of mathematical arguments and analysis code. The author performed human checks
on the AI-assisted material and takes responsibility for the final content and conclusions.

\label{maintext:end}
\appendix
\addtocontents{toc}{\protect\setcounter{tocdepth}{2}}

\section{Black Saturn limits}
\label{app:limits}

This appendix discusses various limits of the general black Saturn solution in the main text.
Removing one component yields Myers-Perry or Pomeransky-Sen'kov black hole, while
removing transverse rotation yields the singly spinning Saturn. We also discuss in detail limits of the black saturn in which only one horizon is extremal.

\subsection{Removing the ring: doubly spinning Myers--Perry}
\label{app:mp}
Let $a_1,a_2,a_3\to0$, put $A=a_4$, $B=a_5$, $C=a_6$ and $L=B-A>0$,
and denote the limiting polarizations by the same symbols. In the
spectral gauge $a_1=0$, define
\begin{gather}
 \eta=1-C_1C_2C_3,\qquad c=C_6=\frac1{\sqrt{2(C-B)}},\\
 h=B_4,\qquad k=-\frac{B_5(B+1)}{2(A-B)(C-B)},\\
 \alpha=\sqrt{2L}\,\frac{k}{c},\qquad
 \beta=\sqrt{2L}\,\frac{1-ch}{h+2cL},\qquad
 n=\eta\,\frac{h+2cL}{2cL},\qquad
 \delta=n^2(1-\alpha^2\beta^2).
 \label{eq:analyticMPaux}
\end{gather}
We take the orientation branch $n>0$ and $|\alpha\beta|<1$.
The formula requires finite polarizations, $\eta\ne0$ and $h+2cL\ne0$.
The exact MP parameters are
\begin{equation}
 m=2Ln^2(1+\alpha^2)(1+\beta^2),\qquad
 a=-\sqrt{2L}\,n\alpha,\qquad b=-\sqrt{2L}\,n\beta.
 \label{eq:analyticMPmap}
\end{equation}
Set $\tau=t+\phi/c$, $\ell=\sqrt{L/2}$ and
\begin{equation}
 \begin{pmatrix}\tau\\\psi\\\phi\end{pmatrix}
 =
 \underbrace{\begin{pmatrix}
 1&2\alpha\ell n&2\beta\ell n\\
 0&n&-\alpha\beta n\\0&-\alpha\beta n&n
 \end{pmatrix}}_{\mathcal B}
 \begin{pmatrix}T\\\Psi\\\Phi\end{pmatrix},
 \qquad
 \rho_{\rm MP}=\delta\rho,\quad
 z_{\rm MP}=\delta\left(z-\frac{A+B}{2}\right).
 \label{eq:analyticMPcoords}
\end{equation}
The determinant of this Killing-coordinate transformation is $\delta$ while
the preceding shift from $t$ to $\tau$ has unit determinant. We eilinate the disappearing poles before taking the limit. Block Gaussian elimination (the Schur complement) of the three
coalescing left poles
leaves the three right poles and a conformal multiplier $\eta^2$.
Put $U=u_A$, $V=u_B$, $W=u_C$ and $R_{UV}=x+UV$. The auxiliary root
satisfies
\begin{equation}
 W^2=\left(\frac{V^2-x}{V}+\frac1{c^2}\right)W+x.
\end{equation}
Eliminating its pole by a second Schur complement and making the
time shift above leaves a two-pole metric
$G_2=H_2-N^{\mathsf T}\Gamma^{-1}N$. With
$p=k/c$ and $q=(1-ch)/(h+2cL)$ its data are
\begin{align}
 H_2&=\diag(-V/U,V,x/U),&
 N&=\begin{pmatrix}1&0&qR_{UV}/V\\1&pR_{UV}/V&0\end{pmatrix},\\
 \Gamma_{00}&=\frac{U(-xV+q^2R_{UV}^2)}{V^2(U^2+x)},&
 \Gamma_{11}&=\frac{-UV+p^2R_{UV}^2/V}{V^2+x},&
 \Gamma_{01}&=\frac{U}{U-V}.
\end{align}
Every occurrence of $W$ cancels exactly. Introduce
\begin{gather}
 R_A=\sqrt{x+(z-A)^2},\quad R_B=\sqrt{x+(z-B)^2},\quad
 P=\frac{R_A+R_B}{L},\quad \xi=\frac{R_A-R_B}{L},\\
 U=\tfrac L2(P-1)(1-\xi),\quad V=\tfrac L2(P+1)(1-\xi),\quad
 x=\tfrac{L^2}{4}(P^2-1)(1-\xi^2).
\end{gather}
Then
\begin{gather}
 H_2=\diag\left(-\frac{P+1}{P-1},
 \ell^2(P+1)(1-\xi),\ell^2(P+1)(1+\xi)\right),\\
 N=\begin{pmatrix}1&0&\beta\ell(P-1)\\1&\alpha\ell(P-1)&0\end{pmatrix},\\
 \Gamma_{00}=\frac{(P-1)[\beta^2(P-1)-1-\xi]}{2(P+\xi)},\quad
 \Gamma_{11}=\frac{(P-1)[\alpha^2(P-1)-1+\xi]}{2(P-\xi)},\\
 \Gamma_{01}=-\frac{P-1}{2}.
\end{gather}
Direct inversion of this $2\times2$ matrix gives
\begin{gather}
 \mathcal D=(1-\alpha^2\beta^2)P+
 (1+\alpha^2)(1+\beta^2)+(\alpha^2-\beta^2)\xi,\\
 (G_2)_{\tau\tau}=-1+
 \frac{2(1+\alpha^2)(1+\beta^2)}{\mathcal D},\qquad
 \Sigma=n^2L\mathcal D,\\
 r^2=\delta LP+\tfrac12(m-a^2-b^2),\qquad
 \cos2\theta=\xi .
\end{gather}
Substitution proves all six independent entries of
$\mathcal B^{\mathsf T}G_2\mathcal B=G_{\rm MP}$.
The conformal-factor determinant ratio gives,
independently,
\begin{equation}
 f_*=\frac{\Sigma}{4R_AR_B}
     =\delta^2 f_{\rm MP}(\rho_{\rm MP},z_{\rm MP}),\qquad
 (m-(a+b)^2)(m-(a-b)^2)=4\delta^2L^2.
\end{equation}
The factor $\eta^2$ from the first Schur complement is included through
$n^2$.
We recover the Myers--Perry metric~\cite{MP} in Boyer--Lindquist coordinates:
\begin{align}
 ds^2_{\rm MP}={}&-\dd T^2+\frac{m}{\Sigma}
 (\dd T-a\sin^2\theta\dd\Psi-b\cos^2\theta\dd\Phi)^2
 +\frac{\Sigma r^2}{\Delta}\dd r^2+\Sigma\dd\theta^2\nonumber\\
 &+(r^2+a^2)\sin^2\theta\dd\Psi^2
 +(r^2+b^2)\cos^2\theta\dd\Phi^2,\\
 \Sigma={}&r^2+a^2\cos^2\theta+b^2\sin^2\theta,\qquad
 \Delta=(r^2+a^2)(r^2+b^2)-mr^2.
\end{align}
The preceding algebra verifies every component of this metric and
its Weyl base factor, without fitting $m,a,b$ to ADM data.
With $D_{\rm MP}=\sqrt{(m-(a+b)^2)(m-(a-b)^2)}$ and
$r_+^2=(m-a^2-b^2+D_{\rm MP})/2$, the limiting charges and
intensive variables are therefore
\begin{gather}
 M_{\rm MP}=\frac{3\pi m}{8G_5},\qquad
 J_\Psi^{\rm MP}=\frac{\pi ma}{4G_5},\qquad
 J_\Phi^{\rm MP}=\frac{\pi mb}{4G_5},\\
 \cA_{\rm MP}=2\pi^2mr_+,\qquad
 S_{\rm MP}=\frac{\pi^2mr_+}{2G_5},\qquad
 \kappa_{\rm MP}=\frac{D_{\rm MP}}{mr_+},\qquad
 \Omega_A^{\rm MP}=\left(\frac{a}{r_+^2+a^2},
                         \frac{b}{r_+^2+b^2}\right).
\end{gather}
The remaining horizon is the MP horizon, and its Komar integrals
coincide with these ADM charges. The regular finite-area isolated extremal limit is
\begin{equation}
 m=(|a|+|b|)^2,\qquad r_+^2=|ab|>0.
\end{equation}
Within the map above it can be reached by a correlated limit
$L\to0$, $n\sim L^{-1/2}$ and $|\alpha\beta|\to1$, keeping
$m,a,b$ finite. Indeed
$m-(|a|+|b|)^2=2Ln^2(1-|\alpha\beta|)^2$ and
$D_{\rm MP}=2\delta L$. This isolated limit is distinct from the interacting
construction in section~\ref{sec:extremal}.
\subsection{Removing the central hole: the Pomeransky--Sen'kov ring}
\label{app:ps}

Use a small parameter $t>0$, with $a_1=0$ and
\begin{gather}
 a_4=A,\quad a_5=A+t^2,\quad a_6=A+(1+q)t^2,\quad q>0,\\
 B_4=b_4t+o(t),\quad B_5=b_5t^3+o(t^3),\quad
 C_6=\frac{\alpha_6}{t}+O(t),\qquad
 \alpha_6^2=\frac{A}{2q(A-a_2)}.
\end{gather}
Take $b_5\ne0$, $\mathcal K=1+q\alpha_6b_4\ne0$, and finite
$C_1,C_2,C_3$. We denote by $G_{\rm CHT}$ and $f_{\rm CHT}$ the Killing block and
base coefficient of the unbalanced doubly spinning black ring
given in~\cite{CHT}. We use the three-soliton representation of
that solution, with four rod endpoints and seed
constant one. The reduction gives
\begin{equation}
 G_6\longrightarrow G_{\rm CHT},\qquad
 f_6\longrightarrow\mathcal K^2 f_{\rm CHT},\qquad
 c'_1=-AC_1,\quad c'_2=-\frac{a_2+1}{A-a_2}C_2,\quad c'_3=C_3.
 \label{eq:analyticPSreduction}
\end{equation}
We can see how the two merging poles cancel without inverting the
full six-by-six matrix. We normalise rows 4 and 6 to time component one and
row 5 to $\psi$ component one. If $U=u_A$, $R=x+U^2$ and
$D_0=2xU/R$, their leading forms are
\begin{equation}
 m_4=(1,0,tb_4r_4),\qquad
 m_6=(1,0,tr_4/\alpha_6),\qquad m_5=(0,1,t l_5).
\end{equation}
In the basis $(m_6-m_4,m_5)$ the limiting Gram block is
\begin{equation}
 \begin{pmatrix}0&B\\ B&C_0\end{pmatrix},\qquad
 B=\frac{r_4l_5H_{\phi\phi}}{q\alpha_6D_0}\mathcal K,\qquad
 H_{\phi\phi}=-u_2u_3/U .
\end{equation}
The difference row has zero limiting $N$ and zero coupling to the
other Gram rows. The lower-right entry of this block's inverse is zero.
Thus its Schur corrections to $G$ and to the remaining Gram data vanish
exactly. Its determinant is $-B^2$. Tracking the raw row factors in
both the numerator and the zero-polarization reference determinant
leaves precisely $\mathcal K^2$ in the conformal-factor ratio.
In particular, one must track those reference factors before setting
polarisations to zero. The normalised row-5 expression is singular
under that substitution.

Row 4 is now a trivial time anti-soliton at $-x/U$. Removing it gives
\begin{equation}
 H_{\rm CHT}=\diag(u_1u_3/x,-u_1u_2U/x,-u_2u_3/U).
\end{equation}
It multiplies the time entry of row $i$ by
$(u_i+x/U)/(u_i-U)$. Rescale row 1 by $-A$.
The resulting three rows are exactly the CHT rows with the parameters
in \eqref{eq:analyticPSreduction}. Coalescing the last three seed
weights also gives the CHT seed conformal factor. These steps prove
the stated reduction of both $G$ and $f$.

Write $s=c'_2$. Impose the CHT regularity and balance conditions
\begin{equation}
 A(A-a_3)=(A-a_2)^2,\qquad
 c'_1=\sqrt{\frac{a_3}{2a_2A}},\qquad
 c'_3=\frac{2c'_1s(a_3-a_2)(A-a_2)}{a_3}.
\end{equation}
The CHT-to-PS transformation of \cite{CHT} then supplies an
analytic identification with the balanced PS metric. Explicitly, put
\begin{gather}
 \bar a=\tfrac12(a_2+a_3),\qquad
 c_s=\frac{a_3-a_2}{2A-a_2-a_3},\quad
 \beta_s=\frac{a_3-a_2}{a_2}s^2,\quad
 Q=1+c_s^2-\beta_s(1-c_s^2),\\
 \mu=\frac{2c_s}{Q+\sqrt{Q^2-4c_s^2}},\quad
 \nu_c=\frac{\mu-c_s}{1-c_s\mu},\\
 \lambda=\mu+\nu_c,\quad \nu_{\rm PS}=\mu\nu_c,\quad
 k^2=\delta_0(A-\bar a).
\end{gather}
Here $\delta_0=\mathcal K^2\delta_{\rm CHT}$,
$\rho_{\rm PS}=\delta_0\rho$ and
$z_{\rm PS}=\delta_0(z-\bar a)$. The explicit ring-coordinate map is given below.
Multiplication of $f_{\rm CHT}$ by $\mathcal K^2$ multiplies both
unit-period axis vectors by $|\mathcal K|$. Thus the physical
Killing basis is $A_{\rm CHT}\diag(1,|\mathcal K|,|\mathcal K|)$.

The reduction fixes the limiting scale analytically. Selecting a
particular path through balanced six-pole solutions additionally
requires following the appropriate balance root.

In the PS angular convention set $(t,\psi,\phi)=(T,-\Psi,-\Phi)$
and $\mathsf G(\xi)=(1-\xi^2)(1+\lambda\xi+\nu\xi^2)$.
The explicit map to ring coordinates is
\begin{equation}
 \rho_{\rm PS}^2=\frac{-4k^4\mathsf G(x_r)\mathsf G(y_r)}
 {(1-\nu)^2(x_r-y_r)^4},\qquad
 z_{\rm PS}=\frac{k^2(1-x_ry_r)
 [2+\lambda(x_r+y_r)+2\nu x_ry_r]}{(1-\nu)(x_r-y_r)^2}.
\end{equation}
Here $x_r,y_r$ are ring coordinates, distinct from the dressing variable
$x=\rho^2$. For brevity write them as $X,Y$ in the next formulae.
The target metric in these coordinates is~\cite{CHT,PS}
\begin{align}
 ds^2_{\rm PS}={}&-\frac{H(Y,X)}{H(X,Y)}
 (\dd t-\omega_\psi\dd\psi-\omega_\phi\dd\phi)^2
 -\frac{F(X,Y)}{H(Y,X)}\dd\psi^2
 -\frac{2J(X,Y)}{H(Y,X)}\dd\psi\dd\phi\notag\\
 &+\frac{F(Y,X)}{H(Y,X)}\dd\phi^2
 +\frac{2k^2H(X,Y)}{(1-\nu)^2(X-Y)^2}
 \left[\frac{\dd X^2}{\mathsf G(X)}-\frac{\dd Y^2}{\mathsf G(Y)}\right],\\
 H(X,Y)={}&1+\lambda^2-\nu^2
 +2\lambda(X+\nu Y)(1-\nu XY)
 +\nu(1-\lambda^2-\nu^2)X^2Y^2,\\
 J(X,Y)={}&\frac{2k^2\lambda\sqrt\nu(1-X^2)(1-Y^2)}{(1-\nu)^2(X-Y)}\notag\\
 &\quad\times\left[1+\lambda^2-\nu^2+2\lambda\nu(X+Y)
 -\nu(1-\lambda^2-\nu^2)XY\right],\\
 F(X,Y)={}&\frac{2k^2}{(1-\nu)^2(X-Y)^2}
 \left\{\mathsf G(X)(1-Y^2)Q_1(Y)
 +\mathsf G(Y)P(X)\right\},\\
 q={}&(1+\nu)[(1-\nu)^2-\lambda^2],\qquad
 Q_1(Y)=q+\lambda(1-\lambda^2+2\nu-3\nu^2)Y,\notag\\
 P(X)={}&2\lambda^2+\lambda[(1-\nu)^2+\lambda^2]X+qX^2
 +\lambda[1-\lambda^2-3\nu^2+2\nu^3]X^3\notag\\
 &+\nu(1-\nu)(1-\lambda^2-\nu^2)X^4,\\
 \omega_\psi={}&\frac{2k\lambda(1+Y)}{H(Y,X)}
 \sqrt{\frac{1+\lambda+\nu}{1-\lambda+\nu}}
 [1+\lambda-\nu+2\nu X(1-Y)+\nu(1-\lambda-\nu)X^2Y],\\
 \omega_\phi={}&\frac{2k\lambda\sqrt{\nu[(1+\nu)^2-\lambda^2]}}
 {H(Y,X)}(1-X^2)Y.
\end{align}
This also specifies the orbit-space metric completely. The comparison
converts it to Weyl form using
$f_{\rm PS}=g_{XX}/[(\partial_X\rho_{\rm PS})^2+
(\partial_Xz_{\rm PS})^2]$, and checks
$f_{\rm Saturn}\to\delta_0^2f_{\rm PS}$.

The PS charges and area, in the Saturn angular orientation, are
\begin{align}
 M_{\rm PS}&=\frac{3\pi k^2\lambda}{G_5(1+\nu-\lambda)},\\
 J_{\Psi}^{\rm PS}&=-\frac{2\pi k^3\lambda
 \sqrt{(1+\nu)^2-\lambda^2}
 [1+\lambda-6\nu+\lambda\nu+\nu^2]}
 {G_5(1-\nu)^2(1+\nu-\lambda)^2},\\
 J_{\Phi}^{\rm PS}&=-\frac{4\pi k^3\lambda
 \sqrt{\nu[(1+\nu)^2-\lambda^2]}}
 {G_5(1-\nu)^2(1+\nu-\lambda)},\\
 \mathcal A_{\rm PS}&=\frac{32\pi^2k^3\lambda(1+\lambda+\nu)}
 {(1-\nu)^2(y_h^{-1}-y_h)},\qquad
 y_h=\frac{-\lambda+\sqrt{\lambda^2-4\nu}}{2\nu}.
\end{align}

The ring domain is $-1\le X\le1$, $y_h\le Y<-1$, with
$0\le\nu<1$, $2\sqrt\nu\le\lambda<1+\nu$. These formulas
include the extremal boundary by continuity, using the nonsingular
expression $y_h=-2/(\lambda+\sqrt{\lambda^2-4\nu})$.
The horizon entropy is $\cA_{\rm PS}/(4G_5)$ and the retained ring
Komar integrals equal the ADM charges.

\subsection{The singly spinning black Saturn}
\label{app:ef}
The six-pole metric admits the general singly spinning Saturn
of~\cite{EF}, including a spinning central hole, as an explicit
auxiliary-pole boundary. ``Singly
spinning'' here means one rotation plane. Both components can carry spin
in that plane.

\subsubsection*{Parameters and balance}

Keep the spectral gauge $a_1=0$, $b_{\rm spectral}=-1$, and take
\begin{gather}\label{eq:EFboundaryscaling}
 0<a_2<a_3<a_4<a_5,\qquad a_6=a_5+\epsilon^2,\qquad\epsilon\to0^+,\\
 B_5=\epsilon\,\mathfrak b+o(\epsilon),\qquad
 C_6=\gamma/\epsilon+O(\epsilon),\qquad
 \gamma^2=\frac{a_5}{2(a_5-a_2)},\\
 C_2\longrightarrow0,\qquad C_3\longrightarrow0,\qquad B_4\longrightarrow0,\qquad
 C_1\longrightarrow-\sqrt{\frac{a_3}{2a_2a_4a_5^2}}.
\end{gather}
$C_1,C_6$ retain their exact phantom-removal values before taking the
limit. The last three zero conditions are imposed on the finite
auxiliary-limit metric. We can also approach them continuously
while the other scaled variables remain finite. Both physical horizon
intervals, $[a_2,a_3]$ and $[a_4,a_5]$, remain nonzero. Only the auxiliary
interval collapses. Setting those three polarizations to zero at a fixed
nonzero auxiliary interval is not the reduction asserted here.

We use the label EF for the solution of~\cite{EF} and denote its
endpoints by $A_i$, to distinguish their ordering from ours.
The complete map is
\begin{gather}\label{eq:EFendpointmap}
 (A_1,A_2,A_3,A_4,A_5)=(0,a_5,a_4,a_3,a_2),\\
 c_1^{\rm EF}=-2a_4a_5 C_1=\sqrt{\frac{2a_3a_4}{a_2}},\qquad
 c_2^{\rm EF}=-\frac{2(a_5+1)(a_5-a_3)}{a_5}\,\mathfrak b\gamma,\\
 L^2=a_5,\qquad
 \kappa_1=\frac{a_4}{a_5},\quad\kappa_2=\frac{a_3}{a_5},\quad
 \kappa_3=\frac{a_2}{a_5},\qquad
 \bar c_2=\frac{c_2^{\rm EF}}{c_1^{\rm EF}(1-\kappa_2)}.
\end{gather}
Thus $0<\kappa_3<\kappa_2<\kappa_1<1$. With
$N=1+\kappa_2\bar c_2\ne0$, the asymptotic constants are
\begin{equation}
 k_{\rm EF}=\frac1{|N|},\qquad
 q_{\rm EF}=\frac{c_1^{\rm EF}\bar c_2}{N}.
\end{equation}
The EF balance condition, fixing $\mathfrak b$ through this map, is
\begin{equation}\label{eq:EFbalanceagain}
 \bar c_2=\frac1{\kappa_2}\left[
 \frac{s(\kappa_1-\kappa_2)}
 {\sqrt{\kappa_1(1-\kappa_2)(1-\kappa_3)(\kappa_1-\kappa_3)}}-1\right],
 \qquad s=\pm1.
\end{equation}
The usual positive-Komar-mass branch is $s=+1$. The metric identification
itself holds before balance and on both balanced branches wherever the
normalization is nonsingular.

\subsubsection*{Metric identity and normalisation}

The permutation~\eqref{eq:EFendpointmap} defines the EF pole
functions $\mu_i$. We use the functions $H_x,H_y,P,G_x,G_y,\omega_\psi$
given in section~2 of~\cite{EF}. In particular,
\begin{equation}\begin{gathered}
 G_x=\frac{\rho^2\mu_4}{\mu_3\mu_5},\qquad
 G_y=\frac{\mu_3\mu_5}{\mu_4},\\
 P=(\rho^2+\mu_3\mu_4)^2(\rho^2+\mu_1\mu_5)(\rho^2+\mu_4\mu_5).
\end{gathered}\end{equation}
Exact cancellation of the Laurent monomials gives, in the intermediate
Killing coordinates,
\begin{gather}\label{eq:EFmetricidentities}
 f_{\rm raw}=H_xP,\qquad
 (G_{tt},G_{t\psi},G_{t\phi},G_{\psi\phi},G_{\phi\phi})
 =\left(-\frac{H_y}{H_x},\frac{\omega_\psi}{H_x},0,0,G_x\right),\\
 G_{\psi\psi}=\frac{H_xG_y}{H_y}
              -\frac{\omega_\psi^2}{H_xH_y}.
\end{gather}
The last identity also follows from $\det G=-\rho^2$ and extends through
$H_y=0$ by the regular rational expression for the metric. One can verify the identity by expanding the published EF functions
and collecting pole monomials with
\begin{equation}
 (u_i-u_j)(\rho^2+u_i u_j)=2(a_i-a_j)u_i u_j.
\end{equation}
The coefficients agree with variable endpoints and polarisations.

In the canonical physical normalisation~\eqref{eq:canonical} the outer-axis matrix has
$\delta=\det A=N^2$, and its unit-determinant version is
\begin{equation}
 B=\begin{pmatrix}1&-q_{\rm EF}&0\\0&1&0\\0&0&1\end{pmatrix},
 \qquad f_{\rm can}=f_{\rm raw}/N^2.
\end{equation}
Then $t_{\rm EF}=T$, $\psi_{\rm EF}=-\Psi$, $\phi_{\rm EF}=\Phi$,
with the same $\rho,z$, gives
\begin{align}
 ds^2={}&-\frac{H_y}{H_x}
 \left[dt_{\rm EF}+\left(\frac{\omega_\psi}{H_y}+q_{\rm EF}\right)
 d\psi_{\rm EF}\right]^2
 +\frac{H_xG_y}{H_y}d\psi_{\rm EF}^2+G_xd\phi_{\rm EF}^2\notag\\
 &+k_{\rm EF}^2H_xP(d\rho^2+dz^2).
\end{align}
Equivalently, $t_{\rm raw}=t_{\rm EF}+q_{\rm EF}\psi_{\rm EF}$ and
$\psi_{\rm raw}=-\psi_{\rm EF}$. This fixes the angular-momentum sign
convention and the conformal normalization explicitly.

\subsubsection*{Charges and entropy}

Here write $b=\bar c_2$ only in the following thermodynamic formulas,
and set $G_5=1$. Define
\begin{align}
 H&=\kappa_3(1-\kappa_1)+\kappa_1\kappa_2(1-\kappa_2)(1-\kappa_3)b^2,\\
 R&=\kappa_3-\kappa_3(\kappa_1-\kappa_2)b
       +\kappa_1\kappa_2(1-\kappa_3)b^2,\\
 S&=\kappa_3-\kappa_2(\kappa_1-\kappa_3)b
       +\kappa_1\kappa_2(1-\kappa_2)b^2.
\end{align}
In the EF orientation the individual charges become exactly
\begin{align}
 M_\Hh&=\frac{3\pi L^2H}{4\kappa_3N},&
 M_\R&=\frac{3\pi L^2\kappa_2[1-(1-\kappa_2)b]R}{4\kappa_3N^2},\\
 J_\Hh^{\rm EF}&=-\frac{\pi L^3bH}{\kappa_3N^2}
                   \sqrt{\frac{\kappa_1\kappa_2}{2\kappa_3}},&
 J_\R^{\rm EF}&=\frac{\pi L^3SR}{\kappa_3N^3}
                   \sqrt{\frac{\kappa_2}{2\kappa_1\kappa_3}},\\
 M_{\rm ADM}&=M_\R+M_\Hh,&J_{\rm ADM}^{\rm EF}&=J_\R^{\rm EF}+J_\Hh^{\rm EF}.
\end{align}
All individual and ADM transverse angular momenta vanish, as do both
transverse horizon angular velocities. Our canonical $J_\Psi$ is minus
the displayed EF $J$, owing to the single common reflection above.
Restoring $G_5$ divides every displayed mass and angular momentum by $G_5$.
The areas are
\begin{align}
 A_\Hh&=4\pi^2L^3
 \sqrt{\frac{2(1-\kappa_1)^3}{(1-\kappa_2)(1-\kappa_3)}}
 \frac{H}{\kappa_3(1-\kappa_1)N^2},\\
 A_\R&=4\pi^2L^3
 \sqrt{\frac{2\kappa_2(\kappa_2-\kappa_3)^3}
 {\kappa_1(\kappa_1-\kappa_3)(1-\kappa_3)}}
 \frac{R}{\kappa_3N^2},\qquad
 S_{\rm total}=\frac{A_\R+A_\Hh}{4G_5}.
\end{align}
These are precisely EF equations~(3.26)--(3.40). The normalized metric
isometry identifies the integration surfaces, Killing fields and volume
form, so it identifies each horizon Komar integral, not just their sum.
Direct asymptotic expansion independently reproduces the EF ADM
formulas. Horizon quadratures at five balanced configurations, including
both signs in~\eqref{eq:EFbalanceagain}, agree with the individual
charges and areas to better than $5\times10^{-31}$ in normalised error.

To approach this limit through solutions that remain in balance,
we fix $(a_1,\ldots,a_5)=(0,1,2,4,5)$ and $C_2=0$, and solve
the axis equations~\eqref{eq:balance} at each nonzero $\epsilon$.
This requires corrections $C_3,B_4=O(\epsilon)$.
The transverse spins vanish as $O(\epsilon)$. Masses, original spins and areas converge as
$O(\epsilon^2)$. Thus the limiting identities and the approach through exactly balanced
finite-pole metrics are consistent.

\subsection{Solutions with one extremal horizon}
\label{app:mixedextremal}
There are two distinct mixed-extremal boundaries: a zero-temperature
ring surrounding a nonextremal hole, and a nonextremal ring surrounding
a zero-temperature hole. We give the metric of each of these cases and regularity checks.

\subsubsection{An extremal ring and a nonextremal central hole}
\label{app:extremalring}

Choose real constants
\begin{equation}
 0<a<b,\qquad 0<A<B<C,\qquad k,q,u,v\in\mathbb R,\qquad q\ne0.
\end{equation}
For an auxiliary parameter $\epsilon>0$, prescribe
\begin{align}
 (a_1,\ldots,a_6)&=(0,a\epsilon,b\epsilon,A,B,C),\nonumber\\
 C_1&=-\sqrt{\frac{b}{2aABC}},\qquad
 C_6=\sqrt{\frac{C}{2(C-a\epsilon)(C-B)}},\nonumber\\
 C_2&=q,\qquad C_3=\frac{1-k\epsilon}{C_1q},\qquad
 B_4=u,\qquad B_5=v .
 \label{eq:mixedRpath}
\end{align}
The values of $C_1$ and $C_6$ satisfy~\eqref{eq:phantoms}, which
remove the apparent turning points at the auxiliary endpoints
$a_1$ and $a_6$. The remaining tuning gives
\begin{equation}
 1-C_1C_2C_3=k\epsilon.
 \label{eq:mixedRcritical}
\end{equation}
The ring rod shrinks to $z=0$, while the hole retains the rod $[A,B]$.
Equation~\eqref{eq:mixedRcritical} retains the ring's finite-area throat.
The common rescaling $(a,b,k)\mapsto s(a,b,k)$ only reparametrises
$\epsilon$, so we may set $a=1$. Eight constants then remain before
balance. Balance is imposed on the limiting metric.
To explicitly construct the metri we use $x=\rho^2$,
$u_i=\sqrt{x+(z-a_i)^2}-(z-a_i)$, the rows $M$ and ratios $r_i$
of~\eqref{eq:normalizedrows}-\eqref{eq:rowratios}, and
\begin{align}
 H&=\diag\left(-\frac{u_1u_3}{u_4u_6},
              -\frac{u_1u_2u_5}{x},-\frac{u_2u_3u_5}{u_4u_6}\right),\qquad\nonumber\\
 (p_1,\ldots,p_6)&=(u_1,u_2,u_3,-x/u_4,u_5,-x/u_6),\nonumber\\
 \Gamma_{ij}&=\frac{M_i H M_j^{\mathsf T}}{x+p_ip_j},\qquad
 N_{i\alpha}=\frac{M_{i\alpha}H_{\alpha\alpha}}{p_i}.
 \label{eq:mixedGram}
\end{align}
All coefficients needed below have degree at most two in $\epsilon$.
For $a_i=a_i^0+s_i\epsilon$, $d_i=z-a_i^0$ and
$D_i=\sqrt{x+d_i^2}$, they follow from
\begin{align}
 u_i&=D_i-d_i+s_i(1-d_i/D_i)\epsilon
            +\frac{s_i^2x}{2D_i^3}\epsilon^2+O(\epsilon^3),\nonumber\\
 C_6&=\frac{1}{\sqrt{2(C-B)}}\left(
 1+\frac{a\epsilon}{2C}+\frac{3a^2\epsilon^2}{8C^2}\right)
 +O(\epsilon^3),\qquad (s_i)=(0,a,b,0,0,0).
 \label{eq:mixedRroots}
\end{align}
We cancel the endpoint differences in the ratios $r_i$ before expanding.
At degree zero we write
\begin{equation}
 M_1^{(0)}=(\alpha,1,0),\quad M_2^{(0)}=(0,\beta,1),\quad
 M_3^{(0)}=(1,0,\gamma),\qquad \alpha\beta\gamma=-1.
\end{equation}
These three symbols are local row coefficients, not the parameters
of section~\ref{sec:extremal}. Let $P$ be the identity except for
$P_{3i}=(-\alpha^{-1},(\alpha\beta)^{-1},1,0,0,0)_i$, and let
$\sigma_i=\delta_{i3}$. Hold $P$ fixed when extracting coefficients:
\begin{align}
 \widehat\Gamma_{ij}
   &=[\epsilon^{\sigma_i+\sigma_j}](P\Gamma P^{\mathsf T})_{ij},\qquad \widehat N_{i\alpha}=[\epsilon^{\sigma_i}](PN)_{i\alpha},\qquad
 \widehat M_{i\alpha}=[\epsilon^{\sigma_i}](PM)_{i\alpha}.
 \label{eq:mixedRdata}
\end{align}
With $\Gamma_{\rm ref}$ formed from the same rows with all six
polarisations zero, the raw limiting metric is
\begin{equation}
\quad
 G_*=H^{(0)}-\widehat N^{\mathsf T}\widehat\Gamma^{-1}\widehat N,
 \qquad
 f_*=f_{\rm seed}^{(0)}
              \frac{\det\widehat\Gamma}{\det\Gamma_{\rm ref}^{(0)}} .
 \quad
 \label{eq:mixedRmetric}
\end{equation}
For completeness, in these row conventions the seed factor is
\begin{gather}
 f_{\rm seed}=\frac{u_2u_5}{u_1}
 \prod_{i\le j}(x+u_i u_j)^{-w_i\cdot w_j/(1+\delta_{ij})},
 \nonumber\\
 w_1=(1,-1,0),\quad w_2=w_5=(0,1,-1),\quad
 w_3=w_6=(-1,0,1),\quad w_4=(1,0,-1).
 \label{eq:mixedseed}
\end{gather}
The matrix inverse in~\eqref{eq:mixedRmetric} is finite and may
equivalently be written as an adjugate divided by a determinant. The row transformation
$\diag(1,1,\epsilon^{-1},1,1,1)P$ explains the construction:
the third row has a first-order zero and its diagonal Gram entry a
second-order zero. Thus $G\to G_*$ and $f/\epsilon^2\to f_*$.
The constant rescaling of $f$ is made before the physical axis
normalisation. This procedure removes the singular matrix inversion
that would result from setting the endpoints equal at the outset.

\subsubsection{An extremal central hole and a nonextremal ring}
\label{app:extremalhole}

To find this branch, we start near the double-extremal solution and
increase the ring area at fixed four local angular momenta. Decreasing
the central rod length then takes an auxiliary endpoint through
infinity. Continuing the physical metric across this coordinate
boundary is essential. In the original BZ representation
$C_1,C_3,B_4,C_6$ become imaginary and the constant transformation
$W=\diag(1,i,i)$ of the Killing coordinates, together with
$f\mapsto-f$, gives a real metric. The purely real formula below
implements this continuation and preserves the continued signed spins. Use eight real constants
\begin{gather}
 a<-1,\qquad C<U<U+\ell<0,\qquad \ell>0,\qquad
 c_2,c_3,b,w\in\mathbb R,\nonumber\\
 \gamma=\sqrt{-\frac{U+\ell}{2Ua(a+1)C}},
\end{gather}
and prescribe the following path for sufficiently small $L>0$:
\begin{gather}
 (a_1,\ldots,a_6)=(0,U,U+\ell,aL,(a+1)L,C),\nonumber\\
 C_1=i\gamma/L,\qquad C_2=c_2,\qquad C_3=ic_3,\qquad B_4=ib,\nonumber\\
 B_5=\frac{L(-b+wL)}{2U\gamma},\qquad
 C_6=i\sqrt{-\frac{C}{2(C-U)(C-(a+1)L)}} .
 \label{eq:mixedHpath}
\end{gather}
The ring occupies $[U,U+\ell]$. The central rod has length $L$ and
collapses to $z=0$. Both phantom equations hold identically. The
critical relation is
\begin{equation}
 \quad C_1=i\gamma/L,\qquad B_4+2UC_1B_5=iwL .\quad
 \label{eq:mixedHcritical}
\end{equation}
Although $B_5$ tends to zero, its first and second coefficients
survive in the limiting metric. At $L\ne0$ these variables describe
all nine parameters before balance. At $L=0$ they leave eight. We use the same $r_i$ as before, with the endpoints in
\eqref{eq:mixedHpath}. The real rows and pre-dressing matrix are
\begin{align}
 M_1&=(r_1\gamma/L,-1,0),\qquad M_2=(0,r_2c_2,1), \qquad M_3=(1,0,r_3c_3),\qquad M_4=(1,0,r_4b),\nonumber\\
 M_5&=(0,r_5B_5,1),\qquad M_6=(r_6c_6,0,-1),\qquad H=\diag\left(-\frac{u_1u_3}{u_4u_6},
                \frac{u_1u_2u_5}{x},\frac{u_2u_3u_5}{u_4u_6}\right).
 \label{eq:mixedHrows}
\end{align}
Here $c_6$ is the positive real square root in
\eqref{eq:mixedHpath}, without $i$. Form $\Gamma,N$ with the same
poles and formulas as in~\eqref{eq:mixedGram}. Define
\begin{align}
 S&=I+(r_5B_5)E_{51},\qquad \det S=1,\qquad \bar\Gamma=[L^0](S\Gamma S^{\mathsf T}),\qquad
 \bar N=[L^0](SN),\nonumber\\
 \bar M&=[L^0](SM),\qquad H_0=[L^0]H,\qquad P_0=[L^0]\left(-\frac{f_{\rm seed}}{\det\Gamma_{\rm ref}}\right).
 \label{eq:mixedHdata}
\end{align}
$E_{51}$ has one unit entry in row five, column one.
The reference matrix sets all six polarisations to zero, retaining
the row signs in~\eqref{eq:mixedHrows}. Coefficients are extracted
after cancelling the known Laurent powers. The result is
\begin{equation}
 \quad G_*=H_0-\bar N^{\mathsf T}\bar\Gamma^{-1}\bar N,
 \qquad f_*=P_0\det\bar\Gamma .\quad
 \label{eq:mixedHmetric}
\end{equation}
This is a full metric at $L=0$. One can see why the tuning is required before inverting the Gram
matrix. Take $q=\lim B_5/L$ and, at $L=0$, denote
$(u_1,u_2,u_3,u_6)=(h_0,v_0,t_0,k_0)$. Then
\begin{equation}
 \lim_{L\to0}L\{\Gamma_{54}+(r_5B_5)\Gamma_{14}\}
 =-\frac{v_0t_0(x+h_0^2)}{2k_0x(x+h_0v_0)}
                   (b+2U\gamma q).
 \label{eq:mixedHcancel}
\end{equation}
It vanishes precisely for the leading relation in
\eqref{eq:mixedHcritical}. The apparent quadratic pole in the
$(5,5)$ entry multiplies a second difference of the Cauchy kernel,
whose zero is quadratic in $p_5-p_1=O(L)$. Thus the displayed
row operation yields finite matrices while retaining the free
subleading coefficient $w$.

\subsubsection{Physical normalisation, balance and thermodynamic charges}
\label{app:mixedcharges}
We now discuss the physical metrics, absense of conical singularities and thermodynamics charges. 
The two raw metrics above must be put in a common asymptotically flat
frame before interpreting their charges. In order to do so, on the outer axes we solve
$G_*v=0$, normalising $v_L^\phi=1$ and $v_R^\psi=1$.
For $i,j$ complementary to the component fixed to one, we set
\begin{equation}
 Q_v=-\left[f_*(G_{*ii}G_{*jj}-G_{*ij}^2)\right]^{-1},
 \qquad
 \mathcal A=(e_t,v_R/\sqrt{Q_R},v_L/\sqrt{Q_L}),\qquad
 \delta=|\det\mathcal A|.
\end{equation}
All these are finite axis coefficients. For the extremal ring
example we choose
$\mathcal B=\mathcal A\diag(1,\delta^{-1/2},\delta^{-1/2})$.
For the continued extremal hole example we choose
\begin{equation}
 \mathcal B=(e_t,-\mathcal A_\psi/\sqrt\delta,
                  -\mathcal A_\phi/\sqrt\delta).
\end{equation}
This choice has determinant $-1$ and fixes the signs of both angular
coordinates. Applying the basis change and rescaling
in~\eqref{eq:canonical} to~\eqref{eq:mixedRmetric}
or~\eqref{eq:mixedHmetric}, we obtain the physical metric
\begin{equation}
 \dd s^2=(\bar G)_{ab}\dd y^a\dd y^b+
          \bar f(\dd\rho^2+\dd z^2),\qquad
 \bar G=\mathcal B^{\mathsf T}G_*\mathcal B,\quad
 \bar f=f_*/\delta,\quad \det\bar G=-\rho^2 .
 \label{eq:mixedphysical}
\end{equation}
Both physical angles have period $2\pi$. On the intermediate axis
the three balance conditions require zero time component, the
same angular direction as the left axis, and unit cone factor.
Equivalently its canonically normalised rod vector is $(0,0,1)$,
up to the fixed orientation. These conditions fix the angular
identifications as well as remove a conical defect. 

All local and ADM charges have finite algebraic expressions. Let $(\mathsf K,\mathsf V,\mathsf M)$ denote
$(\widehat\Gamma,\widehat N,\widehat M)$ for the extremal ring,
or $(\bar\Gamma,\bar N,\bar M)$ for the extremal hole, and define
\begin{align}
 \mathbf s={}&(u_4+u_6-u_1-u_3,\,
              -u_1-u_2-u_5-4z-2,\nonumber\\
             &\hspace{37mm}u_4+u_6-u_2-u_3-u_5-2z),\nonumber\\
 \mathcal Q_*={}&-(\mathsf V^{\mathsf T}\mathsf K^{-1}\mathsf M)^{\mathsf T}
                       -\diag(\mathbf s^{(0)}),\qquad
 \mathcal Q=\mathcal B^{-1}\mathcal Q_*\mathcal B .
 \label{eq:mixedpotential}
\end{align}
Here $(0)$ means the constant coefficient in the respective limit.
The time row of $\mathcal Q$ is constant on each ordinary axis.
Let $\mathcal Q_-,\mathcal Q_0,\mathcal Q_+$ be its values to the
left of the ring, between the horizons, and to the right of the
hole. Then
\begin{align}
 \Delta\mathcal Q_\R&=\mathcal Q_0-\mathcal Q_-,&
 \Delta\mathcal Q_\Hh&=\mathcal Q_+-\mathcal Q_0,\nonumber\\
 M_i&=\frac{3\pi}{8G_5}(\Delta\mathcal Q_i)^t{}_t,&
 J_{iA}&=-\frac{\pi}{4G_5}(\Delta\mathcal Q_i)^t{}_A,\nonumber\\
 M&=M_\R+M_\Hh,& J_A&=J_{\R A}+J_{\Hh A}.
 \label{eq:mixedkomar}
\end{align}
When we add the two horizon contributions, the intermediate-axis
value $\mathcal Q_0$ cancels:
$\Delta\mathcal Q_\R+\Delta\mathcal Q_\Hh=\mathcal Q_+-\mathcal Q_-$.
This gives the final line and agrees with the charges at infinity. Axis values mean constant coefficients of the
finite expressions: expand
$u_i=|z-a_i|-(z-a_i)+x/(2|z-a_i|)+\cdots$,
factor common powers of $x$, and extract $x^0$.

For either non-extremal horizon, let $\Lambda_{AB}=\bar G_{AB}$ and let
$\ell_i$ be its finite rod length ($B-A$ for the hole, $\ell$ for
the ring). At any interior point of that rod,
\begin{gather}
 \Omega_i^A=-(\Lambda^{-1})^{AB}\bar G_{Bt},\qquad
 \kappa_i=(\bar f\det\Lambda)^{-1/2},\qquad T_i=\kappa_i/(2\pi),
 \nonumber\\
 \cA_i=\frac{4\pi^2\ell_i}{\kappa_i},\qquad
 S_i=\frac{\cA_i}{4G_5}.
 \label{eq:mixedwarmthermo}
\end{gather}
At the extremal point use $\rho=R\sin\theta$,
$z=z_i+R\cos\theta$. With
$F_0=[R^0](R^2\bar f)$ and $\Lambda_0=[R^0]\Lambda$,
\begin{gather}
 T_i=\kappa_i=0,\qquad
 \Omega_i^A=-[R^0](\Lambda^{-1})^{AB}\bar G_{Bt},\nonumber\\
 \cA_i=4\pi^2\int_0^\pi
           \sqrt{F_0\det\Lambda_0}\,\dd\theta,\qquad
 S_{\rm tot}=\frac{\cA_\R+\cA_\Hh}{4G_5}.
 \label{eq:mixedcoldthermo}
\end{gather}
On the regular branches the density
$D_i=\sqrt{F_0\det\Lambda_0}/\sin\theta$ is constant, giving
$\cA_i=8\pi^2D_i$ by finite horizon coefficients alone.
Independent area--spin checks give
\begin{equation}
 \cA_\R=8\pi G_5|J_{\R\Phi}|\quad(T_\R=0),\qquad
 \cA_\Hh=8\pi G_5\sqrt{|J_{\Hh\Psi}J_{\Hh\Phi}|}
                          \quad(T_\Hh=0).
 \label{eq:mixedareaspin}
\end{equation}
The local Smarr
relations and the first law along a smooth balanced branch read
\begin{equation}
 \frac23M_i=T_iS_i+\Omega_i^AJ_{iA},\qquad
 \dd M=T_{\rm warm}\dd S_{\rm warm}
              +\sum_{i=\R,\Hh}\Omega_i^A\dd J_{iA},
 \label{eq:mixedmechanics}
\end{equation}
where $T_{\rm warm}$ and $S_{\rm warm}$ are the temperature and entropy, respectively, of the non-extremal horizon. Since one temperature is strictly positive, neither mixed-extremal
branch is in mutual thermal equilibrium. Exact metric components and thermodynamic checks are provided in the \ancillarylink{ancillary Mathematica file}.

\subsubsection{Balanced examples, horizon extension and validation}
\label{app:mixedchecks}
Here we give numerical examples for these configurations with a single extremal horizon. For the extremal ring case take
\begin{gather}
 (a,b,A,B,C,k)=(1,2,4,5,5.1,3.2),\nonumber\\
 q=-2.78734053234366822678,\qquad
 u=1.18604581758086428649,\nonumber\\
 v=0.032580068573907557108.
 \label{eq:mixedRexample}
\end{gather}

For an extremal central hole with a nonextremal ring, we find the
following parameters. Here $\ell=a_3-a_2$ is the ring's Weyl rod
length in~\eqref{eq:mixedHpath}. It is small but strictly positive
and stays fixed when we take the central-hole extremal limit.
The factor $10^{-5}$ below is ordinary scientific notation,
so $\ell\simeq0.0000204936$ in the chosen coordinate units:
\begin{align}
 a&=-3263801.766316202,&U&=-4.796412879959584,\nonumber\\
 \ell&=2.049355872243065\times10^{-5},&C&=-1608957.852727379,\nonumber\\
 c_2&=85867.17426548942,&c_3&=-0.02758903860699587,\nonumber\\
 b&=-0.002097626322389592,&w&=8273.594438716184.
 \label{eq:mixedHexample}
\end{align}

We obtained these examples by solving the balance equations
numerically. The displayed parameters are rounded, while the
calculations use the more precise values supplied in the \ancillarylink{ancillary
file}. This matters especially for~\eqref{eq:mixedHexample}, where
large terms cancel in the metric. Substituting only the printed
digits can therefore produce misleading errors.

\begin{table}[ht]
\centering\small
\begin{tabular}{lrr}
\toprule
&Extremal ring&Nonextremal hole\\\midrule
$M_i$&$10.4851814117$&$3.76050278224$\\
$J_{i\Psi}$&$-27.0871298314$&$-3.01618915937$\\
$J_{i\Phi}$&$-1.98809177025$&$1.72063522681$\\
$\cA_i$&$49.9661960006$&$142.675334550$\\
$T_i$&$0$&$0.0440383429061$\\
$\Omega_i^\Psi$&$-0.187667760010$&$-0.240208135660$\\
$\Omega_i^\Phi$&$-0.959080455508$&$0.123031511830$\\\bottomrule
\end{tabular}
\caption{The example~\eqref{eq:mixedRexample}, with $G_5=1$.
The ADM charges are $M=14.2456841940$,
$J_\Psi=-30.1033189908$, $J_\Phi=-0.267456543440$,
and $S_{\rm tot}=48.1603826378$.}
\label{tab:mixedRcharges}
\end{table}
\begin{table}[ht]
\centering\small
\begin{tabular}{lrr}
\toprule
&Nonextremal ring&Extremal hole\\\midrule
$M_i$&$0.0125362610951$&$2.677692081697$\\
$J_{i\Psi}$&$-0.0230441808417$&$-1.614419215537$\\
$J_{i\Phi}$&$-3.97887357730\,10^{-6}$&$1.078283352215$\\
$\cA_i$&$0.000101$&$33.15999594663$\\
$T_i$&$1.274899277788$&$0$\\
$\Omega_i^\Psi$&$-0.222279558534$&$-0.662982620774$\\
$\Omega_i^\Phi$&$-805.0182503913$&$0.662901982538$\\\bottomrule
\end{tabular}
\caption{The example~\eqref{eq:mixedHexample}, with $G_5=1$.
The ADM charges are $M=2.690228342792$,
$J_\Psi=-1.637463396379$, $J_\Phi=1.078279373341$,
and $S_{\rm tot}=8.290024236659$.}
\label{tab:mixedHcharges}
\end{table}
We checked each example in two independent ways: we extracted the
ADM charges at infinity and integrated the Komar currents around
each horizon. The resulting sums agree to better than $10^{-32}$,
and both horizons satisfy their Smarr relations. We also checked
that the angular velocities are constant over each horizon and
that the extremal area--spin relations hold. At 156 exterior points
for the extremal-ring example and 244 for the extremal-hole example,
we found $\bar f>0$, $\bar G_{\Psi\Psi}>0$ and $\det\Lambda>0$.
These tests support regularity and causality at the sampled points,
but do not establish a uniform bound on the full exterior.

We can extend the coordinates beyond each extremal horizon. Near either extremal point write
\begin{gather}
 \dd s^2=-R^2N\dd t^2+
 \Lambda_{AB}(\dd\phi^A+\beta^A\dd t)(\dd\phi^B+\beta^B\dd t)
 +F\left(\frac{\dd R^2}{R^2}+\dd\theta^2\right),\nonumber\\
 N=\frac{\sin^2\theta}{\det\Lambda},\qquad F=R^2\bar f.
\end{gather}
Let $\beta^A=-\Omega_i^A+k^AR+O(R^2)$, and define
\begin{equation}
 c_{\rm in}^2=N_0/F_0,\qquad
 t_1=\frac{N_1/c_{\rm in}^2-F_1}{2c_{\rm in}F_0}.
\end{equation}
When $c_{\rm in},t_1,k^A$ are angularly constant, the transformation
\begin{equation}
 \dd t=\dd v+\left(-\frac1{c_{\rm in}R^2}+\frac{t_1}{R}\right)\dd R,
 \qquad
 \dd(\phi^A-\Omega_i^At)=
              \dd\psi^A+\frac{k^A}{c_{\rm in}R}\dd R
 \label{eq:mixedingoing}
\end{equation}
cancels the potentially divergent radial terms. This is the
extremal construction used in section~\ref{sec:extremal}, closely
related to the extensions in~\cite{ArmasDipole,DurkeePS}.
At both examples, evaluation at five horizon angles verifies that
the constants required by~\eqref{eq:mixedingoing} are angle independent
and that the subleading radial poles cancel.
Continuing the common roots $u_1,u_4,u_5$ on their signed analytic
branches across $R=0$, three two-sided angular checks give smooth
limits with differences of order $R$ and determinant
$-F^2\sin^2\theta$. These are local extension checks. They do not
construct a maximal interior extension.

\section{Explicit metric coefficients, thermodynamics, regularity and checks}
\label{app:explicit}

We give the formulas needed to evaluate the metric and its physical
properties. In~\ref{app:scalar} we write the general metric as scalar
functions, and in~\ref{app:extscalar} we obtain their doubly extremal
values. In~\ref{app:analyticthermo} we evaluate the metric on the rods
and at infinity to compute the charges, temperatures and entropies.
Finally, in~\ref{app:regularity} we explain how we test conical balance,
smoothness and causality, and state the domains covered by these
tests.\footnote{\label{app:repro}
We checked the formulas using independent matrix and scalar evaluations,
ADM and horizon Komar integrals, Smarr relations and first-law
variations. We also checked the horizon-charge Jacobian, asymptotic
falloffs, axis cone factors, horizon extensions and the plotted phase
curves. The \ancillarylink{ancillary files} contain the precise examples and validation
records, including convergence checks.}

\subsection{General metric coefficients}
\label{app:scalar}

The matrix formula in section~\ref{sec:metric} can be expanded into
scalar products without printing thousands of terms. We do this in three steps: assign each pole a Killing-direction
label, evaluate the resulting Cauchy products, and sum over the finite
set of label assignments. The final seven functions are listed in
\eqref{eq:scalarseven}. The pole functions
are~\eqref{eq:poles}. We use $x=\rho^2$, $p=(u_1,u_2,u_3,-x/u_4,u_5,-x/u_6)$,
and the three scalar coefficients are
\begin{equation}
 h_t=-\frac{u_1u_3}{u_4u_6},\qquad
 h_\psi=-\frac{u_1u_2u_5}{x},\qquad
 h_\phi=-\frac{u_2u_3u_5}{u_4u_6}.
\end{equation}
The normalised dressing rows give a compact set of scalar weights.
We divide each row in~\eqref{eq:rows} by the entry that remains
nonzero when its polarisation constant vanishes. These rescalings depend
on position and endpoints but not on the six polarisation constants.
The resulting $6\times3$ matrix has rows
\begin{equation}\begin{aligned}
 M_1&=(C_1r_1,1,0),&M_2&=(0,C_2r_2,1),&M_3&=(1,0,C_3r_3),\\
 M_4&=(1,0,B_4r_4),&M_5&=(0,B_5r_5,1),&M_6&=(C_6r_6,0,1),
\end{aligned}\label{eq:normalizedrows}\end{equation}
where
\begin{align}
 r_1&=-\frac{4d_{41}d_{61}d_{21}d_{51}}{d_{31}(a_1+1)}
 \frac{u_1^2u_2u_4u_5u_6R_{13}}{u_3R_{14}R_{16}R_{12}R_{15}},\nonumber\\
 r_2&=-\frac{(a_2+1)d_{32}}{2d_{12}d_{42}d_{62}}
 \frac{u_3R_{12}R_{24}R_{26}}{u_1u_2u_4u_6R_{23}},\nonumber\\
 r_3&=\frac{d_{13}}{2d_{23}d_{53}}
 \frac{u_1R_{23}R_{35}}{u_2u_3u_5R_{13}},&
 r_4&=\frac{R_{14}u_4}{R_{24}R_{45}},\nonumber\\
 r_5&=-\frac{(a_5+1)d_{35}}{2d_{15}d_{45}d_{65}}
 \frac{u_3R_{15}R_{45}R_{56}}{u_1u_4u_5u_6R_{35}},&
 r_6&=\frac{R_{26}R_{56}}{u_6R_{16}}.
 \label{eq:rowratios}
\end{align}
We rebuild $\Gamma$ from these rows and put
$N_{ia}=M_{ia}H_{aa}/p_i$. Both $G$ and $\mathcal P$ are invariant
under the row rescaling. The ratio $\det\Gamma/\det\Gamma_0$ is
also unchanged when the same rescaling is used in its reference
matrix, whose six polarisation constants are zero.

\subsubsection{Finite scalar products and sums}

We set $b_i^a=M_{ia}$ using~\eqref{eq:normalizedrows}.
These position-dependent weights differ from the constant angular
basis coefficients $b_{iA}$ in~\eqref{eq:scalarphysicalfunctions}.
Each row has two possible nonzero entries, so choosing one entry
from each of the six rows gives $2^6=64$ assignments. Their set is
$\mathfrak A$, with the label order $t<\psi<\phi$:
\begin{equation}
 \mathfrak A=\{t,\psi\}\times\{\psi,\phi\}\times\{t,\phi\}
             \times\{t,\phi\}\times\{\psi,\phi\}\times\{t,\phi\}.
\end{equation}
For $\alpha\in\mathfrak A$ put
\begin{equation}\begin{aligned}
 I_a(\alpha)&=\{i:\alpha_i=a\},&
 \chi_\alpha&=(-1)^{\sum_{i<j}\boldsymbol1_{\alpha_i>\alpha_j}}
                         \prod_{i=1}^6 b_i^{\alpha_i},&
 r_a(\alpha)&=\#\{i:\alpha_i>a\}.
\end{aligned}\end{equation}
Here $I_a$ groups the poles assigned to direction $a$, while the
sign in $\chi_\alpha$ records the permutations needed to group them.
The integer $r_a(\alpha)$ counts labels and is unrelated to the
rational row functions $r_i$. For increasing index sets $I,J\subset\{1,\ldots,6\}$ and
$\epsilon,\eta\in\{0,1\}$ satisfying
$n=|I|+\epsilon=|J|+\eta$, define the scalar product
\begin{equation}
 \mathcal C_{\epsilon\eta}(I,J)=
 \frac{(-x)^{n(n-1)/2}
       \displaystyle\prod_{\substack{i<i'\\i,i'\in I}}(p_{i'}-p_i)
       \displaystyle\prod_{\substack{j<j'\\j,j'\in J}}(p_{j'}-p_j)}
      {\displaystyle\prod_{i\in I,j\in J}(x+p_i p_j)
       \displaystyle\prod_{j\in J}p_j^\epsilon
       \displaystyle\prod_{i\in I}p_i^\eta}.
 \label{eq:scalarcauchyproduct}
\end{equation}
Every empty product is one. In particular
$\mathcal C_{00}(\varnothing,\varnothing)=
\mathcal C_{11}(\varnothing,\varnothing)=1$.
All pole-pair denominators can be factored by
\eqref{eq:mixedpolefactor}. The differences are equally explicit:
\begin{equation}
 p_i-p_j=\begin{cases}
 2(a_i-a_j)u_i u_j/R_{ij},&i,j\notin\{4,6\},\\
 2x(a_i-a_j)/R_{ij},&i,j\in\{4,6\},\\
 R_{ij}/u_j,&i\notin\{4,6\},\ j\in\{4,6\},\\
 -R_{ij}/u_i,&i\in\{4,6\},\ j\notin\{4,6\}.
 \end{cases}
 \label{eq:scalardifferences}
\end{equation}

The two scalar sums required for all seven functions are
\begin{align}
 \mathcal S_0={}&\sum_{\substack{\alpha,\beta\in\mathfrak A\\
                 |I_k(\alpha)|=|I_k(\beta)|\ \forall k}}
 \chi_\alpha\chi_\beta
 \prod_{k=t,\psi,\phi}h_k^{|I_k(\alpha)|}
 \mathcal C_{00}\bigl(I_k(\alpha),I_k(\beta)\bigr),\nonumber\\
 \mathcal S_{ab}={}&\sum_{\substack{\alpha,\beta\in\mathfrak A\\
 |I_k(\alpha)|+\delta_{ka}=|I_k(\beta)|+\delta_{kb}\ \forall k}}
 (-1)^{r_a(\alpha)+r_b(\beta)}\chi_\alpha\chi_\beta\nonumber\\
 &\hspace{10mm}\times\prod_{k=t,\psi,\phi}
 h_k^{|I_k(\alpha)|+\delta_{ka}}
 \mathcal C_{\delta_{ka},\delta_{kb}}
                   \bigl(I_k(\alpha),I_k(\beta)\bigr).
 \label{eq:scalarsums}
\end{align}
The sums give explicit rational functions of the pole functions and
parameters. Setting all six polarisation constants to zero gives the
reference factor
\begin{equation}\begin{aligned}
 \mathcal S_{\rm ref}={}&h_t^2h_\psi h_\phi^3
 \mathcal C_{00}(\{3,4\},\{3,4\})
 \mathcal C_{00}(\{1\},\{1\})
 \mathcal C_{00}(\{2,5,6\},\{2,5,6\}),\\
 \mathcal L={}&\frac{f_s}{P\mathcal S_{\rm ref}},
\end{aligned}\end{equation}
where the elementary product $f_s$ is given in~\eqref{eq:f} and
$P$ in~\eqref{eq:scalarP}. The seven desired functions are
\begin{equation}\begin{aligned}
 H_x&=\mathcal L\mathcal S_0,&H_y&=-\mathcal L\mathcal S_{tt},\\
 \omega_\psi&=-\mathcal L\mathcal S_{t\psi},&
 \omega_\phi&=-\mathcal L\mathcal S_{t\phi},\\
 V_{\psi\psi}&=\mathcal L\mathcal S_{\psi\psi},&
 V_{\psi\phi}&=\mathcal L\mathcal S_{\psi\phi},&
 V_{\phi\phi}&=\mathcal L\mathcal S_{\phi\phi}.
\end{aligned}\label{eq:scalarseven}\end{equation}
This completes the scalar specification of the general metric.
The product formula~\eqref{eq:scalarcauchyproduct} follows from the
Cauchy identity. Assigning each of the six factors its label
$t,\psi,\phi$ gives~\eqref{eq:scalarsums} by multilinearity.
Consequently $\mathcal S_0=\det\Gamma$,
$\mathcal S_{ab}/\mathcal S_0=G_{ab}$, and
$\mathcal S_{\rm ref}=\det\Gamma_0$. These equalities establish the
identity with the inverse-scattering construction but they are not
needed to evaluate the scalar formulas.

\paragraph{Polynomial form.}

After collecting terms using~\eqref{eq:scalardifferences}, all seven
functions share the positive monomial denominator
\begin{equation}\begin{aligned}
 \mathfrak F={}&4u_1^3u_2^3u_3^4u_4^4u_5^3u_6^4
                  (d_{42}d_{52}d_{54}d_{63})^2\\
 &\times R_{12}R_{23}R_{25}R_{34}^{\,3}R_{36}R_{46}
                         \prod_{1\leq i\leq j\leq6}R_{ij}.
\end{aligned}\label{eq:scalarcommonF}\end{equation}
Specifically,
\begin{equation}
 (H_x,H_y,\omega_\psi,\omega_\phi,V_{\psi\psi},V_{\psi\phi},V_{\phi\phi})
 =\frac1{\mathfrak F}
 (\mathfrak M_x,\mathfrak M_y,\mathfrak W_\psi,\mathfrak W_\phi,
             \mathfrak V_{\psi\psi},\mathfrak V_{\psi\phi},\mathfrak V_{\phi\phi}).
\end{equation}
The numerators are integer-coefficient polynomials in
$x,u_i,R_{ij},d_{ij}$ with $i>j$, $a_i+1$, and
$(C_1,C_2,C_3,B_4,B_5,C_6)$. Their coefficients are specified by the finite
sums~\eqref{eq:scalarsums}--\eqref{eq:scalarseven}. The expanded
polynomials are included in the \ancillarylink{ancillary Mathematica file}.

\subsubsection{Scalar outer-axis normalisation}

We now normalise the two outer axes to primitive angles of period
$2\pi$, as in~\eqref{eq:basis}. To evaluate the constants in
\eqref{eq:scalarphysical}, we set
\begin{equation}
 D=f=PH_x,\quad
 (Q_{tt},Q_{t\psi},Q_{t\phi},Q_{\psi\psi},Q_{\psi\phi},Q_{\phi\phi})
 =P(-H_y,-\omega_\psi,-\omega_\phi,V_{\psi\psi},V_{\psi\phi},V_{\phi\phi}).
\end{equation}
Here $Q_{ab}=fG_{ab}$ are scalar functions and are unrelated to the
Komar-current potential. Evaluate their exact constant coefficients
at $z_L=a_1-1$ and $z_R=a_6+1$. For a collected monomial, assign
\begin{equation}
 \operatorname{ord}_s x=1,\qquad
 \operatorname{ord}_s u_i=\boldsymbol1_{i\leq s},\qquad
 \operatorname{ord}_s R_{ij}=\boldsymbol1_{i\leq s\ \mathrm{or}\ j\leq s},
 \quad s=0,6.
\end{equation}
All other factors have order zero. Every collected term of $D,Q_{ab}$
has nonnegative order. Discard the positive-order terms and replace
$x$ by one, $u_i$ by $\bar u_i$ and $R_{ij}$ by $\bar R_{ij}$
from~\eqref{eq:axisrootpolynomials} and the leading coefficients
following~\eqref{eq:rodblockformulas}. This gives the exact values on
the axes.

On the lower axis we obtain
\begin{equation}\begin{aligned}
 D_L&=Q_{tt}Q_{\psi\psi}-Q_{t\psi}^2,&
 q_L&=-D/D_L,\\
 v_L^t&=\frac{Q_{t\psi}Q_{\psi\phi}-Q_{\psi\psi}Q_{t\phi}}{D_L},&
 v_L^\psi&=\frac{Q_{t\psi}Q_{t\phi}-Q_{tt}Q_{\psi\phi}}{D_L},&v_L^\phi&=1.
\end{aligned}\end{equation}
On the upper axis the corresponding formulas are
\begin{equation}\begin{aligned}
 D_R&=Q_{tt}Q_{\phi\phi}-Q_{t\phi}^2,&q_R&=-D/D_R,\\
 v_R^t&=\frac{Q_{t\phi}Q_{\psi\phi}-Q_{\phi\phi}Q_{t\psi}}{D_R},&
 v_R^\phi&=\frac{Q_{t\phi}Q_{t\psi}-Q_{tt}Q_{\psi\phi}}{D_R},&v_R^\psi&=1.
\end{aligned}\end{equation}
Use the appropriate axis coefficients in each line. Then
\begin{equation}\begin{aligned}
 q_\Psi&=v_R^t/\sqrt{q_R},&q_\Phi&=v_L^t/\sqrt{q_L},\\
 b_{\psi\Psi}&=1/\sqrt{q_R},&b_{\phi\Psi}&=v_R^\phi/\sqrt{q_R},\\
 b_{\psi\Phi}&=v_L^\psi/\sqrt{q_L},&b_{\phi\Phi}&=1/\sqrt{q_L}.
\end{aligned}\label{eq:scalarbasis}\end{equation}
This is the scalar version of~\eqref{eq:basis}. In the canonical
convention, replace $q_A,b_{iA}$ by $q_A/\sqrt\delta,b_{iA}/\sqrt\delta$
and the base coefficient $P$ by $P/\delta$ in the physical line element.

\subsection{Doubly extremal metric coefficients}
\label{app:extscalar}

We obtain the seven functions in~\eqref{eq:extscalarraw} by expanding
the general metric along the correlated limit of
section~\ref{sec:extremal}. The formulas below evaluate the metric
directly at zero temperature. We keep $(Z,c,d,H_*,v_*,w,r_0,\beta)$
free at this stage and impose nonzero denominators and balance
subsequently. We write $H=H_*$ and $v=v_*$.

To see why the relation for $\eta$ in~\eqref{eq:extpolar} is needed,
put $x=\rho^2$ and let
$\eta=E\lambda+O(\lambda^2)$. Eliminating the coupled pole residues
in~\eqref{eq:G} leaves a
coefficient proportional to
\begin{equation}
 \frac{U^2(U-V)}{2V\beta H_*^2(x+U^2)}
           \left[E-c(1+Z^{-1})\right],\qquad
 U=\sqrt{x+z^2}-z,\quad V=\sqrt{x+(z-Z)^2}-(z-Z).
 \label{eq:critical}
\end{equation}
Since $Z=(V-U)(x+UV)/(2UV)$ and $V>U>0$ off the axis, the
prefactor is nonzero. Equation~\eqref{eq:extpolar} therefore enforces
a genuine cancellation required by this limiting construction.
Thus the leading powers of $\lambda$ alone do not determine the
limit: the displayed subleading relation among the polarisations is
also needed.

\subsubsection{Taylor coefficients for the extremal limit}

To obtain the extremal functions it suffices to keep eight
coefficients, through seventh order, of an auxiliary variable
$\epsilon=\tau^2$. Terms of higher order cannot contribute to the
finite metric. We express this truncation by working in the algebra
$\mathcal P=\mathbb R[\epsilon]/(\epsilon^8)$, with coefficients
rational in the parameters and in $x,U,V$. That is, multiply polynomials in the usual way and discard powers
$\epsilon^8$ and higher. For a polynomial $A$ with constant
coefficient $A_0\ne0$, every integer power and every square root below is
specified by the finite formula
\begin{equation}
 \mathcal B_\alpha(A)=A_0^\alpha
 \sum_{k=0}^{7}\binom{\alpha}{k}
 \left(\frac{A-A_0}{A_0}\right)^k\quad\hbox{in }\mathcal P.
 \label{eq:extfinitepower}
\end{equation}
All square-root arguments used below have constant entry one, fixing
the branch. No differential or infinite-series operation is required.
Set $x=\rho^2$, use $U,V$ from~\eqref{eq:critical}, and put
\begin{equation}
 Z=\frac{(V-U)(x+UV)}{2UV},\qquad z=\frac{x-U^2}{2U}.
\end{equation}
We first expand the endpoints:
\begin{align}
 S&=r_0^2\mathcal B_{1/2}\left(1+\frac{\epsilon^2}{4r_0^4}\right),&
 R&=\mathcal B_{1/2}\left(\frac{S+\epsilon/2}{r_0^2}\right),\nonumber\\
 A&=Z-\frac{S-r_0^2}{2}-\frac\epsilon4,&
 B&=Z-\frac{S-r_0^2}{2}+\frac\epsilon4,\nonumber\\
 C&=\frac{H^2}{2}+\epsilon B,&
 (a_1,\ldots,a_5)&=(0,c\epsilon/2,c\epsilon/2+d\epsilon^2/4,A,B).
 \label{eq:extfiniteends}
\end{align}
Here $C$ is the finite factor of the receding endpoint after removing
its divergent power, $a_6=\epsilon^{-1}C$ and
it is distinct from a polarisation constant. Define
\begin{align}
 b_0&=\beta/R,&\alpha&=-1/(\beta R),\nonumber\\
 \pi_1&=-\frac1{ZH}\mathcal B_{1/2}
  \left(\frac{(1+d\epsilon/(2c))Z^2H^2}{2ABC}\right),&
 \pi_6&=\frac1H\mathcal B_{1/2}
  \left(\frac{C}{C-c\epsilon^2/2}\right),\nonumber\\
 \pi_2&=\frac{1-c(1+Z^{-1})\epsilon/2-w\epsilon^2/4}{\pi_1v},&
 \pi_3&=v,\nonumber\\
 \pi_4&=\frac{1-\epsilon\pi_6 b_0}{b_0+\epsilon\pi_6},&
 \pi_5&=\frac{H^2\pi_6\alpha}{2(B+1)}.
 \label{eq:extfinitepolar}
\end{align}
Every quotient here means multiplication by~\eqref{eq:extfinitepower}
with $\alpha=-1$. If $\tau^2=\epsilon$, the six polarisation factors
are products of the finite factors $(\pi_1,\ldots,\pi_6)$ and the
following powers of $\tau$
\begin{equation}
 (\nu_{C_1},\nu_{C_2},\nu_{C_3},\nu_{B_4},\nu_{B_5},\nu_{C_6})
 =(1,-2,1,1,0,1).
 \label{eq:extfinitepolarorders}
\end{equation}
The bookkeeping symbol $\tau$ is not assigned a numerical value. For $i\le5$, put $u_i^0=U$, $a_i^0=0$ for $i\le3$ and
$u_i^0=V$, $a_i^0=Z$ otherwise. With
$r_i=(x+(u_i^0)^2)/(2u_i^0)$ and $\Delta_i=a_i-a_i^0$, set
\begin{align}
 \widehat u_i&=u_i^0+\Delta_i+r_i\left[
  \mathcal B_{1/2}\left(1-\frac{2(z-a_i^0)\Delta_i}{r_i^2}
                       +\frac{\Delta_i^2}{r_i^2}\right)-1\right],\nonumber\\
 D&=C-\epsilon z,&
 \widehat u_6&=D\left[1+\mathcal B_{1/2}
                         (1+x\epsilon^2/D^2)\right].
 \label{eq:extfinitepoles}
\end{align}
Likewise the pole functions have finite factors $\widehat u_i$ and powers
$\nu_{u_i}=0$ for $i\le5$, $\nu_{u_6}=-2$.
All remaining factors of subsection~\ref{app:scalar} are now explicit:
\begin{center}\small
\begin{tabular}{lll}\toprule
Factor&Finite factor in $\mathcal P$&Power $\nu$ of $\tau$\\\midrule
$x$&$x$&$0$\\
$R_{ij}$, $i,j\le5$&$x+\widehat u_i\widehat u_j$&$0$\\
$R_{i6}$, $i\le5$&$\epsilon x+\widehat u_i\widehat u_6$&$-2$\\
$R_{66}$&$\epsilon^2x+\widehat u_6^2$&$-4$\\
$a_i+1$, $i\le5$&$a_i+1$&$0$\\
$a_6+1$&$C+\epsilon$&$-2$\\
$d_{21}$&$c/2$&$2$\\
$d_{31}$&$c/2+d\epsilon/4$&$2$\\
$d_{32}$&$d/4$&$4$\\
$d_{54}$&$1/2$&$2$\\
Other $d_{ij}$, $i\le5$&$a_i-a_j$&$0$\\
$d_{6j}$&$C-\epsilon a_j$&$-2$\\
$s_b$&$(1-\epsilon)/2$&$-1$\\
$c_b$&$(1+\epsilon)/2$&$-1$\\\bottomrule
\end{tabular}
\end{center}
As before $d_{ij}=a_i-a_j$ with $i>j$. The last two rows incorporate
the singular boost $s_b=v_b$, $c_b=u_b$ of section~\ref{sec:extremal}.

\subsubsection{The seven finite scalar sums}

For a monomial $m=q\prod_a a^{n_a}$ from the collected general
expressions, write each factor as $a=\tau^{\nu_a}\widehat a(\epsilon)$, using the
finite factor and power
in the preceding table and~\eqref{eq:extfinitepolarorders}.
Put $N=(6-\sum_a n_a\nu_a)/2$ and define the finite scalar operation
\begin{equation}
 \mathcal E(m)=q
 \sum_{\substack{\sum_a k_a=N\\0\le k_a\le7}}
 \prod_a\left(\mathcal B_{n_a}(\widehat a)\right)_{k_a}.
 \label{eq:extfiniteeval}
\end{equation}
The subscript selects a Taylor coefficient, and an empty sum is zero.
We apply $\mathcal E$ to each monomial and add the results. This
extracts the coefficient that survives in the extremal metric.
We use the general scalar functions of~\ref{app:scalar} to
form $\mathcal D=P H_x$ and
\begin{equation}
 (Q_{tt},Q_{t\psi},Q_{t\phi},Q_{\psi\psi},Q_{\psi\phi},Q_{\phi\phi})
 =P(-H_y,-\omega_\psi,-\omega_\phi,V_{\psi\psi},V_{\psi\phi},V_{\phi\phi}).
\end{equation}
These are the collected monomial expressions for $f$ and $fG_{ab}$.
The desired extremal functions, with every factor evaluated by the
finite table above, are
\begin{equation}\begin{aligned}
 K&=8\mathcal E(\mathcal D),\qquad Y=-8\mathcal E(Q_{tt}),\\
 W_\psi&=8\mathcal E\Bigl(s_b Q_{tt}/C_6-c_bQ_{t\psi}-s_bQ_{t\phi}\Bigr),\\
 W_\phi&=8\mathcal E\Bigl(c_b Q_{tt}/C_6-s_bQ_{t\psi}-c_bQ_{t\phi}\Bigr),\\
 V_{\psi\psi}&=8\mathcal E\Bigl(
 s_b^2Q_{tt}/C_6^2-2s_bc_bQ_{t\psi}/C_6-2s_b^2Q_{t\phi}/C_6\\
 &\hspace{27mm}+c_b^2Q_{\psi\psi}+2s_bc_bQ_{\psi\phi}+s_b^2Q_{\phi\phi}\Bigr),\\
 V_{\psi\phi}&=8\mathcal E\Bigl(
 s_bc_bQ_{tt}/C_6^2-(s_b^2+c_b^2)Q_{t\psi}/C_6-2s_bc_bQ_{t\phi}/C_6\\
 &\hspace{27mm}+s_bc_bQ_{\psi\psi}+(s_b^2+c_b^2)Q_{\psi\phi}+s_bc_bQ_{\phi\phi}\Bigr),\\
 V_{\phi\phi}&=8\mathcal E\Bigl(
 c_b^2Q_{tt}/C_6^2-2s_bc_bQ_{t\psi}/C_6-2c_b^2Q_{t\phi}/C_6\\
 &\hspace{27mm}+s_b^2Q_{\psi\psi}+2s_bc_bQ_{\psi\phi}+c_b^2Q_{\phi\phi}\Bigr).
\end{aligned}\label{eq:extfiniteseven}\end{equation}
The notation $C_6$ in these monomials means the polarisation factor
with finite factor $\pi_6$ and power $1$. It differs from the endpoint
factor $C$.
Their lowest powers of $\tau$ are $(-4,-4,-6,-6,-8,-8,-8)$.
Every contributing $N$ is an integer between zero and seven, which
proves that the stated finite polynomial degree suffices.

To see the identification with the earlier extremal construction,
$\mathcal E$ returns the $\tau^6$ entry of each scalar monomial.
The displayed combinations are precisely the scalar expansion of
$f\mathcal C^{\mathsf T}G\mathcal C$, while
$\lambda^{-3}=8\tau^{-6}$. Cancellation of the lower powers
therefore gives $K=f_e$ and the six densities used in
section~\ref{sec:extscalar}. 

\paragraph{The subfamily used for the global bounds.}
The \ancillarylink{ancillary Mathematica file} evaluates these finite formulas for all
parameters of the correlated ansatz. We then apply the physical
normalisation~\eqref{eq:extscalarphysical} and impose balance.
For $Z=5$, $r_0=37/50$, $\beta=-41/50$, $w=-30$, additional
expanded polynomials retain $(u,y,c,H_*,v_*,d)$ symbolically.
These expanded expressions apply to this subfamily only.

\begin{samepage}
Their common denominator, up to an overall positive constant, is
\begin{equation}
 c^2H^6v^2(uy-y+1)^4(uy-u-y)^4.
 \label{eq:extscalarcommonden}
\end{equation}
\end{samepage}
The denominator is nonzero for $0<u,y<1$ and $cHv\ne0$. This statement concerns
the scalar-density denominator. Regularity of the metric also
requires the already studied zeros and boundary behaviour of $K$.

\subsection{Analytic thermodynamic quantities}
\label{app:analyticthermo}

We compute the thermodynamic quantities in four steps. First, we
write the current potential whose endpoint differences give the
Komar charges. Second, we explain how to take finite values of the
metric and potential on a rod, where direct substitution would give
indeterminate ratios. Third, we evaluate these quantities on the
three axes and two horizons. These values determine the charges,
angular velocities, temperatures and entropies in
\eqref{eq:algebraiccharges}--\eqref{eq:entropy}.
Finally, we extract the ADM charges independently at infinity.
The formulas use only finite algebraic operations. Their denominators
must be nonzero in the chosen chart, and balance and the regularity
conditions must still be imposed.

\subsubsection{Current potential and horizon charges}
\label{app:currentprimitive}

The matrix potential $\mathcal P$ converts each current integral
into a difference between its endpoint values, as in
\eqref{eq:currentpotential}. For this solution it can be written
algebraically. We write $M$
for the $6\times3$ matrix of dressing rows (only in this matrix formula),
$N=K^{\mathsf T}$, and
\begin{align}
 s_t&=u_4+u_6-u_1-u_3,\nonumber\\
 s_\psi&=-u_1-u_2-u_5-4z-2,&
 s_\phi&=u_4+u_6-u_2-u_3-u_5-2z.
\end{align}
Then
\begin{equation}
 \mathcal P=-\diag(s_t,s_\psi,s_\phi)-M^{\mathsf T}\Gamma^{-1}N,
 \qquad
 \partial_z\mathcal P=2xG^{-1}\partial_xG,
 \quad 2\partial_x\mathcal P=-G^{-1}\partial_zG.
 \label{eq:analyticprimitive}
\end{equation}
These identities follow from the first inverse-spectral coefficient
of the dressed generating matrix. We can also check them by
differentiating~\eqref{eq:G}. Constant changes of Killing basis
act by $\mathcal P\mapsto A^{-1}\mathcal PA$. The $T$ row of this
transformed potential is constant on each smooth spacelike axis.

\subsubsection{Taking finite values on the rods}

To resolve the indeterminate ratios on an axis, we use the normalised
rows~\eqref{eq:normalizedrows}, factor the known powers of $x=\rho^2$
and retain the finite coefficients. For an analytic square matrix $B(x)=B_0+xB_1+x^2B_2+\cdots$,
define three explicit functions. Let $B_0[j\leftarrow B_k]$ mean
that only column $j$ of $B_0$ is replaced by column $j$ of $B_k$.
Then
\begin{align}
 \mathfrak D_0(B)&=\det B_0,\nonumber\\
 \mathfrak D_1(B)&=\sum_j\det B_0[j\leftarrow B_1],\nonumber\\
 \mathfrak D_2(B)&=\sum_j\det B_0[j\leftarrow B_2]
 +\sum_{j<k}\det B_0[j\leftarrow B_1,k\leftarrow B_1].
 \label{eq:detcoefficients}
\end{align}
By multilinearity these are exactly $[x^n]\det B$ for $n=0,1,2$.
We use the following evaluation positions:
\begin{equation}\begin{aligned}
 z_0&=a_1-1,&z_2&=(a_2+a_3)/2,&z_3&=(a_3+a_4)/2,\\
 z_4&=(a_4+a_5)/2,&z_6&=a_6+1.
\end{aligned}\label{eq:axispositions}\end{equation}
The unit offset is arbitrary in the fixed reference units. At $z_s$
write $D_i=|z_s-a_i|>0$. The needed input coefficients are the
finite binomial expressions
\begin{equation}
 u_i=2D_i\,\boldsymbol1_{i>s}
       +\sum_{n=1}^{5}\binom{1/2}{n}D_i^{1-2n}x^n+O(x^6),
 \qquad R_{ij}=x+u_i u_j.
 \label{eq:axisrootpolynomials}
\end{equation}
Known powers of $x$ are factored before truncation and division.
Although only quadratic coefficients of the final matrices are
needed, keeping the five root coefficients in
\eqref{eq:axisrootpolynomials} preserves them through intermediate
Laurent divisions. For pole-pair denominators use $R_{ij}$ for two solitons,
$xR_{ij}/(u_i u_j)$ for two antisolitons, and, for a soliton at $i$
and an antisoliton at $j$,
\begin{equation}
 x+p_ip_j=\frac{2x u_i(a_j-a_i)}{R_{ij}}.
 \label{eq:mixedpolefactor}
\end{equation}
This exact identity avoids a subtraction of terms whose leading
coefficients coincide.

\subsubsection{Metric and potential on each rod}

The index $s$ labels the intervals containing the five positions
$z_s$ in~\eqref{eq:axispositions}. On each interval, the table below
lists the powers of $x$ needed to make the matrix entries finite.
Form the diagonal matrices
$\mathsf S=\diag(x^{\alpha_i})$,
$\mathsf R=\diag(x^{\beta_i})$, and set
\begin{equation}
 L=\mathsf R^{-1}\mathsf S\Gamma\mathsf S,\quad
 U=\mathsf R^{-1}\mathsf S N,\quad
 V=\mathsf S N,\quad W=\mathsf S M.
 \label{eq:rodmatrices}
\end{equation}
The powers and determinant orders are
\begin{center}\small
\begin{tabular}{ccllcc}\toprule
$s$ & Rod & $(\alpha_1,\ldots,\alpha_6)$ &
 $(\beta_1,\ldots,\beta_6)$ & $d$ & $\eta$\\\midrule
0 & Lower axis & $(0,0,0,0,0,0)$ & $(-1,-1,-1,-1,-1,-1)$ &1&1\\
2 & Ring horizon & $(0,0,0,0,0,0)$ & $(0,0,0,0,0,0)$ &0&0\\
3 & Middle axis & $(1,0,0,0,0,0)$ & $(1,1,1,1,1,1)$ &2&0\\
4 & Hole horizon & $(0,0,0,0,0,0)$ & $(0,0,0,0,0,0)$ &0&0\\
6 & Upper axis & $(0,1,1,0,1,0)$ & $(1,1,0,0,1,0)$ &2&0\\\bottomrule
\end{tabular}
\end{center}
Here $\det L=x^d(\mathfrak D_d(L)+O(x))$ and
$\mathfrak D_d(L)\ne0$ in this chart. The entries of $L,U,W,x^\eta V$
and $x^\eta H$ are analytic at $x=0$. For $a,b=t,\psi,\phi$ define
two bordered matrices, of size $7\times7$,
\begin{equation}
 B^{ab}=\begin{pmatrix}L&U_{:b}\\
             x^\eta V_{:a}^{\mathsf T}&x^\eta H_{ab}\end{pmatrix},
 \qquad
 E^{ab}=\begin{pmatrix}L&U_{:b}\\W_{:a}^{\mathsf T}&0\end{pmatrix}.
 \label{eq:rodborders}
\end{equation}
The desired metric and potential blocks are the explicit expressions
\begin{equation}
 G^{[s]}_{ab}=\frac{\mathfrak D_{d+\eta}(B^{ab})}
                              {\mathfrak D_d(L)},\qquad
 \mathcal P^{[s]}_{ab}=\frac{\mathfrak D_d(E^{ab})}
                              {\mathfrak D_d(L)}
                       -s_a(0)\delta_{ab}.
 \label{eq:rodblockformulas}
\end{equation}
Here the $s_a$ are the three seed expressions in
\eqref{eq:analyticprimitive}, evaluated by the constant root
coefficients in~\eqref{eq:axisrootpolynomials}. Notice that
$d+\eta\leq2$ for all five rows of the table.

To compute $f^{[s]}$, we remove the known powers of $x$ from $u_i$
and $R_{ij}$ and retain their leading coefficients
\begin{equation}
 \bar u_i=\begin{cases}[2(z_s-a_i)]^{-1},&i\leq s,\\
                         2(a_i-z_s),&i>s,
             \end{cases}\qquad
 \bar R_{ij}=\begin{cases}
 1,&i,j\leq s,\\
 \bar u_i\bar u_j,&i,j>s,\\
 1+\bar u_i\bar u_j,&\text{otherwise}.
 \end{cases}
\end{equation}
The total power of $x$ in $f_s$ is zero on these rods. Thus
\begin{equation}
 f_{s,0}=\frac{\bar u_2\bar u_5}{\bar u_1}
   \prod_i\bar R_{ii}^{-1}\prod_{i<j}\bar R_{ij}^{-w_i\cdot w_j},
 \qquad
 f^{[s]}=f_{s,0}\frac{\mathfrak D_d(L)}
                                  {\mathfrak D_d(L_0)}.
 \label{eq:rodconformal}
\end{equation}
$L_0$ is the same matrix with all six polarisation constants zero.
The subscript $s$ in $f_{s,0}$ denotes the seed, not an extra free
parameter. These formulas together with~\eqref{eq:axisalgebra}
fix the outer basis $A$, $\delta$, both horizon kernels, and every
quantity in~\eqref{eq:algebraiccharges}, \eqref{eq:algebraictemperature},
\eqref{eq:area} and~\eqref{eq:entropy}.

The Schur determinant identity explains the cancellation:
\begin{equation}
 \frac{\det B^{ab}}{\det L}=x^\eta G_{ab},\qquad
 \frac{\det E^{ab}}{\det L}=-M_{:a}^{\mathsf T}\Gamma^{-1}N_{:b}.
\end{equation}
The powers in the table remove all entrywise poles. Exact lower-order
coefficient cancellations then give~\eqref{eq:rodblockformulas}.
This also explains why a large expanded rational function has a
short determinant representation. Individual charges use differences
of the constant $T$ rows on neighbouring axes, so the arbitrary
positions on each connected axis do not enter the answer.

\subsubsection{ADM charges at infinity}
\label{app:asymptoticchecks}

We check the ADM charges independently by expanding the metric at
infinity. Let $F_a=fG_{ta}$, and put
$\xi=1/R_W$, $z=0$, $x=\xi^{-2}$. The collected Cauchy expansion of
\eqref{eq:G}--\eqref{eq:f} is a finite sum of monomials in
$x,u_i,R_{ij}$ and endpoint/polarisation constants. For one monomial
write its spatial part as
$x^{e_x}\prod_i u_i^{e_i}\prod_{i\leq j}R_{ij}^{e_{ij}}$ and
its parameter coefficient as $q$. Define
\begin{equation}\begin{aligned}
 n&=-2e_x-\sum_i e_i-2\sum_{i\leq j}e_{ij},\qquad
 W=q\,2^{\sum e_{ij}},\\
 L_1&=\sum_i e_i a_i+
                \tfrac12\sum_{i\leq j}e_{ij}(a_i+a_j).
\end{aligned}\label{eq:infinityweights}\end{equation}
These follow from $u_i=\xi^{-1}(1+a_i \xi+O(\xi^2))$ and
$R_{ij}=\xi^{-2}(2+(a_i+a_j)\xi+O(\xi^2))$. For each of $f,F_t,F_\psi,F_\phi$
all collected monomials have $n\geq1$ and hence
\begin{equation}\begin{gathered}
 f=f_1\xi+f_2\xi^2+O(\xi^3),\quad
 F_a=h_{a1}\xi+h_{a2}\xi^2+O(\xi^3),\\
 (c_1,c_2)=\left(\sum_{n=1}W,
                     \sum_{n=1}WL_1+\sum_{n=2}W\right),
\end{gathered}
 \label{eq:infinitysums}
\end{equation}
where $(c_1,c_2)$ denotes the coefficient pair of the chosen function.
Only the indicated monomials contribute. We set
\begin{equation}
 b_a=\frac{h_{a2}f_1-h_{a1}f_2}{f_1^2},\qquad
 \widehat b_b=\sum_a b_a A^a{}_b.
\end{equation}
Then the independent ADM expressions in the $A$ convention are
\begin{equation}
 M=\frac{3\pi\delta}{4G_5}\widehat b_T,\qquad
 J_\Psi=-\frac{\pi\delta}{G_5}\widehat b_\Psi,\qquad
 J_\Phi=-\frac{\pi\delta}{G_5}\widehat b_\Phi.
 \label{eq:independentalgebraicadm}
\end{equation}
Equation~\eqref{eq:algebraiccharges} is the more compact standalone
formula, while the monomial form is useful as an independent implementation
and conservation check.

The same expansion can be made at fixed $z/R_W$. The leading
coefficients of $f$ and $fG_{ta}$ are independent of this angle, while
those of the angular densities are affine in $z/R_W$. We also find
$[R_W^{-1}](fG_{tt})=-[R_W^{-1}]f$. Together with the outer-axis
normalisation, these identities give the flat leading metric in the
coordinates of section~\ref{sec:asymptotic}.

\subsection{Regularity criteria and verified domains}
\label{app:regularity}

We express the geometric criteria of section~\ref{sec:nonextregularity}
as explicit sign tests. We first give formulas valid for the full
six-parameter construction, then state which regions have been
verified by interval bounds and which have only been sampled.

\subsubsection{Exterior smoothness and causality}
 All elementary denominators in the raw dressing rows are nonzero
for $x>0$ and ordered endpoints. In particular $u_i>0$,
$x+u_i u_j>0$, and a mixed-pole denominator is
\begin{equation}
 x+u_s(-x/u_t)=\frac{2xu_s(a_t-a_s)}{x+u_su_t}\ne0.
\end{equation}
Let $D=\det\Gamma$ and $D_0=\det\Gamma_0$. The latter splits
into Cauchy blocks on the groups $\{3,4\}$, $\{1\}$ and
$\{2,5,6\}$. Using the Cauchy product~\eqref{eq:scalarcauchyproduct}
and the negative diagonal entries of $H$, we find block signs
$-,-,+$. Thus $D_0>0$ throughout the off-axis domain. Since $f_s>0$, positivity of the physical base factor is
equivalent to $D>0$.

To express angular positivity without inverting the full metric,
use the unit-determinant basis $B$ in~\eqref{eq:canonical} and
write $H'=B^{\mathsf T}HB$, $L=B^{\mathsf T}K$, where $K$ is
the $3\times6$ matrix in~\eqref{eq:G}. Let $L_\Psi$ denote its
$\Psi$ row and $L_{\rm ang}$ its two angular rows. Define
\begin{equation}
 D_\Psi=\det\begin{pmatrix}\Gamma&L_\Psi^{\mathsf T}\\
 L_\Psi&H'_{\Psi\Psi}\end{pmatrix},\qquad
 D_{\mathbb T^2}=\det\begin{pmatrix}\Gamma&L_{\rm ang}^{\mathsf T}\\
 L_{\rm ang}&H'_{\rm ang}\end{pmatrix}.
\end{equation}
Schur complementation gives $D_\Psi=D\bar G_{\Psi\Psi}$ and
$D_{\mathbb T^2}=D\det\Lambda$. An explicit sufficient set of
exterior conditions is therefore
\begin{equation}
 D>0,\quad D_\Psi>0,\quad D_{\mathbb T^2}>0
                    \quad\text{for all }x>0,\ z\in\mathbb R.
 \label{eq:formalineq}
\end{equation}
These determinants have orders six, seven and eight and retain all
six free parameters. They ensure a smooth Lorentzian metric off the
axis and a positive angular block. In particular,
$g^{TT}=-D_{\mathbb T^2}/(xD)<0$, so the physical time $T$ is a time
function and there are no closed causal curves in this domain.
Conical balance is a separate boundary condition, given by
\eqref{eq:balance}. The axes and horizons also require the regular
coordinates described below.
They are sufficient, rather than a minimal characterisation:
removable cancellations could admit a different representation
beyond a domain where this unreduced determinant stays nonzero.

The inequalities must hold at every exterior point. To express this
requirement algebraically, we use
$u_i^2+2(z-a_i)u_i=x$, $u_i>0$, and positive square-root variables
for the normalisation. Multiplying denominators by their squares
preserves signs. This gives a finite system of polynomial equations and inequalities in the
parameters and spatial variables, supplemented by the axis and
horizon conditions. We have not eliminated the spatial variables to obtain
a closed list of elementary parameter inequalities for all six
parameters.

\subsubsection{Balance roots and the nonextremal parameter domain}
\label{app:nonextbounds}
We impose the three balance equations~\eqref{eq:balance}: the middle
axis must have the same degenerating Killing direction and cone
factor as the lower axis. Since both outer angles have period $2\pi$,
these conditions remove the remaining conical defect and align the
angular identifications. We have verified balance in a nonextremal
neighbourhood as follows. Write
$p=(a_2-a_1,\ldots,a_6-a_5,C_2)$ and
$p_*=(1,1,2,1,0.1,-0.1)$. For
\begin{equation}
 |p_i-p_{*i}|\leq10^{-9}\quad(i=1,\ldots,6),
 \label{eq:smallbox}
\end{equation}
the exact balance equations have a unique enclosed analytic root
$(C_3,B_4,B_5)$ near the reference values
\begin{equation}\begin{aligned}
 (a_1,\ldots,a_6)&=(0,1,2,4,5,5.1),\qquad C_2=-0.1,\\
 C_3&\simeq0.833176774260059,\qquad B_4\simeq0.612383796741305,\\
 B_5&\simeq0.058168784076090.
\end{aligned}
 \label{eq:reference}
\end{equation}
These rounded values locate the root. We use the exact balance
solution in the checks. Interval bounds prove
regularity on every finite boundary component: both horizons, all
four poles, all ordinary axes and both auxiliary points. Cone factors
are unity for the primitive $2\pi$ cycles. The proof also covers a
compact exterior rectangle, but does not cover the whole exterior
for every parameter in this nonextremal box.

We use interval arithmetic to verify that the balance root exists
for every parameter choice in~\eqref{eq:smallbox}. We enclose the
three unknown polarisations and the balance Jacobian, including
rounding errors, and apply the interval Newton method in its
Krawczyk form~\cite{Rump}. The resulting map sends the proposed
root box strictly inside itself, with a nonsingular Jacobian
enclosure. This proves existence and uniqueness of the nearby
root. A root-box radius of $10^{-8}$ suffices here. The implicit
function theorem then makes the three polarisations analytic
functions of the six free parameters.

\subsubsection{Nonextremal horizon and endpoint charts}
\label{app:nonextcharts}

For completeness, the ingoing extension~\eqref{eq:nonextingoing} has a
two-sided version. We use the coefficients $\alpha,a,\beta_A$ and
$\lambda_{AB}$ of section~\ref{sec:nonexthorizons}.
For both horizon branches simultaneously, let
\begin{equation}
 U=-\rho e^{-\kappa T},\quad V=\rho e^{\kappa T},\quad
 x=-UV,\qquad \omega=U\dd V-V\dd U.
\end{equation}
The corresponding metric is
\begin{equation}
 ds^2=-f\dd U\dd V+f\dd z^2+\lambda_{AB}\dd\theta^A\dd\theta^B
       -\frac a{4\kappa^2}\omega^2-\frac{\beta_A}\kappa\omega\dd\theta^A,
 \quad\det g=-\frac{\delta^2f^2}{4\kappa^2}.
 \label{eq:kruskal}
\end{equation}

To check the axes, horizon endpoints and auxiliary points, we expand
the exact scalar expressions for $f$ and $fG$ in powers of a local
radial coordinate. This exposes
cancellations at the boundary before any numerical evaluation. On each ordinary axis, positivity of the surviving block,
the exact kernel and the unit cone identity give regular Cartesian
polar tensors. At the two auxiliary seams, the root involutions
remove the nonanalytic square-root branch and the certified
denominators ensure bounded, nondegenerate coefficients.

For a physical endpoint $a_i$ put $y_i=z-a_i$ and
\begin{equation}
 X=\sqrt{\rho^2+y_i^2}-\sigma y_i,\quad
 Y=\sqrt{\rho^2+y_i^2}+\sigma y_i,\quad
 x=XY,\quad F_c=\tfrac14(X+Y)f.
\end{equation}
Choose $(i,\sigma)=(2,+1),(3,-1),(4,+1),(5,-1)$, so that
$X=0$ is the horizon and $Y=0$ the axis. With $\vartheta$ the
degenerating corotating angle and $\zeta$ the surviving one, the
analytic Killing entries have the factors
\begin{align}
 g_{TT}&=-X\mathfrak a,& g_{T\zeta}&=Xb_c,&
 g_{T\vartheta}&=XYe_c,\nonumber\\
 g_{\vartheta\vartheta}&=Yc_c,&
 g_{\vartheta\zeta}&=Yd_c,&g_{\zeta\zeta}&=L_c,\nonumber\\
 \mathfrak a&=4\kappa^2F_c+X\mathfrak a_1,&
 c_c&=4F_c+Yh_c.
\end{align}
The base metric is $F_c(\dd X^2/X+\dd Y^2/Y)$. We factor these
powers in the exact expressions before evaluating $X=Y=0$, so the
cancellations hold at the axis--horizon junction as well as along
each edge. We then set
\begin{equation}
 U_c=-\sqrt X e^{-\kappa T},\quad V_c=\sqrt X e^{\kappa T},\quad
 P=\sqrt Y\cos\vartheta,\quad Q=\sqrt Y\sin\vartheta,
\end{equation}
and define $\omega_c=U_c\dd V_c-V_c\dd U_c$,
$\eta_c=P\dd Q-Q\dd P$. The joint metric is
\begin{align}
 ds^2={}&-4F_c\dd U_c\dd V_c+4F_c(\dd P^2+\dd Q^2)
       +L_c\dd\zeta^2-\frac{\mathfrak a_1}{4\kappa^2}\omega_c^2
       -\frac{b_c}\kappa\omega_c\dd\zeta\nonumber\\
       &-\frac{e_c}\kappa\omega_c\eta_c+h_c\eta_c^2
           +2d_c\eta_c\dd\zeta.
\end{align}
At the joint origin its determinant is $-64F_c^4L_c<0$.
All four $F_c,L_c$ factors have positive interval enclosures on
the small six-parameter box. The compact exterior certificate for
that box covers only
$e^{1/4}\leq\rho\leq e^{3/4}$, $2.75\leq z\leq3.25$,
using 512 cells. It is not an exterior cover to infinity.

\subsubsection{Balanced regular extremal examples}
\label{app:extbalanced}\label{sec:extcert}

To exhibit regular solutions, fix four of the constants in the
limiting ansatz:
\begin{equation}
 Z=5,\qquad r_0=37/50,\qquad\beta=-41/50,\qquad w=-30,
 \label{eq:extfixed}
\end{equation}
and solve the three exact limiting-axis balance equations for
$(c,H_*,v_*)$. For $|d-7/500|\leq10^{-12}$ the root is uniquely
enclosed about
\begin{equation}\begin{aligned}
 c&\simeq0.0048605562993129358582,\\
 H_*&\simeq3.1504581717070573003,\qquad
 v_*\simeq-16.1961121853210728223.
\end{aligned}
 \label{eq:extroot}
\end{equation}
These decimals locate the branch. The metric is defined by the exact
balance equations, whose solution is enclosed by interval arithmetic.
The enclosure proves that for each $d$ in this interval there is one
nearby balanced solution and that it depends analytically on $d$.
The verification method is given in the following subsection.

\begin{proposition}
For~\eqref{eq:extfixed} and $|d-7/500|\leq10^{-12}$, the exactly
balanced coefficient metric has a smooth asymptotically flat vacuum
exterior with two finite-area degenerate horizons of topologies
$S^1\times S^2$ and $S^3$. Its exterior is stably causal and has no
conical or orbifold defects or exterior curvature singularities.
Both horizons, including their poles, have smooth local extensions.
All proper lengths may be multiplied by an arbitrary positive scale.
\label{prop:ext}
\end{proposition}

We prove this by combining exact algebraic cancellations, axis
balance and the local charts of section~\ref{sec:extcharts} with
interval bounds over the whole exterior. The interval calculations
bound the relevant functions at every point in each cell and include
rounding errors. We describe the inequalities and their treatment at
the boundaries below. The result covers this small one-dimensional parameter interval,
together with the free overall scale. It does not establish regularity
for arbitrary values of all constants in the extremal ansatz.

The same complete checks pass at seven separated values of $d$ listed
in table~\ref{tab:extareas}. At each value, interval arithmetic isolates the balance solution and
verifies the same denominator, signature and causality inequalities
throughout the exterior, including the axis, horizon and asymptotic
limits. Their ring areas range from a very small
probe to a component larger than the central horizon. 
\begin{table}[htbp]\centering
\begin{tabular}{rrr}\toprule
 $d$ & $\cA_\R$ & $\cA_\Hh$\\\midrule
 $0.00014$ & $1.0096235\times10^{-8}$ & $33.271729$\\
 $0.0014$ & $1.0096032\times10^{-6}$ & $33.270247$\\
 $0.014$ & $1.0094017\times10^{-4}$ & $33.255455$\\
 $0.14$ & $0.010075599$ & $33.109905$\\
 $1.4$ & $1.0046408$ & $31.871001$\\
 $7$ & $31.809353$ & $29.998732$\\
 $14$ & $291.33197$ & $38.124311$\\\bottomrule
\end{tabular}
\caption{Two nonzero extremal horizon areas in the canonical scale
at seven globally certified balance roots. Both surface gravities
vanish exactly. Displayed areas are rounded from interval enclosures.}
\label{tab:extareas}
\end{table}

These examples establish the existence of a regular extremal central
hole surrounded by a regular extremal ring. 

For the parameters in proposition~\ref{prop:ext} and
 table~\ref{tab:extareas}, interval bounds keep $F_0,C^2$ and the
required denominators positive for every polar angle, including the
pole limits. They therefore establish a regular extension over each
complete horizon. We also checked the signed-radius and polar extensions numerically
(footnote~\ref{app:repro}).

\subsubsection{Complete extremal exterior bounds}

Here we verify the inequalities throughout the unbounded exterior.
The compact coordinates~\eqref{eq:compact} turn this task into a
positivity test on a unit square, with the axes, horizons and infinity
on its boundary. The inverse map is
\begin{equation}
 \rho^2=\frac{4Z^2uy(1-y)}{(1-u)^2},\qquad
 z=Z\frac{(1+u)y-u}{1-u},\qquad
 U=\frac{2Zu(1-y)}{1-u},\quad V=\frac{2Z(1-y)}{1-u}.
\end{equation}
The axes are $y=0,1$ and $u=0$, the two horizons are
$(0,0),(0,1)$, and infinity is $u=1$. For a common polynomial
denominator $Q_e$, write $\bar G_{e,ab}=N_{ab}/Q_e$ and
$\bar f_e=N_f/Q_e$. Sign-independent sufficient polynomial
conditions on the open square are
\begin{equation}
 Q_e^2>0,\quad N_fQ_e>0,\quad N_{\Psi\Psi}Q_e>0,\quad
 N_{\Psi\Psi}N_{\Phi\Phi}-N_{\Psi\Phi}^2>0.
 \label{eq:polyineq}
\end{equation}
Equivalent tests can use the denominator of each component separately
to keep the polynomials smaller. On the boundary, known factors such
as the squared radius of a shrinking circle must first be removed
using the exact axis or horizon expansions.

We express each polynomial in a Bernstein basis on the square.
The basis functions are nonnegative and sum to one, so positive
lower bounds on all coefficients imply positivity at every point.
We enclose the coefficients using outward-rounded Arb
arithmetic~\cite{Arb}, keeping exact zeros distinct from intervals
that contain zero. Near one horizon corner, a potentially negative
mixed coefficient is bounded by positive neighbouring terms.
After removing the known boundary zeros, the remaining terms give
strict positivity in the interior.
We check the signs of the remaining factors on each open axis. The exact
limiting-axis kernels and cone factors fix the global angular
lattice and time coordinate. Closed-angle horizon bounds verify
the denominators, $F_0$, $C^2$, and
$\det\Lambda_0/[s(1-s)]$ with $s=(1+\cos\theta)/2$.
The signed horizon and polar charts of section~\ref{sec:extcharts}
then supply the corner cover. All-angle asymptotic expansions
complete the proof at infinity.

These steps prove proposition~\ref{prop:ext} and the analogous
statement at each root in table~\ref{tab:extareas}. They use the
exact balance roots, not rounded substitutions. They establish exterior regularity and local
horizon extension.

\subsubsection{Scope of the broad numerical search}
\label{app:broadscan}
We performed a wider numerical scan by varying all six parameters independently as
\begin{equation}
 p=p_*\mathbin{\odot}(1+\xi_1,\ldots,1+\xi_6),\qquad
 |\xi_i|\leq0.30.
 \label{eq:widebox}
\end{equation}
The scan comprises 321 balanced states: the centre, all 64 corners
and 256 Halton points, with 4,410 exterior samples per state.
Logarithmic radial and endpoint sampling resolves the rods and
approaches the horizons, axes and infinity, while axis and cap tests use
their algebraic limits. All sampled horizons have positive area
and surface gravity, and all axis, cone and cap tests pass.
Independent evaluations of the original BZ metric resolve
ill-conditioned points and confirm all retained sign minima.
The smallest sampled value of
$\det\Lambda/(\cG_{\Psi\Psi}\cG_{\Phi\Phi})$ is approximately
$0.8768$, with $-g^{TT}>1$ at every sample.
The sampled reduced spins include both signs of the second ADM
angular momentum. This provides numerical evidence for a regular
nonextremal regime with all six parameters varying independently.

\section{Blackfold construction of black Saturns}
\label{app:blackfold}\label{sec:blackfoldcomparison}
We construct a thin black ring around a Myers--Perry hole by bending
a Kerr black string into a circle. We use the blackfold equations
of~\cite{BlackfoldDynamics,BlackfoldHorizons} and the treatment of
transverse rotation in~\cite{AHOExt}.
Keeping the string temperature nonzero gives the thermal configurations
shown in figure~\ref{fig:thermal}. Taking it to zero recovers the
extremal construction of~\cite{AHOExt}.

We first impose radial force balance. We then compute the charges
and, when required, impose equality of the two horizon temperatures
and both angular velocities. These are leading blackfold predictions,
whose range of validity we discuss below.

\subsection{Embedding, conventions and local thermodynamics}

Let the MP mass parameter be $\mu$, with spins $a$ in the ring plane
and $b$ in the transverse plane. In the coordinates of~\cite{AHOExt},
we embed the string by
\begin{equation}
 t=\tau,\qquad r=R,\qquad\theta=0,\qquad\chi=\varphi,\qquad
 0\leq\varphi<2\pi .
\end{equation}
The angle conventions are $(\chi,\psi)_{\rm AHO}=(\Psi,\Phi)$ and
their background mass parameter satisfies $m_{\rm AHO}^2=\mu$.
Their potentials $(\Omega,\widehat\Omega)$ are our
$(\Omega_\R^\Psi,\Omega_\R^\Phi)$. Below, $m$ denotes the
\emph{local Kerr-string} mass scale and must not be confused with
$m_{\rm AHO}$. The induced metric on the string and its useful functions are
\begin{align}
 \gamma_{ab}\dd\sigma^a\dd\sigma^b
   &=-N^2\dd\tau^2+\mathcal R^2(\dd\varphi-\varpi\dd\tau)^2,
 &\sqrt{-\gamma}&=N\mathcal R,\nonumber\\
 \Sigma_R&=R^2+b^2,&
 \Delta_R&=(R^2+a^2)(R^2+b^2)-\mu R^2,\nonumber\\
 \mathcal R^2&=R^2+a^2+\frac{\mu a^2}{\Sigma_R},&
 N^2&=\frac{\Delta_R}{\Sigma_R\mathcal R^2},\qquad
 \varpi=\frac{\mu a}{\Sigma_R\mathcal R^2}.
 \label{eq:bfembedding}
\end{align}
Thus $\mathcal R$ is the proper radius of the circular orbit and
$\varpi$ is the background angular shift. For
$\boldsymbol k=\partial_\tau+\Omega\partial_\varphi$, put
\begin{equation}
 v=\frac{\mathcal R(\Omega-\varpi)}N,\qquad
 \Gamma=(1-v^2)^{-1/2},\qquad k=|\boldsymbol k|=\frac N\Gamma.
\end{equation}
Equivalently,
\begin{equation}
 k^2=1-\Omega^2(R^2+a^2)
           -\frac{\mu(1-a\Omega)^2}{R^2+b^2}.
 \label{eq:bfnorm}
\end{equation}
We require $R>r_{+,\Hh}$ and $k^2>0$. The local seed has Kerr rotation parameter $b_s=m\alpha$,
$|\alpha|\leq1$, and horizon radius
\begin{equation}
 \sigma=m\sqrt{1-\alpha^2},\qquad r_s=m+\sigma.
\end{equation}
Its pressure, local temperature and transverse angular velocity are
\begin{equation}
 P=-\frac{m}{2G_5},\qquad
 \mathcal T=\frac{\sigma}{4\pi m r_s},\qquad
 \omega_{\rm loc}=\frac{\alpha}{2r_s}.
 \label{eq:bfseed}
\end{equation}
In~\cite{AHOExt} the local horizon radius $r_0$ and Kerr parameter
$\widehat b$ are therefore $(r_s,m\alpha)$. Stationarity redshifts
the local potentials as
$\mathcal T=T/k$ and $\omega_{\rm loc}=\widehat\Omega/k$.
For $\kappa=2\pi T$ and $D=\sqrt{\kappa^2+\widehat\Omega^2}$,
inverting these relations gives
\begin{equation}
 \alpha=\frac{\widehat\Omega}{D},\qquad
 m=\frac{k}{2(D+\kappa)},\qquad r_s=\frac{k}{2D}.
 \label{eq:bfseedinverse}
\end{equation}
Here $\kappa\geq0$ and the sign of $\widehat\Omega$ fixes the transverse
spin orientation.

\subsection{Stationary action and mechanical balance}

At leading order we use the stationary free-energy functional
of~\cite{BlackfoldDynamics,AHOExt}. For our circular embedding it is
the pressure integral
\begin{equation}
 \mathcal F=-\int_0^{2\pi}\dd\varphi\,\sqrt{-\gamma}\,P
   =\frac{\pi N\mathcal Rm}{G_5}
   =\frac{\pi N\mathcal R k}{2G_5(D+\kappa)}.
 \label{eq:bffunctional}
\end{equation}
At $T=0$ it becomes
$\mathcal F=\pi N\mathcal Rk/(2G_5|\widehat\Omega|)$,
the leading extremal functional of~\cite{AHOExt}. Spin-connection
and self-gravity corrections are omitted consistently at this order. We vary $R$ while holding $(T,\Omega,\widehat\Omega)$ and
$(\mu,a,b)$ fixed. Since $D+\kappa$ is then constant, the embedding
equation is simply
\begin{equation}
 \frac{\dd}{\dd R}(N\mathcal R k)=0 .
 \label{eq:bfradialvariation}
\end{equation}
We see that radial balance is independent of the local
Kerr spin at leading order. To solve it explicitly, define
\begin{equation}
 K_{00}=\frac{N'}N,\qquad K_{11}=-\frac{\mathcal R'}{\mathcal R},
 \qquad K_{01}=\frac{\mathcal R\varpi'}{2N},
 \qquad {}'=\frac{\dd}{\dd R}.
\end{equation}
Using $k=N\sqrt{1-v^2}$ in~\eqref{eq:bfradialvariation} gives
\begin{equation}
 A_vv^2+2K_{01}v+C_v=0,\qquad
 A_v=2K_{11}-K_{00},\qquad C_v=2K_{00}-K_{11}.
 \label{eq:bfbalance}
\end{equation}
Thus the mechanically balanced branches are
\begin{equation}
 v_\pm=\frac{-K_{01}\pm\sqrt{K_{01}^2-A_vC_v}}{A_v},
 \qquad \Omega_\pm=\varpi+\frac{Nv_\pm}{\mathcal R}.
 \label{eq:bforbits}
\end{equation}
We keep only real roots with $|v_\pm|<1$. If $A_v=0$, we use the
linear equation~\eqref{eq:bfbalance} instead. In flat space these reduce
to $v=\pm1/\sqrt2$ and $\Omega=\pm1/(\sqrt2R)$.
Substitution of an extremal MP background gives the circular-ring
balance equation used in section~3.2 of~\cite{AHOExt}.
Choosing an extremal central background and $\kappa>0$ constructs
an extremal hole with a nonextremal probe ring. Conversely,
$\kappa=0$ in a nonextremal background constructs an extremal ring
around a nonextremal hole. 

\subsection{Thermodynamic charges and the mass convention}

On a stationary embedding, the free-energy derivatives do not contain
a contribution from the change of $R$, because
$\partial_R\mathcal F=0$. Therefore
\begin{equation}
 J_{\R\Psi}=-\frac{\partial\mathcal F}{\partial\Omega},\qquad
 J_{\R\Phi}=-\frac{\partial\mathcal F}{\partial\widehat\Omega},
 \qquad S_\R=-\frac{\partial\mathcal F}{\partial T},
\end{equation}
with the background held fixed. Equations~\eqref{eq:bfseedinverse}
and~\eqref{eq:bffunctional} give
\begin{align}
 J_{\R\Psi}&=\frac{\pi m\mathcal R^2\Gamma^2v}{G_5},&
 J_{\R\Phi}&=\frac{2\pi\mathcal R\Gamma m^2\alpha}{G_5},
 \label{eq:bfcharges}\\
 S_\R&=\frac{2\pi\mathcal F}{D}
      =\frac{4\pi^2\mathcal R\Gamma m r_s}{G_5},&
 \cA_\R&=16\pi^2\mathcal R\Gamma m r_s .
 \label{eq:bfentropy}
\end{align}
The intensive quantities can also be written in local-seed form:
\begin{equation}
 \kappa_\R=\frac{k\sigma}{2mr_s},\qquad
 \Omega_\R^\Phi=\frac{k\alpha}{2r_s},\qquad
 \Omega_\R^\Psi=\varpi+\frac{Nv}{\mathcal R},\qquad
 T_\R=\frac{\kappa_\R}{2\pi}.
 \label{eq:bfpotentials}
\end{equation}
In particular $T_\R S_\R+\Omega_\R^\Phi J_{\R\Phi}=\mathcal F$.
The energy conjugate to the background time at infinity is
\begin{equation}
 E_\R=\mathcal F+T_\R S_\R+
               \Omega_\R^\Psi J_{\R\Psi}+\Omega_\R^\Phi J_{\R\Phi}
     =2\mathcal F+\Omega_\R^\Psi J_{\R\Psi}.
\end{equation}
It is important to distinguish this probe energy from a horizon
Komar mass in a background containing another gravitating object.
Using the binding-tension convention of~\cite{AHOExt,ASOTension},
we have
\begin{equation}
 \widehat{\mathcal T}
   =2E_\R-3(T_\R S_\R+\Omega_\R^\Psi J_{\R\Psi}
                         +\Omega_\R^\Phi J_{\R\Phi})
   =\mathcal F-\Omega_\R^\Psi J_{\R\Psi}.
 \label{eq:bftension}
\end{equation}
Their intrinsic mass prescription
$M_{\R,K}=E_\R-\widehat{\mathcal T}/2$ then gives
\begin{equation}
 M_{\R,K}=\frac32(\mathcal F+\Omega_\R^\Psi J_{\R\Psi}),\qquad
 E_\R-M_{\R,K}=\frac{\widehat{\mathcal T}}2,\qquad
 \ell_\R=2N\mathcal R\sigma .
 \label{eq:bfall}
\end{equation}
Here $\ell_\R$ is the leading canonical Weyl rod length.
The local Smarr relation follows immediately. The mass distinction
and its relation to background response are discussed
in~\cite{AHOExt,ASOTension}.

The entropy must be differentiated at finite temperature before
taking $\kappa\to0$. For $\widehat\Omega\ne0$ this limit is finite:
\begin{equation}
 \alpha=\operatorname{sgn}\widehat\Omega,\qquad
 m=r_s=\frac{k}{2|\widehat\Omega|},\qquad
 \cA_\R=8\pi G_5|J_{\R\Phi}|.
 \label{eq:bfextremalentropy}
\end{equation}
Setting $T=0$ in the action before differentiating would lose this
entropy.

\subsection{Solutions in mutual thermal equilibrium}
We now look for solutions of the blackfold construction that satisfy thermodynamic equilibrium.
Write $h=r_{+,\Hh}^2$ for the squared MP horizon radius. Its parameters
and potentials satisfy
\begin{align}
 \mu&=\frac{(h+a^2)(h+b^2)}h,\qquad h>|ab|,\nonumber\\
 \kappa_\Hh&=\frac{h^2-a^2b^2}
 {\sqrt h\,(h+a^2)(h+b^2)},&
 \Omega_\Hh^\Psi&=\frac{a}{h+a^2},\qquad
 \Omega_\Hh^\Phi=\frac{b}{h+b^2}.
 \label{eq:bfcentralpotentials}
\end{align}
The strict inequality selects a nonextremal outer horizon.
For mutual equilibrium set
$(\kappa,\Omega,\widehat\Omega)=
(\kappa_\Hh,\Omega_\Hh^\Psi,\Omega_\Hh^\Phi)$.
The transverse seed is then fixed:
\begin{equation}
 D_\Hh=\sqrt{\kappa_\Hh^2+(\Omega_\Hh^\Phi)^2},\qquad
 \alpha=\frac{\Omega_\Hh^\Phi}{D_\Hh},\qquad
 m=\frac{k}{2(D_\Hh+\kappa_\Hh)}.
 \label{eq:bfthermal}
\end{equation}
The remaining radial equation has a particularly simple analytic
form. Let $x=R^2$, assume $a\ne0$, and define
\begin{equation}
 x_-=\frac{a^2b^2}{h},\qquad
 x_L=\frac{h(h+a^2)}{a^2}-b^2 .
\end{equation}
Then $x_-<h<x_L$, and direct substitution
in~\eqref{eq:bfnorm} gives
\begin{equation}
 k_\Hh^2=
 \frac{a^2(x-h)(x_L-x)}{(h+a^2)^2(x+b^2)},\qquad
 (N\mathcal Rk_\Hh)^2=
 \frac{a^2(x-h)^2(x-x_-)(x_L-x)}
 {(h+a^2)^2(x+b^2)^2}.
 \label{eq:bfthermalnorm}
\end{equation}
Thus the orbit must lie between the horizon and the outer light
surface, $h<x<x_L$. Taking the logarithmic derivative of the
second expression reduces radial balance to
\begin{equation}
 \frac2{x-h}+\frac1{x-x_-}
       -\frac1{x_L-x}-\frac2{x+b^2}=0.
 \label{eq:bfthermalradius}
\end{equation}
Equivalently, its physical root solves the cubic
\begin{equation}
 2(h+b^2)(x-x_-)(x_L-x)
 +(x_L+x_--2x)(x-h)(x+b^2)=0,\qquad h<x<x_L .
 \label{eq:bfthermalcubic}
\end{equation}
This gives an analytic specification of the equilibrium radius
without a simultaneous fit of the three potentials. The left side
of~\eqref{eq:bfthermalradius} decreases strictly from $+\infty$
to $-\infty$: the first and last terms combine to
$2(h+b^2)/[(x-h)(x+b^2)]$, which decreases, and each remaining
term also decreases. Hence there is exactly one physical radius
for each nonextremal background with $a\ne0$ at this order.
Equations~\eqref{eq:bfthermal}, \eqref{eq:bfcharges}
and~\eqref{eq:bfentropy} then determine its thermodynamics.
The existence of this formal root does not ensure that the resulting
ring is thin enough for the approximation as we shall see below.

This calculation also exposes the obstruction to finding doubly extremal black Saturns in thermodynamic equilibrium.
For a finite-area extremal centre, $h=|ab|>0$ and $x_L=h$, so
\begin{equation}
 k_\Hh^2=-\frac{a^2(x-h)^2}{(h+a^2)^2(x+b^2)}<0
 \quad(x>h).
 \label{eq:bfextremeobstruction}
\end{equation}
A circular ring with the central hole's orbital angular velocity is
then spacelike outside the horizon. No separated, timelike ring in
this leading MP-background construction can be in mutual equilibrium
with that extremal centre. This agrees with the obstruction in
section~3.5 of~\cite{AHOExt}. The same obstruction for the exact Saturn metric obtained in this paper is proved
separately in appendix~\ref{app:extremequilibrium}. 

\subsection{Matching to exact Saturn charges}

For comparison with an exact solution, we choose an isolated MP
having the central horizon's area and its two Komar angular momenta.
With
\begin{equation}
 X=\frac{\cA_\Hh}{2\pi^2},\qquad
 p=\frac{4G_5J_{\Hh\Psi}}\pi,\qquad
 q=\frac{4G_5J_{\Hh\Phi}}\pi,
\end{equation}
its parameters are fixed algebraically:
\begin{equation}
 \mu^3=\frac{(X^2+p^2)(X^2+q^2)}{X^2},\qquad
 (a,b,r_{+,\Hh})=(p,q,X)/\mu,\qquad
 M_{\rm MP}=\frac{3\pi\mu}{8G_5}.
\end{equation}
We use $M_{\rm MP}$ for the mass of this isolated reference hole.
The quantity $M_{\Hh,K}$ instead comes from the Komar integral around
the central horizon of the full Saturn, where the ring's gravitational
field is present. These masses need not agree, even though we have
chosen the same central area and angular momenta.

We denote the blackfold ring's probe energy by $E_\R^{\rm bf}$ and
its horizon Komar mass by $M_{\R,K}^{\rm bf}$. They are the quantities
$E_\R$ and $M_{\R,K}$ defined above, with the superscript identifying
the approximation. Holding the central entropy and spins fixed,
we compare the masses through
\begin{align}
 M_{\rm ADM}-M_{\rm MP}&\simeq E_\R^{\rm bf},\nonumber\\
 M_{\Hh,K}-M_{\rm MP}&\simeq E_\R^{\rm bf}-M_{\R,K}^{\rm bf}.
 \label{eq:bfresponse}
\end{align}
The second relation follows from the first and the exact
ADM--Komar sum~\eqref{eq:sums}.
Equivalently, the sum of the two physical horizon masses includes
the central response relative to the isolated reference. Replacing
that responding central mass by $M_{\rm MP}$ while retaining only
$M_{\R,K}^{\rm bf}$ would omit this contribution.

The canonical rod separation fixes the leading orbital radius through
\begin{equation}
 d_z=\frac{a_4+a_5-a_2-a_3}{2},\qquad
 R^2=2d_z+\frac{\mu-a^2-b^2}{2}.
\end{equation}
The exact ring's $\kappa_\R,\Omega_\R^\Phi$ then determine its local
$m,\alpha$ by~\eqref{eq:bfseedinverse}. Its orbital potential,
Komar mass, spins and area provide independent comparisons, as does
the total mass increment. Along a thin nonextremal sequence, decreasing
the ring rod length from $10^{-4}$ to $10^{-16}$ reduces the relative
orbital-spin error from $1.93\times10^{-5}$ to $1.93\times10^{-17}$.
For the extremal-hole example~\eqref{eq:mixedHexample}, the
blackfold surface gravity differs from the exact ring value by
$4.7\times10^{-4}$ in relative terms. Its thickness-to-radius and
thickness-to-separation ratios are $1.04\times10^{-4}$ and
$1.42\times10^{-4}$, respectively. We discuss the doubly extremal
comparison in section~\ref{sec:phases}.

The thermal curves use the same reference ensemble:
$M^{\rm bf}=M_{\rm MP}+E_\R$,
$J_A^{\rm bf}=J_A^{\rm MP}+J_{\R A}$ and
$S^{\rm bf}=S_{\rm MP}+S_\R$. These leading expressions do not
include a higher-order calculation of the central deformation or
ring self-field. The reference MP quantities are
$J_A^{\rm MP}=\pi\mu(a,b)/(4G_5)$ and
$S_{\rm MP}=\pi^2\mu\sqrt h/(2G_5)$.

\subsection{Validity and the plotted comparisons}
\label{app:blackfoldvalidity}
Here we discuss the validity of the blackfold curves presented in the phase diagrams. The local string thickness must be small compared with the bending
radius, background-curvature radii and separation from the central
horizon. Transverse spin-connection corrections must also be small
compared with the local rotation scale~\cite{AHOExt}.
We monitor $2m/\mathcal R$ and $2m/(R-r_{+,\Hh})$, where the latter uses a
coordinate separation proxy.
A small ring-to-centre mass ratio is not required by itself:
a distant ring can be heavier than the centre while producing a
weak potential there. For thermal matching one generically has
$m\sim L_\Hh$, $E_\R/M_{\rm MP}\sim R/L_\Hh$, but
$G_5E_\R/R^2\sim m/R\ll1$, where $L_\Hh$ is a central length scale.
The local scale ratios and omitted response terms must be assessed
together. The flat-background limit supplies a useful analytic check. For
positive spins the leading isolated-ring relations are
\begin{equation}
 \alpha=8j_\Psi j_\Phi,\qquad
 a_{\rm ring,bf}=\frac{\sqrt2}{4j_\Psi}
       \left(1+\sqrt{1-64j_\Psi^2j_\Phi^2}\right),\qquad
 0\leq8j_\Psi j_\Phi\leq1 .
 \label{eq:isolatedblackfold}
\end{equation}
Since $m/R=1/(16j_\Psi^2)$, a fixed $j_\Phi>0$ implies
\begin{equation}
 j_\Psi\leq\frac1{8j_\Phi},\qquad
 \frac{2m}{R}=\frac1{8j_\Psi^2}\geq8j_\Phi^2 .
 \label{eq:blackfoldthinnessfloor}
\end{equation}
A nonzero fixed transverse spin therefore prevents this isolated
ring from becoming arbitrarily thin. Its leading extremal endpoint
is $j_\Psi=1/(8j_\Phi)$, whereas exact PS gives
$j_\Psi=1/(8j_\Phi)+j_\Phi$, with limiting area
$2\sqrt2j_\Phi$. At $j_\Phi=0.10$ the endpoints are $1.25$ and
$1.35$, respectively. This illustrates sensitivity to corrections
near transverse extremality. The isolated-ring thickness bound
does not apply directly to Saturn, whose reduced spins include
the central contribution.

\paragraph{Accuracy of the thermal blackfold comparison.}
\label{app:thermalbfaccuracy}
At $j_\Psi=1.2$, linear interpolation along each plotted continuation
at fixed $j_\Phi$ gives the following rounded values. The shortfall
is $100(a_{\rm metric}-a_{\rm bf})/a_{\rm metric}$. The last two
columns refer to the Saturn blackfold at the same two total reduced
spins, with $\alpha$ as in~\eqref{eq:bfthermal}.
\begin{center}\small
\begin{tabular}{rrrrrr}\toprule
 $j_\Phi$&$a_{\rm metric}$&$a_{\rm bf}$&Shortfall (\%)&$2m/\mathcal R$&$\alpha$\\\midrule
 0.01&0.578&0.541&6.4&0.066&0.086\\
 0.05&0.548&0.508&7.3&0.067&0.438\\
 0.10&0.417&0.330&21.0&0.076&0.965\\
\bottomrule\end{tabular}
\end{center}
These values diagnose the plotted approximation. On the sampled
$j_\Phi=0.10$, $j_\Psi\geq1$ blackfold segment, the smallest
$2m/\mathcal R$ is approximately $0.0758$. At $j_\Psi=1.2$, its
separation proxy is $2m/(R-r_{+,\Hh})\simeq0.0794$, while
$\alpha\simeq0.965$ is close to the transverse Kerr bound. The
corresponding minimum thickness ratios on the sampled $j_\Phi=0.01$
and $0.05$ segments are approximately $0.000801$ and $0.0198$.

The $j_\Phi=0.10$ Saturn slice is only moderately thin and is close
to the transverse Kerr bound. Its roughly $21\%$ area mismatch at
$j_\Psi=1.2$ is consistent with omitted finite-thickness and
interaction effects, but the leading calculation does not predict
the magnitude or sign of the correction. In five dimensions,
self-gravity enters at the same order as spin-orbit coupling so
adding the latter alone is not a complete next-order result. The exact solution therefore offers a way to test the spin--orbit
contribution derived in~\cite[section~3.2.5]{AHOExt}, provided that
self-gravity effects at the same order are accounted for.
The thick-regime blackfold cusps are therefore extrapolations.

\section{Obstruction to doubly extremal mutual equilibrium}
\label{app:extremequilibrium}

Zero temperature at both horizons does not imply mutual equilibrium:
the two angular velocities must also agree in a common physical
frame. We first prove that this is impossible for a regular member of
the finite double-extremal family in section~\ref{sec:extremal}.
The proof is uniform in its remaining parameters and sign sectors,
before imposing balance. We then describe exact boundary exclusions
and the numerical searches with independently retuned BZ data.
The latter strengthen the evidence that such thermodynamic equilibrium is not possible to achieve for doubly spinning extremal black Saturns.

\subsection{Conventions and the two exact matching equations}

We use the rational version of the coefficient construction in
section~\ref{sec:extremal}, with
\begin{equation}
 \lambda=\tau^2/2,\qquad h=H^2/4,\qquad
 C_3=v\tau,\qquad q_{\rm boost}=\tau^{-1}.
 \label{eq:eqnogo-conventions}
\end{equation}
Here $H=H_*$ and $v=v_*$ in the notation of section~\ref{sec:extremal}, in particular $v$ multiplies $\tau$, rather than $\sqrt\lambda$. The constants
$(Z,c,d,H,v,w,r_0,\beta)$ remain free before balance, and the horizon
points are $z_\R=0$ and $z_\Hh=Z$. This overparameterized description
contains the finite doubly extremal boundary of the six-parameter
balanced family. We do not fix
$Z$, $r_0$, $\beta$, or the value of $w$ used in earlier examples.
Throughout this calculation
\begin{equation}
 Z>0,\qquad r_0>0,\qquad cHv\beta\ne0,
 \qquad |c|,|H|,|v|,|\beta|<\infty .
 \label{eq:eqnogo-domain}
\end{equation}
All signs of $c,H,v,\beta$ are retained. Write the intermediate coefficient metric as
\begin{equation}
 \dd s^2=G_{ab}\dd x^a\dd x^b+f(\dd\rho^2+\dd z^2),
 \qquad x^a=(t,\psi,\phi),\qquad \det G=-\rho^2.
\end{equation}
The symbols in this section refer to this raw coefficient metric.
The physical asymptotic frame is obtained by one constant, nonsingular
change of Killing basis and a finite positive normalization of $f$.
Equality of the two horizon generator rays is invariant under that
common basis change. Therefore, when the raw time components are
nonzero, equality of both physical angular velocities is equivalent to
\begin{equation}
 \omega_{\R\psi}=\omega_{\Hh\psi},\qquad
 \omega_{\R\phi}=\omega_{\Hh\phi}.
 \label{eq:eqnogo-match}
\end{equation}
Individual raw angular velocities need not equal the physically
normalized ones. Exact extraction from the horizon kernels gives
\begin{align}
 \omega_{\R\psi}&=\frac1\beta-
             \frac{c/Z+r_0^{-2}}{4\beta},&
 \omega_{\R\phi}&=\frac1\beta+
             \frac{c/Z+r_0^{-2}}{4\beta},
 \label{eq:eqnogo-ringomega}\\
 \omega_{\Hh\psi}&=-\frac{Hcv}{ZD_\Hh}
                  +\frac{N_\Hh}{8HD_\Hh},&
 \omega_{\Hh\phi}&=-\frac{Hcv}{ZD_\Hh}
                  -\frac{N_\Hh}{8HD_\Hh},
 \label{eq:eqnogo-holeomega}
\end{align}
where
\begin{align}
 D_\Hh&=2cv-Hd,\\
 N_\Hh&=\frac{2(Z+1)}Z H^2c^2v-\frac{H^2dv}Z
       +2wH^2v+\frac{4Hcv^2}Z+4Hd-4cv.
\end{align}
The sum of equations~\eqref{eq:eqnogo-match} fixes $d$, and their
difference fixes $w$:
\begin{align}
 d&=cv\left(\frac2H+\frac\beta Z\right),
 \label{eq:eqnogo-d}\\
 w&=\frac{c^2}{Z^2}+\frac{c}{Zr_0^2}
     -\frac{(Z+1)c^2}Z-\frac{cv}{HZ}
     +\frac{\beta cv}{2Z^2}-\frac{2c}{H^2}
     -\frac{2\beta c}{HZ}.
 \label{eq:eqnogo-w}
\end{align}
Substitution makes both angular-velocity differences vanish
identically. In particular,
\begin{equation}
 D_\Hh=-\frac{\beta Hcv}Z\ne0
 \label{eq:eqnogo-Dnonzero}
\end{equation}
on the matching locus. The locus $D_\Hh=0$ is not an omitted matching
branch: for finite nonzero $Hcv$ the unnormalized hole generator has
zero time component and a nonzero angular sum, whereas the ring
generator in~\eqref{eq:eqnogo-ringomega} has nonzero time component.
Their rays cannot agree. This also explains why clearing a vanishing
normalization denominator must not create spurious solutions.

\subsection{A uniform obstruction on the middle axis}

We can establish the obstruction by examining the exterior axis at
$z=Z/2$. A convenient exact
rational approach is
\begin{equation}
 z=Z/2,\qquad \rho^2=\frac{Z^2s}{(1-s)^2},\qquad s\downarrow0.
 \label{eq:eqnogo-midcurve}
\end{equation}
Insert~\eqref{eq:eqnogo-d}--\eqref{eq:eqnogo-w} directly into the
full Laurent-coefficient metric, before expanding in $s$. Exact
rational arithmetic gives
\begin{equation}
 f(s,Z/2)=
 \frac{s\,P(s\mid c,H,v,Z,r_0,\beta)}
 {8H^4Z^6r_0^4\beta^4(1+s)^7},
 \label{eq:eqnogo-fullf}
\end{equation}
where $P$ is a polynomial. The degree is at most six. Its coefficients are obtained by the finite scalar construction in subsection~\ref{app:extscalar}. We supply the exact coefficient
calculation with the \ancillarylink{ancillary material}. Crucially, the denominator
at $s=0$ is nonzero throughout~\eqref{eq:eqnogo-domain}. Hence
\begin{equation}
 \displaystyle\lim_{\rho\to0}f(\rho,Z/2)=0.
 \label{eq:eqnogo-fzero}
\end{equation}
No sign assumption, numerical minimization, balance equation, or
division by a potentially vanishing leading numerator is used here. For a shorter description of the generic behavior, define
\begin{align}
 k&=\frac{2Z}{H},& C&=2c+\frac{2Z}{r_0^2},\\
 \mathcal E&=C+1-v(\beta+k)-2\beta k-k^2,\\
 \Delta_{\rm eq}&=(\beta v+1)^2-4\beta^4
       -2\beta^2kv-4\beta^3k+2\beta k-3\beta^2k^2.
\end{align}
This $\Delta_{\rm eq}$ is unrelated to the earlier Killing-frame
determinant. The first coefficient of~\eqref{eq:eqnogo-fullf} is
\begin{equation}
 f(s,Z/2)=\frac{c^2\Delta_{\rm eq}\mathcal E^2}
                   {2Z^4k^2\beta^4}s+O(s^2).
 \label{eq:eqnogo-fleading}
\end{equation}
On $\mathcal E=0$ or $\Delta_{\rm eq}=0$ this leading coefficient
vanishes. Equation~\eqref{eq:eqnogo-fullf}, whose full denominator
contains neither factor, still gives~\eqref{eq:eqnogo-fzero}.
Thus these exceptional loci cannot evade the obstruction.

\begin{proposition}[No regular mutual equilibrium in the finite chart]
\label{prop:eqnogo-nogo}
Within~\eqref{eq:eqnogo-domain}, the direct correlated doubly extremal
metric cannot simultaneously have two distinct regular horizons,
a regular spacelike axis between them, a finite nonsingular common
asymptotic Killing frame, and equality of both horizon angular
velocities. The conclusion includes every real sign sector of the
finite chart and therefore every balanced subfamily contained in it.
\end{proposition}

\begin{proof}
At an interior point of a smooth spacelike axis, let $q$ be proper
radial distance. The vanishing rotational Killing field has squared
norm proportional to $q^2$, with a finite positive proportionality
constant. The complementary two-dimensional Killing block has a
finite nonzero Lorentzian determinant. The identity
$\det G=-\rho^2$ then implies $\rho=a q+O(q^3)$ with $a\ne0$.
Consequently the radial base coefficient has a finite strictly
positive axis limit, $f\to a^{-2}>0$. The same necessary property
holds with a finite conical deficit. A divergent or degenerate
complementary block would itself violate the assumed smooth
spacelike axis.

Equations~\eqref{eq:eqnogo-match} force
\eqref{eq:eqnogo-d}--\eqref{eq:eqnogo-w}, and the exact identity
\eqref{eq:eqnogo-fullf} instead gives $f\to0$ at the interior axis
point $z=Z/2$. A finite positive conformal normalization and a
nonsingular Killing-basis change cannot remove that zero. This
contradicts the necessary axis behavior, independently of balance.
\end{proof}

There is also a generic invariant symptom of the failure. For the
fixed raw Killing vector $\xi=\partial_\psi-\partial_\phi$, the exact
metric gives, when $\mathcal E\Delta_{\rm eq}\ne0$,
\begin{equation}
 g(\xi,\xi)=\frac{16Z^2\beta^2}{\Delta_{\rm eq}s}+O(1)
           =\frac{16Z^4\beta^2}{\Delta_{\rm eq}\rho^2}+O(1).
 \label{eq:eqnogo-normpole}
\end{equation}
The norm of a finite fixed linear combination of smooth Killing
fields cannot diverge at a regular axis point. We do not use this
generic residue on its exceptional loci, and do not assume that
$\xi$ is itself a periodic physical rotation. The uniform conformal
identity already excludes those loci without either assumption.

\subsection{Checks and boundary limits}
\label{app:extremeqsigns}
\label{app:extremeqboundaries}
\label{app:extremeqindependent}

We checked the matching and axis identities with 13 exact algebraic
tests, including the exceptional loci. Independent Laurent
calculations at two points in each of the 16 sign sectors of
$(c,H,v,\beta)$ reproduce the horizon generators and axis coefficients
with normalised errors below $4\times10^{-14}$. These checks support
the symbolic identities. They do not give interval bounds on a wider
parameter region. For example, $Z=r_0=c=H=\beta=1$, $v=6$, $d=18$
and $w=-7$ give equal raw angular velocities $(1/2,3/2)$, but
$f=275.625\,\rho^2+O(\rho^4)$ at $z=1/2$. The middle axis is singular.

\paragraph{Limits that remain matched.}
We can extend the obstruction analytically to convergent limits that
satisfy angular matching at every nonzero value of the limiting
parameter. After rescaling the separation to keep the horizon points
distinct, the middle-slice base coefficient belongs to
\begin{equation}
 \mathcal V=\operatorname{span}\left\{
 \frac{s}{(1+s)^7},\frac{s^2}{(1+s)^7},\ldots,
 \frac{s^7}{(1+s)^7}\right\}.
 \label{eq:nogoclosure}
\end{equation}
Constant parameter-dependent normalisations preserve this space.
Finite limits at seven distinct off-axis values of $s$ force the
seven coefficients to converge, since the evaluation matrix is
invertible. The limit therefore still vanishes at $s=0$ and fails
the regular-axis condition. This includes correlated zero or infinite
parameters without exchanging parameter and axis limits. Paths that
become matched only at their endpoint require separate tests.

We examined 53 exact rational boundary paths near
$Z=c=H=r_0=\beta=1$, $v=6$. Seventeen remain matched and 36 perturb
$d,w$ independently. Forty-four retain the axis zero. The other nine
fail the primitive angular identification required for a Saturn.
In a further 192 simultaneous limits extracted directly from the
six-pole metric, we scaled one of $(c,H,v,\beta,r_0,Z)$ as
$\tau^{\pm1}$ in all 16 sign sectors, with $d,w$ fixed by the matching
relations. Ninety-six retain a zero middle-axis coefficient, 80 have
a negative base coefficient and 16 lose the central extremal double
pole. None retains two regular extremal horizons.

\paragraph{Independent retuning of the BZ coefficients.}
We also allowed the physical BZ parameters $(\alpha,b_0,C_3,\eta)$
to vary independently and wrote $\chi=-\alpha b_0$. Two direct
boundary calculations yield exact obstructions. For
$S=1/r_0^2$, $\alpha=(-1+SL)/\beta$, $b_0=\beta$ and $L=\tau^2/2$,
the matching equations are~\eqref{eq:eqnogo-d}--\eqref{eq:eqnogo-w}
with $r_0^{-2}$ replaced by $S$. We obtain
\begin{equation}
 f_{\rm match}=\frac{sP_6(s)}{8H^4Z^6\beta^4(1+s)^7}.
 \label{eq:nogoSboundary}
\end{equation}
This calculation includes $S=0$ imposed before eliminating the Gram
matrix. For the simultaneous scaling $H=H_0/\tau$, matching instead
requires
\begin{equation}
 d=\frac{cv\beta}{Z},\qquad
 w=\frac{c^2}{Z^2}+\frac{cS}{Z}-\frac{(Z+1)c^2}{Z}
                      +\frac{\beta cv}{2Z^2},
\end{equation}
and the rescaled base coefficient is
\begin{equation}
 \widehat f_{\rm match}=\lim_{\tau\to0}\tau^2f_{\rm match}
 =\frac{s\widetilde P_6(s)}{8Z^6\beta^4(1+s)^7}.
 \label{eq:nogoHboundary}
\end{equation}
Both polynomials have degree at most six. Thus the regular-axis
obstruction persists in these two boundary classes. This conclusion
assumes the leading Gram rank used in the extraction. Further rank
degenerations require a separate calculation.

We followed one exceptional cancellation through two more orders,
allowing independent subleading BZ coefficients. The ring then has
at most a simple radial pole, while the central double pole survives.
This excludes a finite-area extremal ring on that branch. A separate
finite-auxiliary-pole ansatz also fails its necessary leading-order
cancellation condition. These are exclusions of specific mechanisms.
The \ancillarylink{ancillary material} retains the ansatz and exact coefficient
calculations.

The wider numerical searches give the following results. We extract
the limiting metric coefficients before testing the horizons.
\begin{center}\small
\begin{tabular}{p{.34\textwidth}rp{.49\textwidth}}\toprule
Search & Paths & Outcome\\\midrule
Independent corrections to $\chi$ and $\eta$, through order $\tau^6$
&98&
30 pass necessary base tests. All belong to the finite or $S=0$
classes already excluded by the matched-axis obstruction.\\[3pt]
Also scale one of $c,H,v,\beta,Z$ as $\tau^{\pm1}$
&200&
15 pass base tests. Four have divergent central angular velocities
and eleven retain finite mismatches.\\[3pt]
Finite auxiliary pole with independently varied leading BZ parameters
&64&
None retains a central extremal double pole. The measured radial
orders of $f$ are zero or $-1$.\\\bottomrule
\end{tabular}
\end{center}
We checked representative exclusions by increasing the precision
and truncation order and comparing exact rational calculations with
the direct six-pole metric. Full parameter lists are in the \ancillarylink{ancillary
material}.

No searched branch retains two regular extremal horizons with equal
angular velocities. The exact no-go result covers the finite family
of section~\ref{sec:extremal} and the boundary classes specified above.
These wider searches provide numerical evidence for the nonexistence of doubly spinning extremal Saturns in thermodynamic equilibrium.

\end{document}